\documentclass{aa}    

\usepackage{upgreek}
\usepackage{color}

\usepackage[varg]{txfonts}
\usepackage{microtype}
\usepackage{booktabs}
\usepackage{xcolor}
\usepackage{hyperref}
\usepackage{graphicx}
\usepackage{multirow}
\usepackage{float}

\hypersetup{%
  colorlinks=true, 
  linkcolor=blue, 
  citecolor=blue
}

\usepackage{etoolbox}

\let\oldcitet=\citet

\renewcommand{\citet}[1]{\textcolor[rgb]{0,0,1}{\oldcitet{#1}}}

\newcommand{\SI}{S\,{\sc i}}

\newcommand{\HII}{H\,{\sc ii}}

\begin{document} 

\title{Bottlenecks to interstellar sulfur chemistry}

\subtitle{Sulfur-bearing hydrides in UV-illuminated gas and grains}

 \titlerunning{Bottlenecks to interstellar sulfur chemistry} 
\authorrunning{Goicoechea et al.} 
                                      
 \author{J.\,R.\,Goicoechea\inst{1}
          \and
        A. Aguado\inst{2}
         \and
        S. Cuadrado\inst{1} 
        \and
        O. Roncero\inst{1}
        \and
        J. Pety\inst{3}
        \and
        E. Bron\inst{4}
        \and\\
        A. Fuente\inst{5}
        \and         
        D. Riquelme\inst{6}
        \and
        E. Chapillon\inst{3,7}
        \and
        C. Herrera\inst{3}
        \and
        C. A. Duran\inst{6,8}
}

 \institute{Instituto de F\'{\i}sica Fundamental
     (CSIC). Calle Serrano 121-123, 28006, Madrid, Spain.
              \email{javier.r.goicoechea@csic.es}
 \and   
 Facultad de Ciencias. Universidad Aut\'onoma de Madrid, 28049 Madrid, Spain. 
 \and
Institut de Radioastronomie Millim\'etrique (IRAM), Grenoble, France.
 \and
 LERMA, Observatoire de Paris, PSL Research University, CNRS, Sorbonne Universit\'es, 92190 Meudon, France.
 \and
 Observatorio Astron\'omico Nacional (OAN), Alfonso XII, 3, 28014 Madrid, Spain.
 \and
 Max-Planck-Institut für Radioastronomie, Auf dem Hügel 69, 53121 Bonn, Germany.
 \and
 OASU/LAB-UMR5804, CNRS, Universit\'e Bordeaux, 33615 Pessac, France.
 \and
European Southern Observatory, Alonso de Cordova 3107, Vitacura, Santiago, Chile.}
   
   \date{Received 25 October 2020 / Accepted 23 December 2020}


\abstract{Hydride molecules lie at the base of interstellar chemistry, but the
 synthesis  of sulfuretted hydrides  is poorly understood and their abundances often crudely constrained. Motivated by new observations of the  \mbox{Orion Bar}  photodissociation region (PDR) -- 1$''$ resolution ALMA images of  SH$^+$; IRAM\,30m detections of  bright H$_{2}^{32}$S, H$_{2}^{34}$S, and H$_{2}^{33}$S lines;  H$_3$S$^+$ (upper limits); and SOFIA/GREAT observations of SH \mbox{(upper limits)} -- we perform a systematic study of the chemistry of sulfur-bearing hydrides.  We self-consistently  determine their column densities  using \mbox{coupled} excitation, radiative transfer as well as chemical formation and destruction models. We revise some of the key gas-phase reactions that lead to their chemical synthesis. This includes \mbox{\textit{ab initio}} quantum calculations of the vibrational-state-dependent reactions \mbox{$\rm SH^++H_2({\it{v}}) \rightleftarrows H_2S^++H$} and \mbox{$\rm S\,+\,H_2\,({\it{v}}) \rightleftarrows SH \,+\, H$}. We find that reactions of \mbox{UV-pumped} H$_2$($v$\,$\geq$\,2) molecules with S$^+$ ions explain the presence of SH$^+$  in a high thermal-pressure gas component, $P_{\rm th}/k$\,$\approx$\,10$^8$\,cm$^{-3}$\,K, close 
to the H$_2$ dissociation front (at \mbox{$A_V$\,$<$\,2\,mag}). These PDR layers 
 are characterized by no or very little depletion of elemental sulfur from the gas. However, subsequent hydrogen abstraction reactions of SH$^+$, H$_2$S$^+$, and S atoms with  vibrationally excited H$_2$, fail to form enough H$_2$S$^+$, H$_3$S$^+$, and SH to  ultimately explain the observed H$_2$S column density  ($\sim$2.5$\times$10$^{14}$\,cm$^{-2}$, with an ortho-to-para ratio of 
 \mbox{2.9\,$\pm$\,0.3}; consistent with the high-temperature statistical value).
\mbox{To overcome} these bottlenecks, we build PDR models that include a simple network of \mbox{grain surface}  reactions leading to the formation of solid H$_2$S  \mbox{(s-H$_2$S)}. The higher adsorption binding energies of S and SH  
suggested by recent studies imply that S atoms adsorb on  grains (and form \mbox{s-H$_2$S})  at warmer dust temperatures (\mbox{$T_d$\,$<$\,50\,K}) and closer to the UV-illuminated edges of molecular clouds. 
We show that  everywhere \mbox{s-H$_2$S}  mantles form(ed), gas-phase H$_2$S emission lines will be detectable.  Photodesorption and,  to a lesser extent, chemical desorption,  produce roughly the same H$_2$S column density \mbox{(a few  10$^{14}$\,cm$^{-2}$)} and abundance peak 
\mbox{(a few 10$^{-8}$)} nearly independently of $n_{\rm H}$ and $G_0$. This agrees with the  observed H$_2$S column density in the Orion Bar as well as at the edges of dark clouds without  invoking substantial depletion of elemental sulfur abundances.
 
}


   \maketitle
%

\section{Introduction}

 Hydride molecules play a pivotal role in \mbox{interstellar chemistry} \citep[e.g.,][]{Gerin16}, being among the first molecules to form in diffuse interstellar  clouds and at the \mbox{UV-illuminated} edges of dense star-forming clouds,  
so-called photodissociation regions \citep[PDRs;][]{Hollenbach97}. 
\mbox{Sulfur} is on the top ten list of  most abundant cosmic elements and it is particularly relevant for astrochemistry and star-formation studies. Its low 
\mbox{ionization potential} (10.4\,eV) makes the photoionization of S atoms a dominant source of electrons in molecular gas  at   \mbox{intermediate} visual  extinctions  \mbox{$A_V$\,$\simeq$\,2\,-\,4\,mag} \citep[][]{Sternberg95,Goico09,Fuente16}. 

The sulfur abundance, \mbox{[S/H]}, in diffuse clouds \citep[e.g.,][]{Howk06} is
very close to the \mbox{[S/H]} measured in the \mbox{solar} photosphere 
\mbox{\citep[\mbox{${\rm [S/H]}_\odot$\,$\simeq$\,1.4$\times$10$^{-5}$};][]{Asplund09}}.
Still, the observed abundances of S-bearing molecules in diffuse  and translucent molecular clouds ($n_{\rm H}$\,$\simeq$\,$10^2$\,$-$\,$10^3$\,cm$^{-3}$) make up a very small fraction, \mbox{$<$\,1\,$\%$}, of the sulfur nuclei 
\citep[mostly locked as~S$^+$;][]{Tieftrunk94,Turner96,Lucas02,Neufeld15}. 
In colder dark clouds and  dense cores shielded from stellar UV radiation,  most sulfur is expected  in molecular form. However, the result of adding the abundances of all detected gas-phase \mbox{S-bearing} molecules is typically a factor of 
\mbox{$\sim$10$^2$-10$^3$} lower than [S/H]$_\odot$
\mbox{\citep[e.g.,][]{Fuente19}}. Hence, it is \mbox{historically} assumed that sulfur species deplete on grain mantles at cold temperatures and high densities \citep[e.g.,][]{Graedel82,Millar90,Agundez13}. 
However, recent chemical models predict that the major sulfur reservoir in dark clouds can
be either gas-phase neutral S atoms \mbox{\citep[][]{Vidal17,Almaida20}} or organo-sulfur species trapped on grains \mbox{\citep{Laas19}}. Unfortunately, it is difficult to overcome this
 \mbox{dichotomy} from an observational perspective.
In particular, no  ice carrier of an abundant sulfur reservoir other than solid OCS 
 \citep[hereafter \mbox{s-OCS}, with an abundance of $\sim$10$^{-8}$ with respect to
  H~nuclei;][]{Palumbo97} has been convincingly  identified. Considering the large abundances of water ice (\mbox{s-H$_2$O}) grain mantles  in dense  molecular clouds and cold protostellar envelopes \citep[see reviews by][]{vD04,Gibb04,Dartois05}, one may also expect  hydrogen \mbox{sulfide} (\mbox{s-H$_2$S}) to be the dominant sulfur reservoir.  \mbox{Indeed}, \mbox{s-H$_2$S} is the most abundant \mbox{S-bearing} ice in comets such as \mbox{67P/Churyumov–Gerasimenko} \citep[][]{Calmonte16}.
However, only upper limits to the \mbox{s-H$_2$S} abundance of $\lesssim$1\,\% relative to water ice have so far been  estimated toward a few interstellar sightlines \mbox{\citep[e.g.,][]{Smith91,Escobar11}}.
These values  imply a maximum \mbox{s-H$_2$S} ice abundance of \mbox{several 10$^{-6}$} with respect to H nuclei. Still, this upper limit could be higher if  \mbox{s-H$_2$S} ices are well mixed with \mbox{s-H$_2$O} and \mbox{s-CO} ices \citep{Brittain20}.

The bright rims of molecular clouds \mbox{illuminated} by nearby massive stars are intermediate environments between  diffuse and cold dark clouds. Such environments host the transition from ionized S$^+$ to neutral atomic S, as well as the gradual formation of S-bearing molecules \citep[][]{Sternberg95}. In one prototypical low-illumination PDR, the edge of the \mbox{Horsehead} nebula, \citet{Goico06} inferred very modest  gas-phase sulfur depletions.  In addition, the detection of narrow sulfur \mbox{radio} recombination lines in dark clouds \citep[implying the presence of S$^+$;][]{Pankonin78} is an argument against large sulfur depletions in the mildly illuminated surfaces of these clouds.  \mbox{The presence} of new S-bearing molecules such as S$_2$H, the first (and so far only) doubly sulfuretted species detected in a PDR \citep[][]{Fuente17},  
suggests that the chemical pathways leading to the synthesis of sulfuretted species are not well constrained; and  that the list of  S-bearing molecules is likely not complete.

\begin{figure*}[ht]
\centering   
\includegraphics[scale=0.47, angle=0]{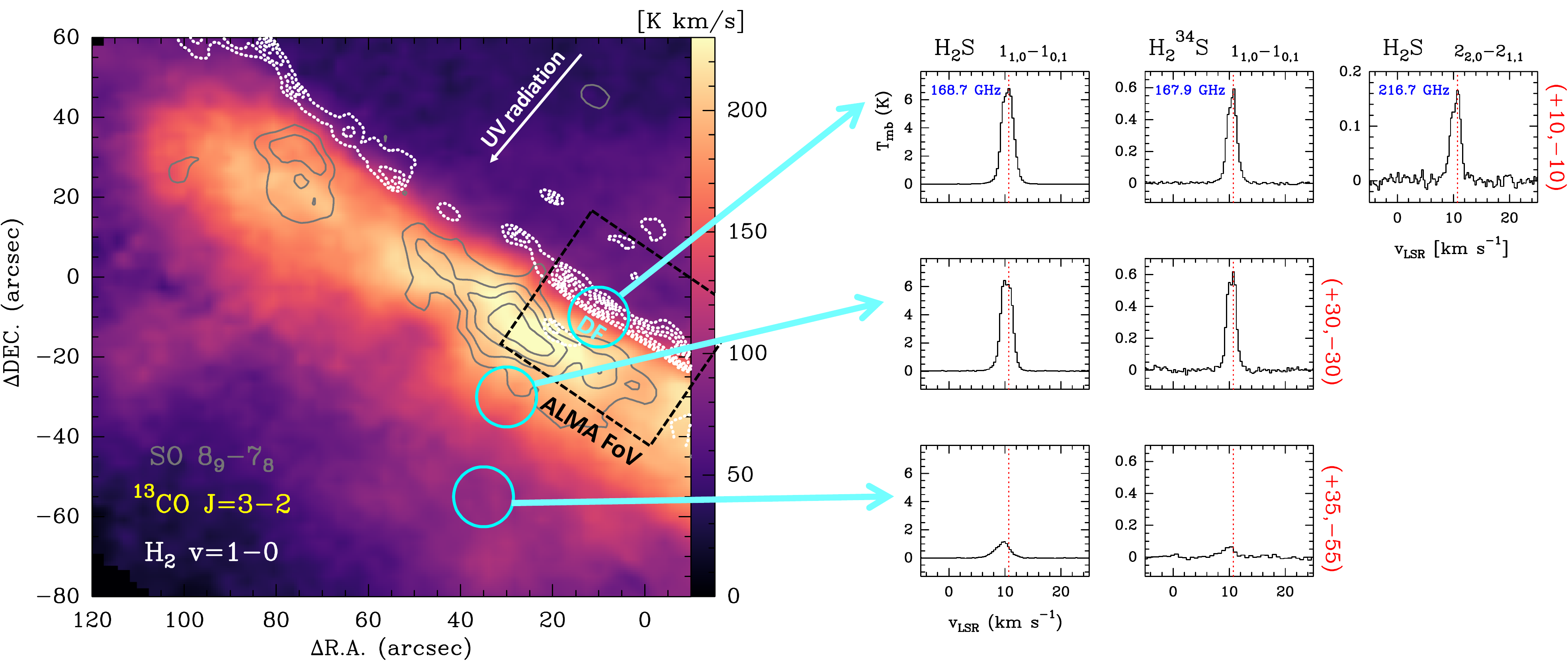}
\caption{Overview of the Orion Bar. The (0$''$, 0$''$) position 
corresponds to \mbox{$\mathrm{\alpha_{2000}=05^{h}\,35^{m}\,20.1^{s}\,}$}; \mbox{$\mathrm{\delta_{2000}=-\,05^{\circ}25'07.0''}$}. \textit{Left panel}:
Integrated line intensity maps in the $^{13}$CO\,$J$\,=\,3-2  (color scale) and SO 8$_9$-7$_8$ emission (gray contours; from 6 to \mbox{23.5\,K\,km\,s$^{-1}$} in steps of \mbox{2.5\,K\,km\,s$^{-1}$}) obtained with the IRAM\,30\,m telescope at 8$''$ resolution.
The white dotted contours delineate the position of the H$_2$ dissociation front as traced by the infrared H$_2$ $v$\,=\,1–0 $S$(1) line \citep[from 1.5 to \mbox{4.0\,$\times$\,10$^{-4}$\,erg\,s$^{-1}$\,cm$^{-2}$\,sr$^{-1}$} in steps of \mbox{0.5\,$\times$\,10$^{-4}$\,erg\,s$^{-1}$\,cm$^{-2}$\,sr$^{-1}$}; from][]{Walmsley_2000}.
The black-dashed rectangle shows the smaller FoV imaged with ALMA (Fig.~\ref{fig:SHp-ALMA}). The DF position has been observed with SOFIA, IRAM\,30\,m, and Herschel. Cyan circles represent the $\sim$15$''$ beam  at 168\,GHz. 
\textit{Right panel}:  H$_2$S lines lines detected toward three positions of the Orion Bar.} 
\label{fig:IRAM-map}
\end{figure*}

Interstellar sulfur chemistry is unusual compared to that of  other elements in that 
none of the simplest species, \mbox{X=S, S$^+$, SH, SH$^+$, or H$_2$S$^+$}, 
react exothermically with \mbox{H$_2$\,($v$\,=\,0)} in the initiation reactions \mbox{X + H$_2$ $\rightarrow$ XH + H} (so-called \mbox{hydrogen} abstraction reactions). 
Hence, one would expect a slow sulfur chemistry and very low abundances of  \mbox{SH$^+$ (sulfanylium)} and  \mbox{SH (mercapto)} radicals in cold interstellar gas. However,  H$_2$S \citep{Lucas02},  SH$^+$ \citep{Menten11,Godard12}, and
 SH \citep{Neufeld12,Neufeld15}  have been detected 
in low-density diffuse clouds \mbox{($n_{\rm H}\lesssim100$\,cm$^{-3}$)} through \mbox{absorption} measurements of their ground-state rotational lines\footnote{SH was first reported by IR spectroscopy toward the cirumstellar envelope around the evolved star \mbox{R Andromedae} \citep{Yamamura20}.}.  In \mbox{UV-illuminated gas}, most sulfur atoms are ionized, but the very high \mbox{endothermicity} of 
reaction
\begin{equation}
\label{reac-1}
{\rm S^+\,(^4{\it{S}}) + H_2\,(^1\Sigma^+, \nu=0) \rightleftarrows SH^+\,(^3\Sigma^-) + H\,(^2{\it{S}})}   
\end{equation}    
\citep[$E\,/\,k$\,=\,9860\,K, e.g.,][]{Zanchet13a,Zanchet19} prevents this reaction from 
 being efficient unless the gas is heated to very high temperatures. 
In  diffuse molecular clouds \mbox{(on average at \mbox{$T_{\rm k}$\,$\sim$\,100\,K)}}, the formation of SH$^+$ and SH only seems possible in the context of local regions of overheated gas \mbox{subjected} to magnetized shocks \citep[][]{PdF86} or in dissipative vortices  of the interstellar turbulent cascade    \citep{Godard12,Godard14}. In these tiny pockets ($\sim$100\,AU in size), the gas would attain the hot temperatures
($T_{\rm k}\,$$\simeq$\,1000\,K) and/or ion-neutral drift needed to overcome the endothermicities of the 
above hydrogen abstraction reactions \mbox{\citep[see, e.g.,][]{Neufeld15}}.

Dense PDRs \mbox{($n_{\rm H}$\,$\simeq10^3-10^6$\,cm$^{-3}$)}  offer a  complementary environment to study the first steps of sulfur chemistry. Because of their higher densities  and more quiescent gas, 
fast shocks or turbulence dissipation do not contribute
to the gas heating. Instead, the molecular gas is heated to \mbox{$T_{\rm k}\lesssim$\,500\,K} by mechanisms  that depend on the flux of \mbox{far-UV} photons 
(FUV; $E$\,$<$13.6\,eV). A different perspective of the \mbox{H$_2$\,($v$)}  reactivity emerges because certain  endoergic reactions  
become exoergic and fast when a significant fraction of the H$_2$
reagents  are radiatively pumped to  vibrationally excited states \mbox{$v\geq1$}
\citep{Stecher72,Freeman82,Tielens_1985a,Sternberg95}. In this case, state-specific reaction rates for \mbox{H$_2$\,($\nu,J$)} 
are needed to make realistic predictions of the abundance of the product XH  \citep[][]{Agundez10,Zanchet13b,Faure17}. 
The presence of abundant \mbox{FUV-pumped} \mbox{H$_2$\,($v$\,$\geq$\,1)} triggers a  \mbox{nonthermal} ``hot'' chemistry. 
Indeed, CH$^+$ and SH$^+$ emission lines have been detected in the  Orion Bar PDR \mbox{\citep{Nagy13,Goico17}}
where H$_2$ lines  up to \mbox{$v$\,=\,10} 
 have been detected as well \citep{Kaplan17}.

In this study we present a systematic (observational and modeling) study of the chemistry of S-bearing \mbox{hydrides} in  \mbox{FUV-illuminated} gas. We try to answer the question of whether gas-phase reactions of S atoms and SH$^+$ molecules with vibrationally excited H$_2$ can ultimately explain the presence of abundant H$_2$S, or if grain surface chemistry has to  be invoked.

The paper is organized as follows. In Sects.~\ref{sec:observations} and \ref{sec-results}  we report on new observations of  H$_{2}^{32}$S, H$_{2}^{34}$S, H$_{2}^{33}$S, SH$^+$, SH, and H$_3$S$^+$  \mbox{emission} lines  toward the Orion Bar. In Sect.~\ref{sec:MTC_mods} we study their excitation and derive their column densities.
In Sect.~\ref{sec:pdr-mods} we discuss their abundances
in the context of updated PDR models, with emphasis on the 
role of hydrogen abstraction reactions\\
\begin{equation}
\label{reac-2}
{\rm SH^+\,(^3\Sigma^-) + H_2\,(^1\Sigma^+) \rightleftarrows H_2S^+\,(^2A') + H\,(^2{\it{S}})}, 
\end{equation}
\begin{equation}  
\label{reac-3} 
 \rm{H_2S^+\,(^2{\it{A'}}) + H_2\,(^1\Sigma^+)  \rightleftarrows H_3S^+\,(X^1{\it{A_{\rm 1}}}) + H\,(^2{\it{S}})}, 
\end{equation} 
\begin{equation}
\label{reac-4}
 \rm{S\,({\it{^3P}}) + H_2\,(^1\Sigma^+) \rightleftarrows SH\,(X^2\Pi) + H\,({\it{^2S}})}, 
\end{equation} 
 photoreactions, and grain surface chemistry. In Sect.~\ref{sec-rates} we summarize the 
\mbox{\textit{ab initio}} quantum calculations we carried out to
 determine the state-dependent rates of reactions (\ref{reac-2}) and (\ref{reac-4}). Details of these calculations are given in Appendices~\ref{Apendix:ab-initio} and \ref{Apendix:S_H2}.

\section{Observations of S-bearing hydrides}
\label{sec:observations}

\subsection{The Orion Bar} 

At an adopted distance of $\sim$414\,pc, the Orion Bar is an interface of the Orion molecular cloud and the Huygens \HII~region that surrounds the Trapezium cluster \citep[][]{Genzel89,Odell01,Bally08,Goico19,Goico20,Pabst19,Pabst20}. 
The Orion Bar is a prototypical strongly illuminated dense PDR. The impinging flux of stellar FUV photons ($G_0$) is a few 10$^4$ times the mean interstellar radiation field 
\citep[][]{Habing68}. The Bar is seen nearly edge-on with respect to the FUV illuminating sources, mainly $\theta^1$ Ori\,C, the most  massive star in the Trapezium. This favorable orientation allows observers to spatially resolve
the \mbox{H$^+$-to-H} transition  \citep[the ionization front or IF; see, e.g.,][]{Walmsley_2000,Pellegrini09} from the \mbox{H-to-H$_2$} transition \citep[the dissociation front or DF; see, e.g.,][]{Allers_2005,vanderWerf96,vanderWerf_2013,Wyrowski97,Cuadrado19}. It also allows one to study the stratification of different molecular species as
a function of cloud depth 
\citep[i.e., as the flux of FUV photons is attenuated; see, e.g.,][]{Tielens_1993,Wiel09,Habart10,Goico16,Parikka17,Andree17}.
  
Regarding sulfur\footnote{Sulfur has four stable isotopes, in decreasing order of abundance: \mbox{$^{32}$S ($I_{\rm N}$\,=\,0)}, 
\mbox{$^{34}$S ($I_{\rm N}$\,=\,0)}, 
\mbox{$^{33}$S ($I_{\rm N}$\,=\,3/2)}, and 
\mbox{$^{36}$S ($I_{\rm N}$\,=\,0)}, where $I_{\rm N}$ is the \mbox{nuclear} spin. The most abundant isotope is here simply  referred to as S.}, several studies  previously reported the detection of S-bearing 
molecules in the Orion Bar.
These include CS, C$^{34}$S, SO, SO$_2$, and H$_2$S \citep{Hoger95,Jansen95}, 
SO$^+$ \citep{Fuente03}, C$^{33}$S, HCS$^+$, H$_2$CS, and NS \citep{Leurini06}, and
SH$^+$ \citep{Nagy13}. These detections refer to modest angular resolution pointed observations using single-dish  telescopes.  
Higher-angular-resolution interferometric imaging of 
SH$^+$, SO, and SO$^+$ \citep{Goico17} was possible thanks to the  \textit{Atacama Compact Array} (ACA). 

\subsection{Observations of H$_2$S isotopologues and H$_3$S$^+$} 

We observed the Orion Bar with the IRAM\,30\,m telescope at Pico Veleta (Spain). We used the EMIR receivers in combination with the Fast Fourier Transform Spectrometer (FTS) backends at 200\,kHz resolution ($\sim$0.4\,km\,s$^{-1}$,
$\sim$0.3\,km\,s$^{-1}$, and $\sim$0.2\,km\,s$^{-1}$ at $\sim$168\,GHz, $\sim$217\,GHz, and $\sim$293\,GHz, respectively). These observations are part of a complete line survey covering the frequency range \mbox{80\,$-$\,360\,GHz} \mbox{\citep{Cuadrado15,Cuadrado16,Cuadrado17,Cuadrado19}} and include deep integrations at 168~GHz toward  three positions of the PDR located at a distance of 14$''$, 40$''$, and 65$''$ from the IF (see Fig.~\ref{fig:IRAM-map}). Their offsets  with respect to the IF position at \mbox{$\mathrm{\alpha_{2000}=05^{h}\,35^{m}\,20.1^{s}\,}$}, 
\mbox{$\mathrm{\delta_{2000}=-\,05^{\circ}25'07.0''}$}
are \mbox{(+10$''$, -10$''$)}, \mbox{(+30$''$, -30$''$')}, and \mbox{(+35$''$, -55$''$)}. The first position is the DF.  

We carried out these observations in the position switching mode taking a distant reference position at \mbox{($-$600$''$, 0$''$)}.  The half power beam width (HPBW) at $\sim$168\,GHz, $\sim$217\,GHz, and $\sim$293\,GHz  is $\sim$15$''$, $\sim$11$''$, and $\sim$8$''$, respectively.  The latest observations (those at 168\,GHz) were performed in \mbox{March 2020}. The data were first calibrated in the antenna temperature scale $T^{*}_{\rm A}$ and then  converted to the main beam temperature scale, $T_{\rm mb}$, using \mbox{$T_{\rm mb}$ = $T^{*}_{\rm A}/ \upeta_{\rm mb}$}, where $\upeta_{\rm mb}$ is the antenna efficiency \mbox{($\upeta_{\rm mb}$ = 0.74} at $\sim$168\,GHz). We reduced and analyzed the data using the GILDAS software  
as described in \citet{Cuadrado15}. The typical rms noise of the spectra  is  
\mbox{$\sim$3.5, 5.3, and 7.8\,mK} per velocity channel at $\sim$168\,GHz, $\sim$217\,GHz, and $\sim$293\,GHz, respectively. \mbox{Figures~\ref{fig:IRAM-map}} and \ref{fig:H233S_detection} show the detection of \mbox{$o$-H$_2$S $1_{1,0}-1_{0,1}$} (168.7\,GHz), \mbox{$p$-H$_2$S $2_{2,0}-2_{1,1}$} (216.7\,GHz), and \mbox{$o$-H$_2$$^{34}$S $1_{1,0}-1_{0,1}$} lines (167.9\,GHz) (see Table~\ref{Table_H2S} for the line parameters), as well as several \mbox{$o$-H$_2$$^{33}$S $1_{1,0}-1_{0,1}$}  hyperfine lines (168.3 GHz). 

 We complemented our  dataset with higher frequency \mbox{H$_2$S}   lines  detected by the \textit{Herschel Space Observatory} \citep{Nagy17}
 toward the  ``CO$^+$ peak'' position \mbox{\citep{Stoerzer95}}, which is located at only $\sim$4$''$ from our DF position (i.e., within the HPBW of these observations). These observations were carried out with the HIFI receiver \citep{deGra10} at a spectral-resolution of 1.1\,MHz (0.7\,km\,s$^{-1}$ at 500\,GHz). HIFI's HPBW 
 range from  $\sim$42$''$ to $\sim$20$''$
 in the \mbox{500\,-\,1000\,GHz}  window  \citep[][]{Roelfsema12}.
The list of additional hydrogen sulfide lines detected by Herschel includes the \mbox{$o$-H$_2$S $2_{2,1}-2_{1,2}$} \mbox(505.5\,GHz),
\mbox{$2_{1,2}-1_{0,1}$} \mbox(736.0\,GHz), and \mbox{$3_{0,3}-2_{1,2}$} \mbox(993.1\,GHz), as well as the \mbox{$p$-H$_2$S} \mbox{$2_{0,2}-1_{1,1}$} \mbox(687.3\,GHz) line. We used the  line intensities, in the $T_{\rm mb}$ scale, shown in \mbox{Table A.1} of \citet{Nagy17}. 

In order to get a global view of the Orion Bar, we also obtained  \mbox{2.5$'$\,$\times$\,2.5$'$} maps of the region observed by us with the IRAM\,30\,m telescope using the 330\,GHz EMIR receiver and the FTS backend at 200 kHz spectral-resolution ($\sim$0.2\,km\,s$^{-1}$). \mbox{On-the-fly} (OTF) scans were obtained along and perpendicular to the Bar. The resulting spectra were gridded to a data cube through convolution with a Gaussian kernel providing a final resolution of $\sim$8$''$.
The total integration time was $\sim$6\,h. The achieved rms noise is $\sim$1\,K per resolution channel. Figure~\ref{fig:IRAM-map} shows the spatial distribution of the $^{13}$CO\,$J$=3-2 (330.5\,GHz) and
SO\,8$_9$-7$_8$ (346.5\,GHz) integrated line intensities.

\subsection{ALMA imaging of Orion Bar edge in SH$^+$ emission}\label{sec-ALMA-obs}

We carried out  mosaics of a small field of the Orion Bar using twenty-seven ALMA \mbox{12 m} antennas in \mbox{band 7} (at $\sim$346\,GHz).  These unpublished observations belong to project 2012.1.00352.S (\mbox{P.I.: J. R. Goicoechea}) and consisted of a 27-pointing mosaic centered at \mbox{$\alpha$(2000) = 5$^h$35$^m$20.6$^s$};  \mbox{$\delta$(2000) = -05$^o$25$'$20$''$}.
 The total field-of-view (FoV) is 58$''$$\times$52$''$ 
 (shown in Fig.~\ref{fig:IRAM-map}). The two hyperfine line components of the 
 \mbox{SH$^+$ $N_J$\,=\,$1_0-0_1$} transition 
  were observed with correlators providing $\sim$500\,kHz resolution (0.4~km\,s$^{-1}$) over a 937.5~MHz bandwidth. The total observation time with the ALMA\,12\,m array was $\sim$2h. In order to recover the large-scale extended emission filtered out by the interferometer, we used deep and fully sampled single-dish maps, obtained with the total-power (TP) antennas at 19$''$ resolution,
as zero- and short-spacings. Data calibration procedures and image synthesis steps are described in \cite{Goico16}. The synthesized beam  is $\sim$1$''$. 
This  is a factor of $\sim$4 better than previous interferometric SH$^+$ observations \citep{Goico17}. Figure~\ref{fig:SHp-ALMA}  shows the resulting image of the 
\mbox{SH$^+$ $1_0-0_1$} \mbox{$F$\,=\,1/2-3/2} hyperfine emission line at 345.944\,GHz.
We rotated this image 37.5$^o$ clockwise to bring the FUV illumination in the horizontal direction. The typical rms noise of the final cube is $\sim$\,80\,mK per velocity channel and \mbox{1$''$-beam}.
As expected from their Einstein coefficients, the other  \mbox{$F$\,=\,1/2-1/2} hyperfine line component at 345.858\,GHz is a factor of $\sim$2 fainter 
(see Table~\ref{Table_SHp}) and the resulting image has low signal-to-noise (S/N).

We complemented the SH$^+$ dataset with the higher frequency 
 lines  observed by HIFI \citep[][]{Nagy13,Nagy17}  at $\sim$526\,GHz and $\sim$683\,GHz 	\mbox{(upper limit)}. These pointed observations  have HPBWs of $\sim$41$''$ and $\sim$32$''$ respectively, thus they do not spatially
resolve the  SH$^+$ emission. To determine their beam coupling factors ($f_{\rm b}$), we smoothed the \mbox{bigger 4$''$-resolution} ACA\,+\,TP SH$^+$ image shown in \cite{Goico17} to the different HIFI's  HPBWs. We obtain 
$f_{\rm b}$\,$\simeq$\,0.4 at $\sim$526\,GHz and $f_{\rm b}$\,$\simeq$\,0.6 at $\sim$683\,GHz.  
The corrected intensities are computed as \mbox{$W_{\rm corr}$\,=\,$W_{\rm HIFI}$\,/\,$f_{\rm b}$}.
These correction factors are only a factor of $\lesssim$\,2 lower than simply assuming uniform SH$^+$ emission from a 10$''$ width filament.

\subsection{SOFIA/GREAT search for SH emission} 

We finally used the GREAT receiver
\citep[][]{Heyminck12} on board the \textit{\mbox{Stratospheric} Observatory For Infrared Astronomy} \citep[SOFIA;][]{Young12} to search for the lowest-energy rotational lines of SH \mbox{($^2\Pi_{3/2}$\,$J$\,=\,3/2-1/2)} at
1382.910 and 1383.241\,GHz \citep[e.g.,][]{Klisch96,Martin-Drumel12}. These lines lie in a frequency gap that Herschel/HIFI could not observe from space.
These SOFIA observations belong to project $07\_0115$ (\mbox{P.I.: J. R. Goicoechea}). The SH lines were searched on the lower side band of  4GREAT band\,3. 
We employed the 4GREAT/HFA frontends and 4GFFT spectrometers as backends.
The HPBW of SOFIA at 1.3\,THz is $\sim$20$''$, thus comparable with IRAM\,30\,m/EMIR and Herschel/HIFI observations.  We also employed the total power mode with a reference position at ($-$600$''$,0$''$). The original plan was to observe during two flights in November 2019 but due to bad weather conditions, only $\sim$70\,min of observations were carried out in a single flight. 

After calibration, data reduction included: removal of a first order spectral baseline, dropping scans with problematic receiver response, rms weighted average of the spectral scans, and calibration to  $T_{\rm mb}$ intensity scale \mbox{($\upeta_{\rm mb}$ = 0.71)}. 
The final spectrum, smoothed to a velocity-resolution of 1\,km\,s$^{-1}$  has a rms noise of $\sim$50\,mK (shown in Fig.~\ref{fig:SH-spectra}). Two emission peaks are seen at the  frequencies of the \mbox{$\Lambda$-doublet} lines. Unfortunately, the achieved rms  is not enough to assure the unambiguous detection of each component of the doublet. 
Although the stacked spectrum does display a single line (suggesting a tentative detection) the resulting line-width \mbox{($\Delta$v\,$\simeq$\,7\,km\,s$^{-1}$)} 
is a factor of $\sim$3 broader than expected in the Orion Bar (see Table~\ref{Table_SH}). Hence,  this spectrum provides stringent upper limits to the SH column density but deeper integrations would be needed to confirm the detection.

\section{Observational results}\label{sec-results}

\begin{figure}[t]
\centering   
\includegraphics[scale=0.50, angle=0]{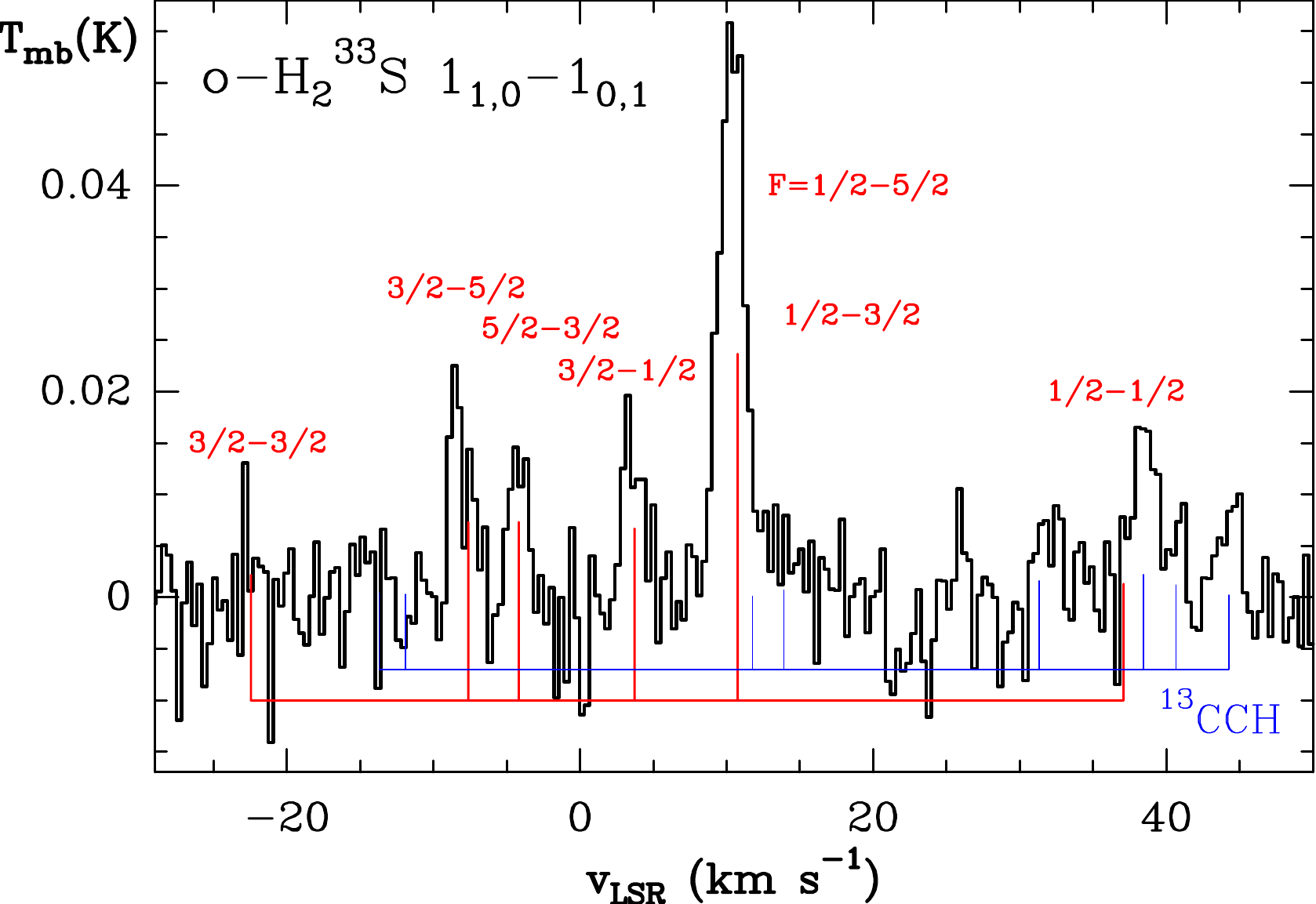}
\caption{Detection of H$_{2}$$^{33}$S (at $\sim$168.3 GHz) toward the DF position
of the Orion Bar. Red lines indicate hyperfine components.
Blue lines show interloping lines from $^{13}$CCH. The length of each line
is proportional to the transition line strength \citep[taken from the Cologne Database for Molecular Spectroscopy, CDMS;][]{Endres16}.}
\label{fig:H233S_detection}
\end{figure}

\begin{figure*}[ht]
\centering   
\includegraphics[scale=0.55, angle=0]{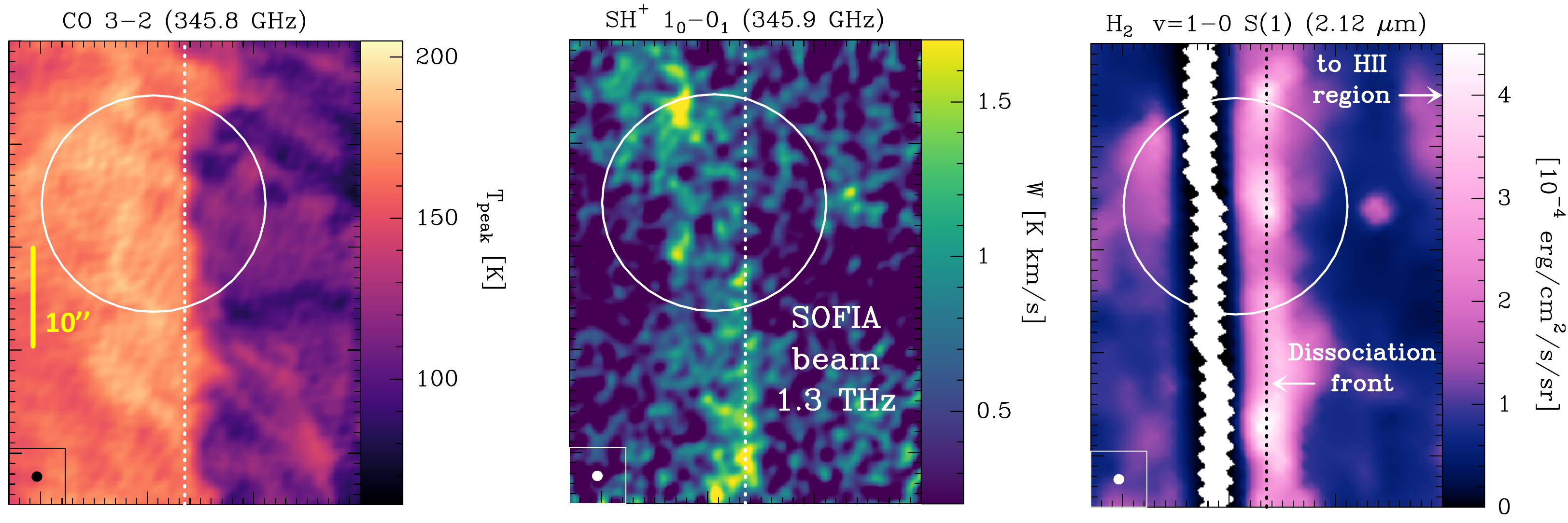}
\caption{ALMA  1$''$-resolution images zooming into the edge of the Orion Bar  in
\mbox{$^{12}$CO 3-2} \citep[\textit{left panel},][]{Goico16} and \mbox{SH$^+$ 1$_0$-0$_1$ $F$\,=\,1/2-3/2} line (\textit{middle panel}, integrated line intensity). The \textit{right panel} shows the  H$_2$ $v$\,=\,1–0 $S$(1) line \citep{Walmsley_2000}.
We rotated these images (all showing the same FoV) with respect to Fig.~\ref{fig:IRAM-map} to bring the FUV illuminating direction  in the horizontal direction (from the right). The circle shows the DF position targeted with SOFIA in SH (20$''$ beam) and with the IRAM\,30m telescope in H$_2$S and H$_3$S$^+$.}
\label{fig:SHp-ALMA}
\end{figure*}

\subsection{H$_2$$^{32}$S, H$_2$$^{34}$S, and H$_2$$^{33}$S across the PDR} 

Figure~\ref{fig:IRAM-map} shows an expanded view of the Orion Bar in the \mbox{$^{13}$CO\,($J$\,=\,3-2)} emission. FUV radiation from the Trapezium stars
comes from the upper-right corner of the image. The FUV radiation field is attenuated in the direction perpendicular to the Bar.  The infrared \mbox{H$_2$ $v$\,=\,1–0 $S$(1)} line emission (white contours) delineates the position of the \mbox{H-to-H$_2$ transition}, the DF. Many molecular species, such as SO, specifically emit from deeper inside the PDR  where the flux of FUV photons has considerably decreased. In contrast, H$_2$S, and even its isotopologue H$_{2}^{34}$S, show bright \mbox{1$_{1,0}$-1$_{0,1}$} line emission toward the DF \citep[right panels in Fig.~\ref{fig:IRAM-map}; see also][]{Jansen95}. Rotationally excited H$_2$S lines have been also detected toward this position \citep{Nagy17}, implying the presence of
warm H$_2$S close to the irradiated cloud surface (i.e., at relatively low extinctions).
 The presence of moderately  large H$_2$S column densities in the PDR  is also  demonstrated by the unexpected detection of the rare isotopologue  H$_{2}^{33}$S toward the DF (at the
correct LSR velocity of the PDR: \mbox{v$_{\rm LSR}$\,$\simeq$\,10.5\,km\,s$^{-1}$}). Figure~\ref{fig:H233S_detection} shows the H$_{2}^{33}$S  \mbox{1$_{1,0}$-1$_{0,1}$} line and its hyperfine splittings (produced by the $^{33}$S nuclear spin).
To our knowledge, H$_{2}^{33}$S lines had only been reported toward the hot cores in
Sgr~B2 and Orion~KL  before \citep{Crockett14}.

The observed  \mbox{$o$-H$_2$S/$o$-H$_{2}^{34}$S 1$_{1,0}$-1$_{0,1}$} line intensity ratio toward the DF is \mbox{15\,$\pm$\,2}, below the solar isotopic ratio of \mbox{$^{32}$S/$^{34}$S\,$=$\,23} \citep[e.g.,][]{Anders89}. The observed ratio thus  implies optically thick  $o$-H$_2$S  line emission at $\sim$168\,GHz. However, the observed \mbox{$o$-H$_{2}^{34}$S/$o$-H$_{2}^{33}$S 1$_{1,0}$-1$_{0,1}$}  intensity ratio is \mbox{6\,$\pm$\,1}, thus compatible with the solar isotopic ratio (\mbox{$^{34}$S/$^{33}$S\,$=$\,5.5})  and with H$_{2}^{34}$S and H$_{2}^{33}$S optically thin emission.

\begin{figure}[b]
\centering   
\includegraphics[scale=0.5, angle=0]{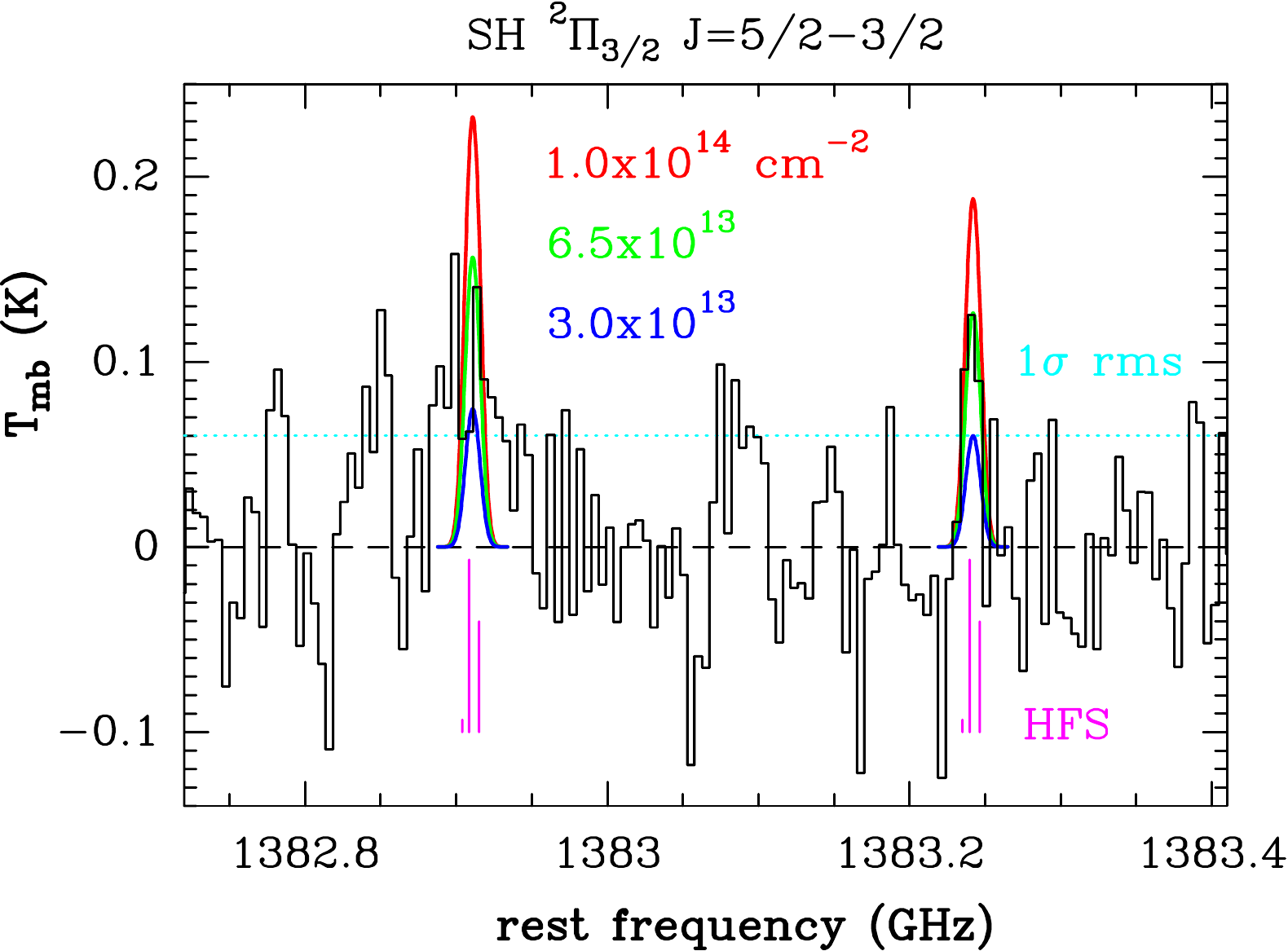}
\caption{Search for the SH $^2$$\Pi_{3/2}$ $J$=5/2-3/2 doublet \mbox{(at $\sim$1383\,GHz)} toward the DF position with SOFIA/GREAT. Vertical magenta lines indicate the position of hyperfine splittings taken from CDMS.}  
\label{fig:SH-spectra}
\end{figure}

\subsection{SH$^+$ emission from the PDR edge} 

Figure~\ref{fig:SHp-ALMA} zooms into a small field of the Bar edge. The ALMA image of the CO\,$J$\,=\,3-2 line peak temperature was first presented by \cite{Goico16}. Because the CO\,$J$\,=\,3-2 emission is
nearly thermalized and optically thick from the DF to the  molecular cloud interior, the line peak temperature scale ($T_{\rm peak}$) is a good proxy of the gas temperature ($T_{\rm k}$\,$\simeq$\,$T_{\rm ex}$\,$\simeq$\,$T_{\rm peak}$).
The CO image implies small temperature variations  around \mbox{$T_{\rm k}$\,$\simeq$\,200\,K}. The middle panel in Fig.~\ref{fig:SHp-ALMA} shows the ALMA image of the SH$^+$ \mbox{$N_J$\,=\,1$_0$-0$_1$ $F$\,=\,1/2-3/2}  hyperfine  line at 345.944\,GHz. Compared to CO, the SH$^+$ emission follows the edge of the molecular PDR, akin to a  filament of $\sim$10$''$ width 
\citep[for the spatial distribution of other molecular ions, see,][]{Goico17}. 
The SH$^+$ emission shows localized small-scale emission
peaks (density or column density enhancements) that match, or are very close to, the vibrationally excited \mbox{H$_2$ ($v$\,=\,1-0)} emission (Fig.~\ref{fig:SHp-ALMA}). We note that while some \mbox{H$_2$ ($v$\,=\,1-0)} emission  peaks likely coincide
 with gas density enhancements \mbox{\citep[e.g.,][]{Burton90}}, the region also shows extended emission from  \mbox{FUV-pumped} \mbox{H$_2$ ($v$\,=\,2-1)}
 \citep[][]{vanderWerf96} that does not necessarily coincide with the \mbox{H$_2$ ($v$\,=\,1-0)} emission peaks.

\subsection{Search for SH, H$_3$S$^+$, and H$_2$S~$\nu_2$\,=\,1 emission}

 We used SOFIA/GREAT to search for \mbox{SH $^2$$\Pi_{3/2}$ $J$=5/2-3/2} lines toward the DF (Fig.~\ref{fig:SH-spectra}). This would have been the first time that interstellar  SH rotational lines were seen in emission. 
\mbox{Unfortunately}, the achieved rms of the observation does not allow a definitive confirmation of these lines, so here we will only discuss upper limits to the SH column density. 
The red, green, and blue curves in Fig.~\ref{fig:SH-spectra} show radiative transfer models for  $n_{\rm H}$\,=\,10$^6$\,cm$^{-3}$, $T_{\rm k}$\,=\,200\,K, and different SH column densities (see Sect.~\ref{sec:MTC_mods} for more details).

\begin{figure}[h]
\centering   
\includegraphics[scale=0.62, angle=0]{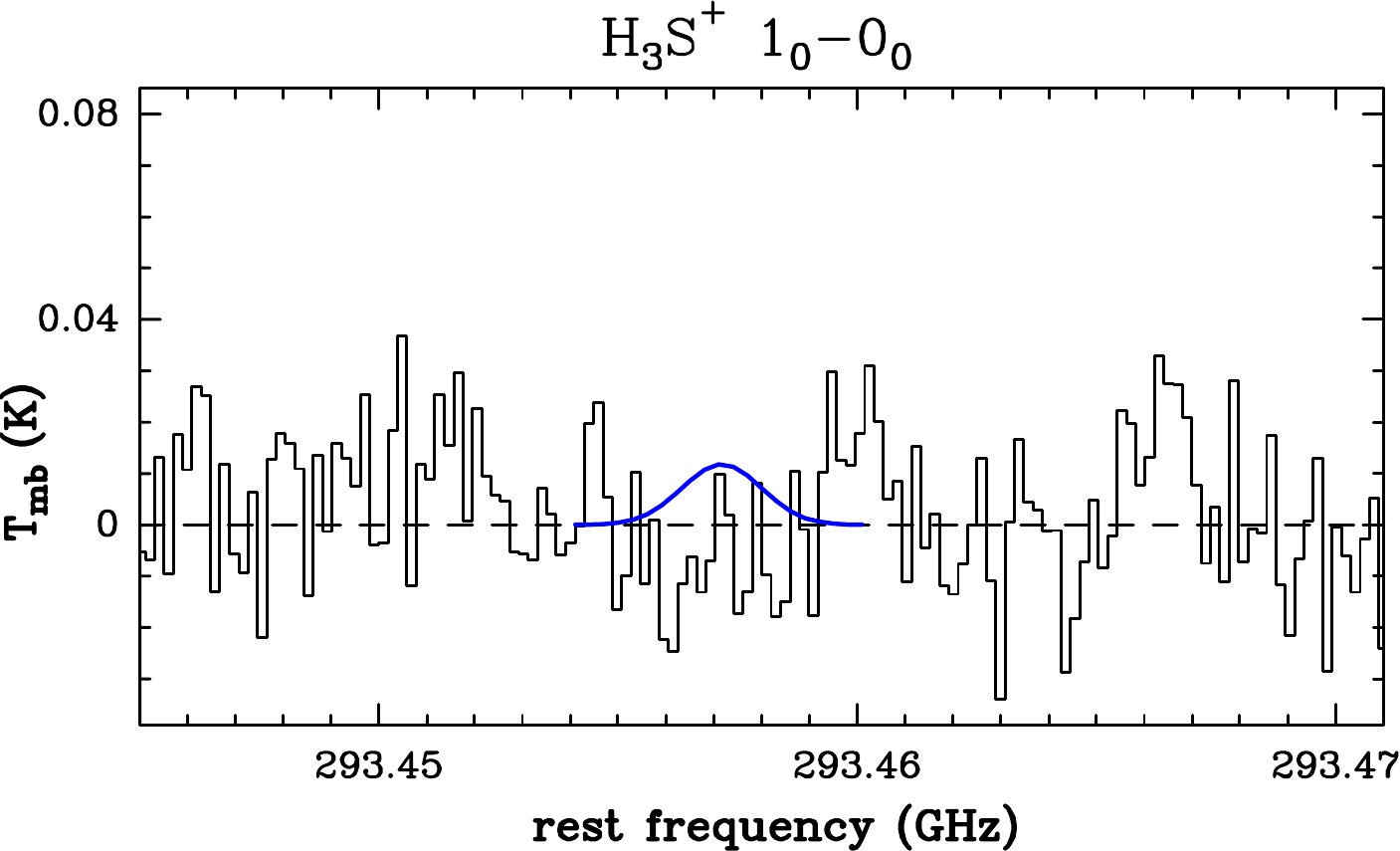}
\caption{Search for H$_3$S$^+$  toward the Orion Bar 
with the IRAM\,30\,m telescope. The blue curve shows the expected position of the line.}
\label{fig:h3sp}
\end{figure}

Our IRAM\,30\,m observations toward the DF  neither resulted in a detection of 
\mbox{H$_3$S$^+$}, a key gas-phase precursor of H$_2$S.
The $\sim$293.4\,GHz spectrum around the targeted \mbox{H$_3$S$^+$\,$1_0$-$0_0$} line is shown in Fig.~\ref{fig:h3sp}. Again, the achieved low rms allows us to provide a sensitive upper limit to the H$_3$S$^+$ column density. This results in  $N$(H$_3$S$^+$)\,=(5.5-7.5)$\times$10$^{10}$\,cm$^{-2}$ (5$\sigma$) assuming an excitation temperature range \mbox{$T_{\rm ex}$\,=\,10-30\,K} and extended emission.
Given the bright H$_2$S emission close to the edge of the Orion Bar, and because H$_2$S formation at the DF might be driven by very exoergic processes, we also searched for the \mbox{1$_{1,0}$-1$_{0,1}$}  line of vibrationally excited H$_2$S (in the bending mode $\nu_2$). The frequency of this line lies at $\sim$181.4\,GHz \citep{Azzam13}, thus at the 
end  of our 2\,mm-band observations of the DF (\mbox{rms\,$\simeq$\,16\,mK}). 
However, we do not detect this line either.

\section{Coupled nonlocal excitation and chemistry}\label{sec:MTC_mods}

In this section we study the rotational excitation of the observed 
\mbox{S-bearing} hydrides\footnote{Readers interested only in the chemistry of these species and in depth-dependent PDR models could directly jump to Section~\ref{sec:pdr-mods}.}. We determine the SH$^+$, SH (upper limit), and H$_2$S \mbox{column} densities in the Orion Bar, and the ``average'' gas physical conditions in the sense that we search for the combination of single $T_{\rm k}$, $n_{\rm H}$, and $N$ that better reproduces the observed line intensities (so-called \mbox{``single-slab''} approach). 
In \mbox{Sect.~\ref{sec:pdr-mods}}  
we expand these excitation models to \mbox{multi-slab} calculations that take into account the expected steep gradients in a PDR.

In the ISM, rotationally excited levels are typically populated by inelastic collisions. 
However, the lifetime of very reactive molecules can  be so short that the details of their formation and destruction need to be taken into account when determining how these levels are actually populated \citep[][]{Black98}.
Reactive collisions (collisions that lead to  a  reaction and thus to molecule destruction) influence the excitation of these species when their timescales become comparable to those of nonreactive collisions. The lifetime of  
 reactive molecular ions observed in PDRs \citep[e.g.,][]{Fuente03,Nagy13,Tak13,Goico17,Goico19} can be so short that they do not get thermalized  by nonreactive collisions 
or by absorption of the background radiation field \citep{Black98}. In these cases, a proper treatment of the molecule excitation requires including chemical formation and destruction rates in the statistical equilibrium equations \mbox{(d$n_i$\,/\,d$t$\,=\,0)} that determine the level populations:
\begin{eqnarray}
\sum_{j>i} n_j\,A_{ji} + \sum_{j \neq i} n_i \left(B_{ji}\,\bar{J}_{ji} + C_{ji} \right) + F_i =\\
= n_i \left( \sum_{j<i} A_{ij} + \sum_{j \neq i} \left(B_{ij}\,\bar{J}_{ij} + C_{ij}  \right) \,+\, D_i \right),
\end{eqnarray}
where \mbox{$n_i$ $[\rm cm^{-3}]$} is the population of rotational level $i$, $A_{ij}$ and $B_{ij}$ are the Einstein coefficients for spontaneous
and induced emission, 
\mbox{$C_{ij}$ $[\rm s^{-1}]$} is the rate of inelastic collisions\footnote{We use the following inelastic collision rate coefficients $\gamma_{ij}$:\\ 
$\bullet$ SH$^+$--\,$e^-$, including hyperfine splittings \citep{Hamilton18}.\\
$\bullet$ SH$^+$--\,$o$-H$_2$ and $p$-H$_2$, including hyperfine splittings \citep{Dagdigian19}.\\
$\bullet$ SH$^+$--\,H, including hyperfine splittings \citep{Lique20}.\\
$\bullet$ $o$-H$_2$S and $p$-H$_2$S with $o$-H$_2$ and $p$-H$_2$ \citep{Dagdigian20}.\\
$\bullet$ SH--\,He, including fine-structure splittings \citep{Klos09}.\\} (\mbox{$C_{ij}$\,=\,$\sum_{k} \gamma_{ij,\,k}\,n_k $}, where \mbox{$\gamma_{ij,\,k}(T)$ $[\rm cm^3 s^{-1}]$} are the collisional rate coefficients  and $k$ stands for  H$_2$, H, and $e^-$), and  $\bar{J}_{ij}$ is the  mean \mbox{intensity} of the total radiation field  over the line profile. In these equations, \mbox{$n_i\,D_i$} is the destruction rate per unit volume of the molecule in level $i$, and \mbox{$F_i$} its formation rate 
per unit volume (both in $\rm cm^{-3} s^{-1}$). When \mbox{state-to-state}  formation rates are not available, and assuming that the destruction rate is the same in every level 
\mbox{($D_i$\,=\,$D$)}, one can use the total destruction rate  \mbox{$D\,[{\rm s^{-1}}]$ ($=\,\sum_{k} n_k\,k_{k}(T)$\,+\, \rm{photodestruction rate,}}
where \mbox{$k_{k}$ $[\rm cm^3 s^{-1}]$} is the state-averaged rate of the two-body chemical reaction with \mbox{species $k$}) and consider that the level populations of the nascent molecule follow a Boltzmann distribution at an effective formation temperature $T_{\rm form}$:\\
\begin{equation}
F_i = F\, g_i \, e^{-E_i/kT_{\rm form}} \,/\, Q(T_{\rm form}).
\end{equation}
In this formalism, \mbox{$F$ $[\rm cm^{-3}\,s^{-1}]$} is the state-averaged formation rate per unit volume, $g_i$ 
the degeneracy of level $i$, and $Q(T_{\rm form})$ is the partition function at $T_{\rm form}$ \citep[][]{Tak07}.

This \mbox{``formation pumping''} formalism has been previously implemented in  large velocity gradient codes to treat, for example, the local excitation of  the very reactive ion CH$^+$ \citep[][]{Nagy13,Godard13,Zanchet13b,Faure17}. 
However, interstellar clouds are inhomogeneous and gas velocity gradients are typically modest at small spatial scales. This means that line photons 
can be absorbed and reemitted several times before leaving the cloud. 
Here we implemented this formalism in a \mbox{Monte Carlo} code that explicitly  models the \mbox{nonlocal behavior}  of the excitation and radiative transfer problem \mbox{\citep[see Appendix of][]{Goico06}}. 

Although radiative pumping  by dust continuum photons does not generally dominate in PDRs,  for completeness we also included radiative excitation by
a modified  blackbody at a dust temperature of $\sim$50\,K and a dust opacity  \mbox{$\tau_{\lambda}$\,=\,0.03\,(150/$\lambda [\upmu{\rm m}]$)$^{1.6}$}
\citep[which reproduces the observed intensity and wavelength dependence of the dust emission in the Bar;][]{Arab_2012}.
The molecular gas \mbox{fraction}, \mbox{$f$(H$_2$)\,=\,2$n$(H$_2$)/$n_{\rm H}$},
is set to 2/3, where \mbox{$n_{\rm H}$\,=\,$n$(H)\,+\,2$n$(H$_2$)} is the total density of H \mbox{nuclei}. This choice is appropriate for the dissociation front and implies \mbox{$n$(H$_2$)\,=\,$n$(H)}. As most electrons in the DF come from the ionization of carbon atoms, the electron density $n_e$ is 
set to \mbox{$n_e$\,$\simeq$\,$n$(C$^+$)\,=\,1.4$\times$10$^{-4}$\,$n_{\rm H}$}
\citep[e.g.,][]{Cuadrado19}. For the inelastic collisions with $o$-H$_2$ and $p$-H$_2$, we assumed that the H$_2$ \mbox{ortho-to-para} (OTP) ratio is thermalized to the gas temperature.

\subsection{SH$^+$ excitation and column density}\label{sec-MTC-SHp}

We start by assuming that the main destruction pathway of SH$^+$    are  reactions with H atoms and recombinations with electrons (see Sect.~\ref{sec:gas_models}).  Hence, the SH$^+$ destruction rate
is \mbox{$D$\,$\simeq$\,$n_e$\,$k_{\rm e}$($T$)\,+\,$n$(H)\,$k_{\rm H}$($T$)} (see \mbox{Table~\ref{rates}} for
the relevant chemical destruction rates). For \mbox{$T_{\rm k}$\,=\,$T_{\rm e}$\,=\,200\,K} and 
$n_{\rm H}$\,=\,10$^6$\,cm$^{-3}$ \mbox{\citep[e.g.,][]{Goico16}} this implies $D$\,$\simeq$\,10$^{-4}$\,s$^{-1}$
(i.e., the lifetime of an SH$^+$ molecule in the Bar is less than 3\,h).
At these temperatures and densities, $D$ is about ten times smaller than the
rate of \mbox{radiative} and inelastic collisional transitions that \mbox{depopulate} the lowest-energy rotational levels of SH$^+$. Hence,  \mbox{formation pumping} does not significantly alter the excitation of the observed SH$^+$ lines, but it does influence the population of 
higher-energy  levels. Formation pumping effects have been readily  seen in CH$^+$ because this
species is more reactive\footnote{CH$^+$ is more reactive than SH$^+$ because 
CH$^+$ does react with H$_2$($v$=0) exothermically producing CH$_{2}^{+}$ at
 $k$\,=\,1.2$\times$10$^{-9}$\,cm$^3$\,s$^{-1}$ \citep{Anicich03} and also because
reaction of CH$^+$ with H is faster, $k$\,=\,7.5$\times$10$^{-10}$\,cm$^3$\,s$^{-1}$.}
and its rotationally excited levels lie at higher-energy 
\citep[i.e., their inelastic collision pumping rates are slower, e.g.,][]{Zanchet13b} 

\begin{figure}[t]
\centering   
\includegraphics[scale=0.6, angle=0]{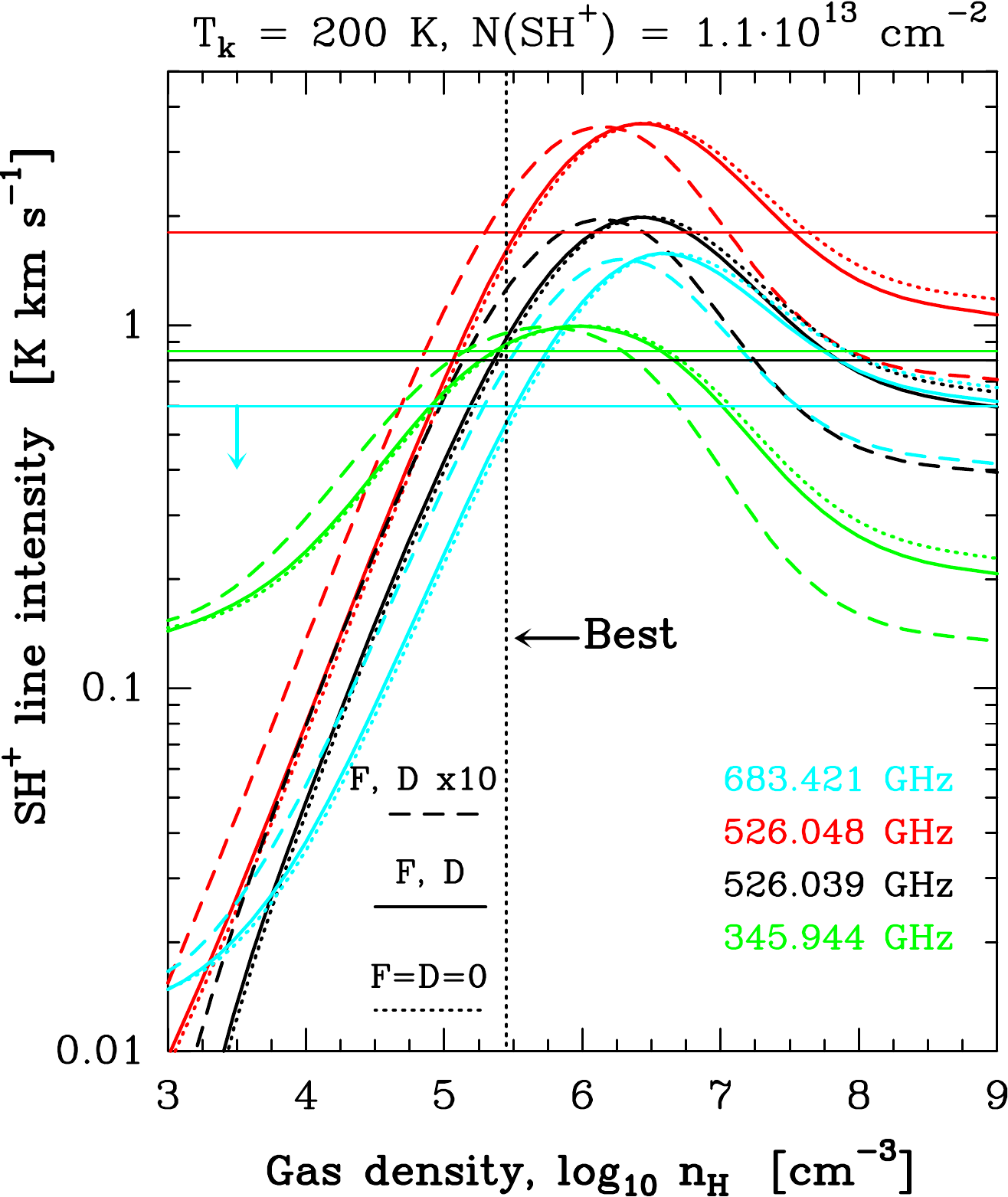}
\caption{Non-LTE excitation models of SH$^+$. 
The horizontal lines mark the observed line intensities in the Orion Bar. 
Dotted curves are for a standard model \mbox{($F=D=0$)}. Continuous curves are for  a model that includes chemical destruction by H atoms and $e^-$
\mbox{(model $F, D$)}. Dashed lines are for
a model in which destruction rates are multiplied by ten (model \mbox{$F, D$ $\times$10}).
The vertical black line marks the best model.} 
\label{fig:MTC-SHp}
\end{figure}

 \mbox{Figure~\ref{fig:MTC-SHp}} shows results of  several models: without \mbox{formation pumping} (dotted curves for model \mbox{``$F=D$\,=\,0''}), adding formation pumping 
 with SH$^+$ destruction by H and $e^-$ (continuous curves for model \mbox{``$F, D$''}), and using a factor of  ten higher SH$^+$  
 destruction rates (simulating a dominant role of SH$^+$ photodissociation or destruction by reactions with vibrationally excited H$_2$; dashed curves for model
  \mbox{``$F, D$ $\times$10''}).
Since the formation of SH$^+$ is driven by \mbox{reaction~(\ref{reac-1})} when H$_2$ molecules are in \mbox{$v$\,$\geq$\,2}, here we adopted  \mbox{$T_{\rm form}$\,$\simeq$\,$E$($v$\,=\,2,\,$J$\,=\,0)\,/\,$k$\,$-$\,9860\,K\,$\approx$\,2000\,K}. 
\mbox{Because} these are constant column density $N$(SH$^+$) excitation and radiative transfer models, we used a normalized formation rate \mbox{$F$\,=\,$\sum F_i$} that assumes \mbox{steady-state} SH$^+$ abundances consistent with the varying gas density in each model. 
That is, \mbox{$F$\,=\,$\sum F_i$\,=\,$x$(SH$^+$)\,$n_{\rm H}$\,$D$} \mbox{$[\rm cm^{-3} s^{-1}]$}, where $x$ refers to the abundance with respect to H nuclei.

The detected SH$^+$ rotational   lines connect the fine-structure levels \mbox{$N_J$\,=\,1$_0$-0$_1$} (345\,GHz) and \mbox{1$_2$-0$_1$} (526\,GHz). Upper limits also exist for the 1$_1$-0$_1$ (683\,GHz) lines. 
SH$^+$ critical densities \mbox{($n_{\rm cr}$\,=\,$A_{ij}$\,/\,$\gamma_{ij}$)} for inelastic collisions with H or H$_2$ are of the same order and equal to several 10$^6$\,cm$^{-3}$. 
As for many molecular ions \citep[e.g.,][]{Desrousseaux21}, \mbox{SH$^+$--H$_2$} (and \mbox{SH$^+$--H}) inelastic collisional rate coefficients$^4$ are large (\mbox{$\gamma_{ij}$\,$\gtrsim$\,10$^{-10}$\,cm$^3$\,s$^{-1}$}). Thus, collisions with H (at low $A_V$) and H$_2$ (at higher $A_V$) generally  dominate over collisions with electrons (\mbox{$\gamma_{ij}$ of a few 10$^{-7}$\,cm$^3$\,s$^{-1}$}).
At low densities (meaning \mbox{$n_{\rm H}$\,$<$\,$n_{\rm cr}$})  formation pumping  increases the population of the higher-energy levels (and their $T_{\rm ex}$), but there are only  minor effects in the  low-energy submillimeter  lines. At high densities, \mbox{$n_{\rm H}$\,$>$\,10$^7$\,cm$^{-3}$},  formation pumping with \mbox{$T_{\rm form}$\,=\,2000\,K} produces lower intensities in these lines  because the lowest-energy levels \mbox{($E_{\rm u}/k$\,$<$\,$T_{\rm k}$\,$<$\,$T_{\rm form}$)} are less populated.

The best fit to the observed lines in model \mbox{F, D} is for \mbox{$N$(SH$^+$)\,$\simeq$\,1.1$\times$10$^{13}$\,cm$^{-3}$}, \mbox{$n_{\rm H}$\,$\simeq$\,3$\times$10$^5$\,cm$^{-3}$}, and 
\mbox{$T_{\rm k}$\,$\simeq$\,200\,K}. This is shown by the  vertical dotted line in \mbox{Fig.~\ref{fig:MTC-SHp}}. This model is consistent with the  upper limit intensity of the 683\,GHz line \mbox{\citep{Nagy13}}.
In this comparison, and following the morphology of the SH$^+$ emission revealed by ALMA (Fig.~\ref{fig:SHp-ALMA}), we corrected the line intensities of the SH$^+$ lines detected by \mbox{Herschel/HIFI} with the beam coupling factors discussed in Sec.~\ref{sec-ALMA-obs},
The observed \mbox{1$_2$-0$_1$/1$_0$-0$_1$} line ratio (\mbox{$R$\,=\,$W$(526.048)/$W$(345.944)\,$\simeq$\,2}) is sensitive to the gas density. In these models, $R$ is \mbox{1.1} for \mbox{$n_{\rm H}$\,=\,10$^5$\,cm$^{-3}$}
and \mbox{3.0} for \mbox{$n_{\rm H}$\,=10$^6$\,cm$^{-3}$}. 
We note that $n_{\rm H}$ could be lower if SH$^+$ formation/destruction rates
were faster, as in the  \mbox{$F, D$ $\times$10} model. This could happen if 
SH$^+$ photodissociation or destruction \mbox{reactions} with \mbox{H$_2$($v$\,$\geq$2)} 
were faster than  reactions of SH$^+$ with H atoms or with electrons.
In \mbox{Sec.~\ref{sec:pdr-mods}} we show that this is not the case.

\subsection{SH excitation and column density}

SH is a $^2\Pi$ open-shell radical with fine-structure, \mbox{$\Lambda$-doubling}, and hyperfine splittings \citep[e.g.,][]{Martin-Drumel12}.
However, the frequency separation of the SH \mbox{$^2\Pi_{3/2}$\,$J$\,=\,5/2-3/2} hyperfine components is too small to be spectrally resolved in observations of the Orion Bar (see Fig.~\ref{fig:SH-spectra}).The available rate coefficients for \mbox{inelastic} collisions of SH with helium atoms do not resolve the hyperfine splittings. Hence, we first determined line frequencies, level degeneracies, and Einstein coefficients of an SH molecule without hyperfine structure. To do this, we took the
complete set of hyperfine levels tabulated in CDMS.
Lacking specific inelastic collision rate coefficients,
we scaled the available SH--\,He rates of \citet{Klos09}
by the square root of the reduced mass ratios and estimated the SH--\,H and SH--\,H$_2$ collisional rates. 

\begin{figure}[b]
\centering   
\includegraphics[scale=0.62, angle=0]{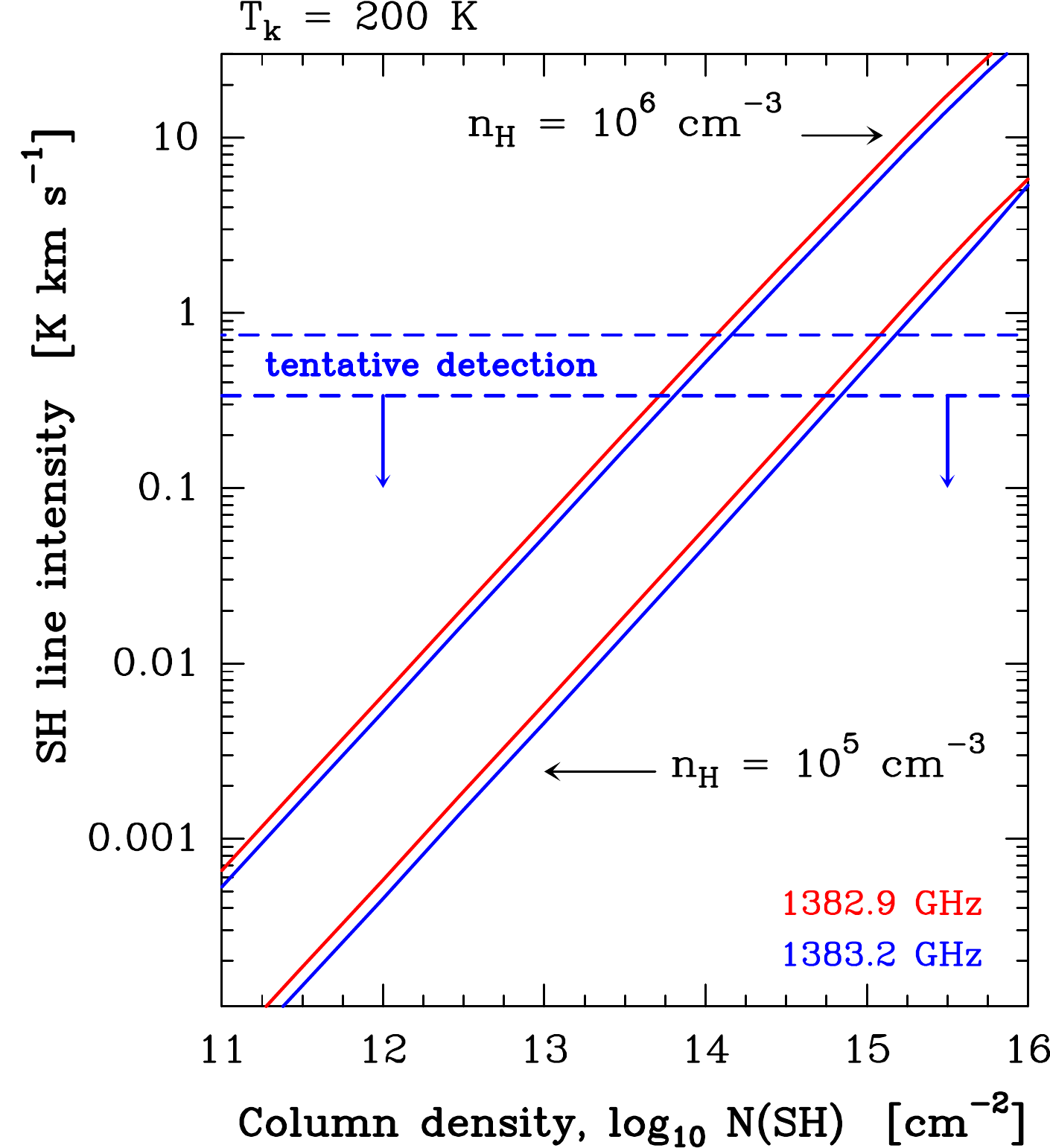}
\caption{Non-LTE excitation  models of SH emission lines targeted with SOFIA/GREAT.  Horizontal dashed lines refer to observational limits, assuming extended emission (lower intensities) and for a 10$''$ width emission filament
at the PDR surface (higher intensities).} 
\label{fig:MTC-SH}
\end{figure}

The scaled rate coefficients are about an order of magnitude smaller
than those of SH$^+$. However, the chemical destruction rate of SH at the PDR edge 
(reactions with H, photodissociation, and photoionization, see Sect.~\ref{sec:gas_models})
is also slower \citep[we take the rates of SH--H reactive collisions  from ][]{Zanchet19}. We determine $D$\,$\simeq$\,3$\times$10$^{-6}$\,s$^{-1}$ for $n_{\rm H}$\,=\,10$^6$\,cm$^{-3}$,
 $T_{\rm k}$\,=\,200\,K, and $A_V$\,$\simeq$\,0.7\,mag.
Models in Fig.~\ref{fig:MTC-SH} include these chemical rates for $T_{\rm form}$\,=$T_{\rm k}$ (a lower limit to the unknown formation temperature). Formation pumping  enhances the intensity of the  \mbox{$^2\Pi_{3/2}$\,$J$\,=\,5/2-3/2} ground-state lines
by a few percent only.

To estimate the SH column density in the Orion Bar we compare with the upper limit
intensities of the SH lines targeted by SOFIA. 
If SH and SH$^+$ arise from roughly the same gas at similar physical conditions 
(\mbox{$n_{\rm H}$\,$\simeq$\,10$^6$\,cm$^{-3}$} and \mbox{$T_k$\,$\simeq$\,200\,K}) the best model column density is for \mbox{$N$(SH)\,$\leq$\,(0.6-1.6)$\times$10$^{14}$\,cm$^{-2}$}. If densities were lower, around \mbox{$n_{\rm H}$\,$\simeq$\,10$^5$\,cm$^{-3}$}, the upper limit $N$(SH) column
densities will be a factor ten higher.

\begin{figure*}[th]
\centering   
\includegraphics[scale=0.41, angle=0]{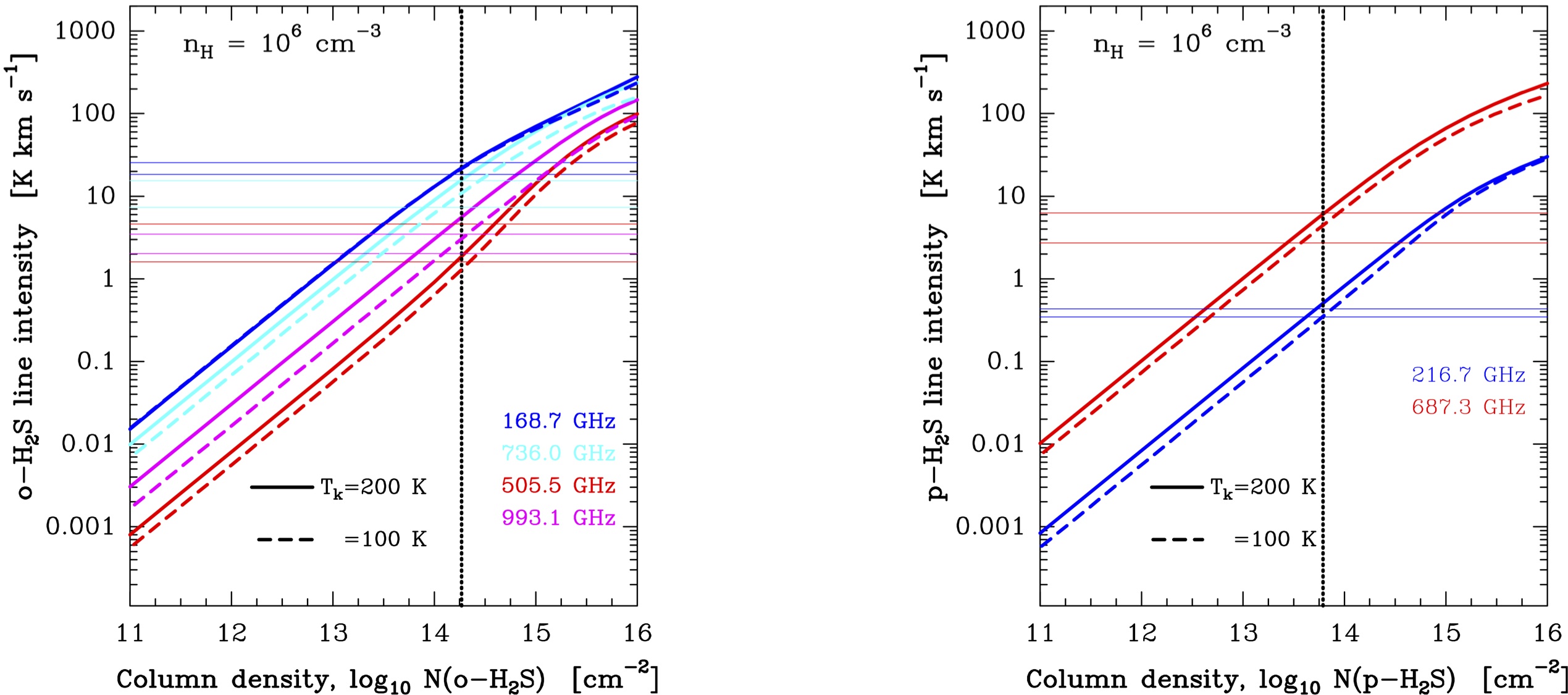}
\caption{Non-LTE excitation models for $o$-H$_2$S and $p$-H$_2$S.
Thin horizontal lines show the observed  intensities assuming either 
extended emission (lower limit) or emission that fills the 15$''$ beam  at
168.7\,GHz.  The vertical line marks the best model, resulting in an OTP ratio of \mbox{2.9\,$\pm$\,0.3}.}
\label{fig:MTC-H2S}
\end{figure*}

\subsection{H$_2$S excitation and column density}

H$_2$S has a $X^2A$ ground electronic state and
 two nuclear spin symmetries that we treat separately, $o$-H$_2$S and $p$-H$_2$S. Previous studies of the H$_2$S line excitation  have used \mbox{collisional} rates coefficients scaled from those of the \mbox{H$_2$O\,--\,H$_2$} system.
\mbox{\citet{Dagdigian20}} recently carried out specific  calculations of the
  cross sections of $o$-H$_2$S and $p$-H$_2$S inelastic collisions with \mbox{$o$--H$_2$} and \mbox{$p$-H$_2$} at different temperatures. The behavior of the new and the scaled rates is different
and it depends on the H$_2$ OTP ratio (e.g., on gas temperature) because the collisional
cross sections are different for \mbox{$o$-H$_2$--H$_2$S} and \mbox{$p$-H$_2$--H$_2$S} systems. At the warm temperatures of the PDR, collisions with $o$-H$_2$ dominate, 
resulting in  rate coefficients for the $\sim$168\,GHz $o$-H$_2$S line  that are a factor up to $\sim$2.5 smaller than those scaled from \mbox{H$_2$O--H$_2$}.

H$_2$S is not a reactive molecule. At the edge of the PDR its destruction is driven by  photodissociation. We determine that the radiative and collisional pumping rates  are typically a factor of $\sim$100 higher than \mbox{$D$\,$\approx$\,2$\times$10$^{-6}$\,s$^{-1}$} (for $n_{\rm H}$\,=\,10$^6$\,cm$^{-3}$, $T_{\rm k}$\,=\,200\,K, $G_0$\,$\simeq$10$^4$, and $A_V$\,$\simeq$\,0.7\,mag). 
\mbox{Figure~\ref{fig:MTC-H2S}} shows \mbox{non-LTE}  \mbox{$o$-H$_2$S} and \mbox{$p$-H$_2$S} excitation and radiative transfer models. As H$_2$S may have its abundance peak deeper inside the PDR and display more extended emission than SH$^+$ \mbox{\citep[e.g.,][]{Sternberg95}}, we show results for
 $T_{\rm k}$\,=\,200 and 100\,K. When comparing with the observed line intensities, we considered either emission that fills all beams, or a correction that assumes that the H$_2$S emission only fills the 15$''$ beam of the IRAM\,30m telescope at 168\,GHz. The vertical dotted lines in Fig.~\ref{fig:MTC-H2S} show the best model, \mbox{$N$(H$_2$S)\,=\,$N$($o$-H$_2$S)+$N$($p$-H$_2$S)\,=\,2.5$\times$10$^{14}$\,cm$^{-2}$}, with an OTP ratio of {\mbox{2.9\,$\pm$\,0.3}, thus consistent with the
 \mbox{high-temperature} \mbox{statistical} ratio of 3/1 (see discussion at the end of Sect.~\ref{line-models}). Models with lower densities, 
 $n_{\rm H}$\,$\simeq$\,10$^5$\,cm$^{-3}$, 
show worse agreement, and would translate into even higher $N$(H$_2$S) of $\gtrsim$\,10$^{15}$\,cm$^{-2}$. In either case, these calculations 
imply large columns of warm H$_2$S  toward the PDR. They result in a limit to the SH to H$_2$S column density ratio of  \mbox{$\leq$\,0.2-0.6}. 
This upper limit is already lower than the \mbox{$N$(SH)/$N$(H$_2$S)\,=\,1.1-3.0} ratios observed in diffuse clouds \citep[][]{Neufeld15}. This difference  suggests an enhanced H$_2$S formation mechanism in \mbox{FUV-illuminated} dense gas.

\begin{figure}[h]
\centering   
\includegraphics[width=8.3cm, angle=0]{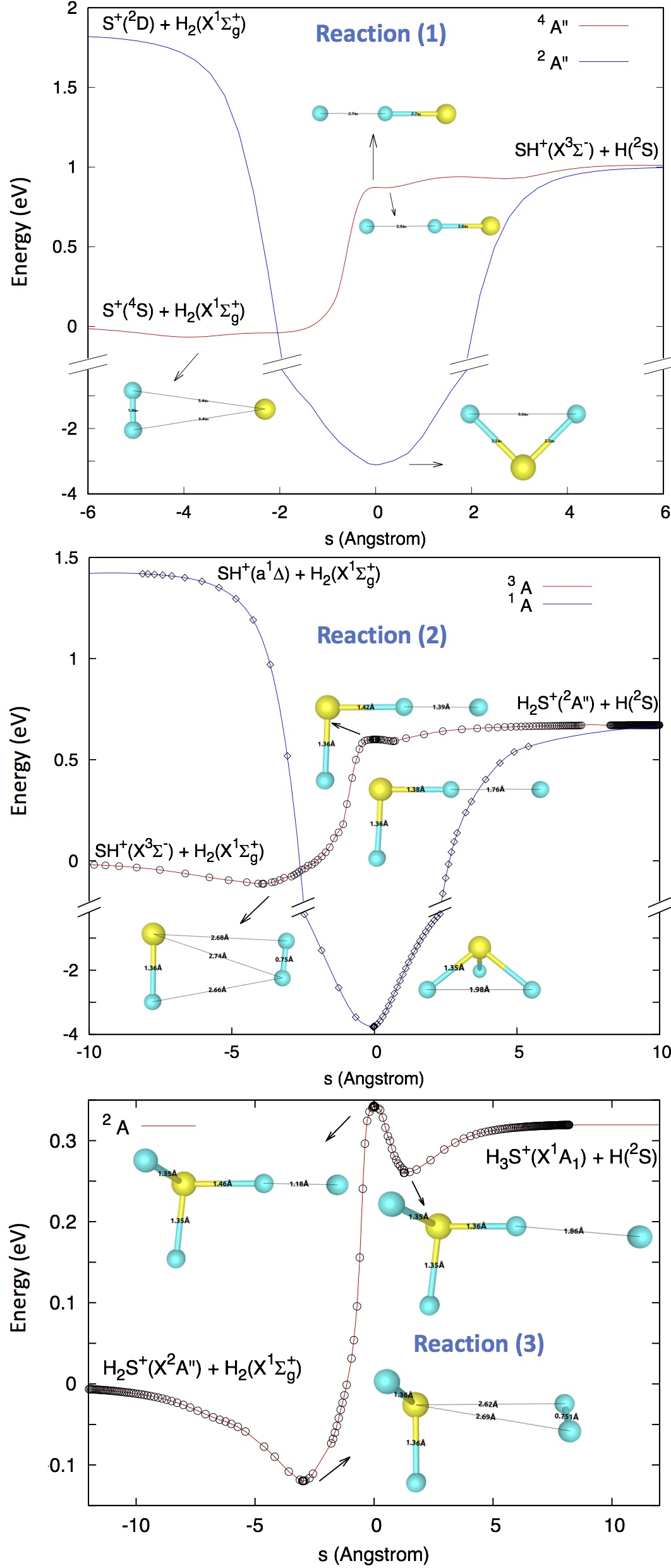}
\caption{Minimum energy paths for reactions~(\ref{reac-1}), (\ref{reac-2}), and
(\ref{reac-3}). 
Points correspond to \mbox{RCCSD(T)-F12a} calculations and lines to fits
(\mbox{Appendix~\ref{Apendix:ab-initio}}).
The reaction coordinate, $s$, is defined independently for each path.
The geometries of each species at $s$=0 are different.}
\label{fig:PES}
\end{figure}

\section{New results on sulfur-hydride reactions}\label{sec-rates}

In this section we summarize the \textit{ab initio} quantum calculations we carried out
to determine the vibrationally-state-dependent rates of  gas-phase reactions 
of \mbox{H$_2$($v$\,$>$\,0)} with several \mbox{S-bearing} species.
We recall that all hydrogen abstraction
reactions,\vspace{0.1cm}\\
  ${\rm S}^+ \xrightarrow[(1)]{{{+{\rm H_2}}}} {\rm SH}^+ \xrightarrow[(2)]{{+{\rm H_2}}} {\rm H_2S^+} \xrightarrow[(3)]{{+{\rm H_2}}} {\rm H_3S^+,}\hspace{1cm}{\rm S} \xrightarrow[(4)]{{{+{\rm H_2}}}} {\rm SH,}\vspace{0.1cm}$\\
are very endoergic for H$_2$\,($v$\,=\,0), with  endothermicities  in Kelvin units  that are significantly higher than $T_{\rm k}$ even in PDRs. This is  markedly different to O$^+$ chemistry, for which all  hydrogen abstraction reactions
leading to H$_3$O$^+$ are exothermic and fast   \citep[][]{Gerin10,Neufled10,Hollenbach12}.

The endothermicity of reactions involving H$_n$S$^+$ ions decreases as the number of hydrogen atoms increases. The potential energy surfaces (PES) of these reactions possess  shallow wells at the entrance and products channels (shown in Fig.~\ref{fig:PES}). In addition, these PESs show  \mbox{saddle} points between the energy walls of reactants
and products whose heights increase with the number of H atoms. 
For \mbox{reaction~~(\ref{reac-2})}, the saddle point has an energy of 0.6\,eV
($\simeq$7,000\,K) and is slightly below the energy of the products. However, for \mbox{reaction~(3)}, the saddle point is above the energy of the products and is a reaction barrier. These saddle points act as a bottleneck in the gas-phase hydrogenation of S$^+$.

If one considers the state dependent reactivity of vibrationally excited H$_2$, the formation of SH$^+$ through reaction~(\ref{reac-1}) becomes 
exoergic\footnote{If one considers H$_2$ rovibrational levels, reaction~(\ref{reac-1}) becomes exoergic for $v$\,=\,0, $J$\,$\geq$\,11 and for  $v$\,=\,1, $J$\,$\geq$\,7 \citep{Zanchet19}.} when $v$\,$\geq$\,2 \citep[][]{Zanchet19}. 
The detection of bright H$_2$S emission in the Orion Bar 
(\mbox{Figs.~\ref{fig:IRAM-map} and \ref{fig:SH-spectra}}) might suggest that subsequent  hydrogen abstraction reactions with \mbox{H$_2$\,($v$\,$\geq$\,2)} proceed  as well. 
Motivated by these \mbox{findings}, and before carrying out any PDR  model, we studied \mbox{reaction~(\ref{reac-2})} and the reverse process in detail.
This  required to build a full dimensional quantum PES
of the H$_3$S$^+$\,(X$^1$A$_1$) system (see Appendix~\ref{Apendix:ab-initio}). 

In addition, we studied reaction  (\ref{reac-4}) (and its reverse) through quantum calculations. Details of these \mbox{\textit{ab initio}} calculations 
and of the resulting reactive cross sections 
are given in \mbox{Appendix~\ref{Apendix:S_H2}}.
Table~\ref{rates} summarizes the  updated reaction rate coefficients that we will include later in our PDR models.

\begin{table}[t]
  \begin{center}
    \caption{\label{rates} Relevant rate coefficients from a fit of the Arrhenius-like form
     $k\,(T)$\,=\,$\alpha\,(T/300\,{\rm K})^\beta \,{\rm exp}(-\gamma/T)$
     to the calculated reaction rates.}
    \begin{tabular}{lccr}
      \hline \hline
      Reaction & $\alpha$            & $\beta$ & $\gamma$\\
               &  (cm$^3$\,s$^{-1}$) &         &  (K)\\      
      \hline
      SH$^+$ + H$_2$\,($v$=1) $\rightarrow$ H$_2$S$^+$ + H  & 4.97e-11  & 0       & 1973.4 $^a$ \\
      SH$^+$ + H$_2$\,($v$=2) $\rightarrow$ H$_2$S$^+$ + H  & 5.31e-10  &  -0.17  & 0 $^a$ \\
      SH$^+$ + H$_2$\,($v$=3) $\rightarrow$ H$_2$S$^+$ + H  & 9.40e-10  & -0.16   & 0 $^a$  \\
      \hline
      SH$^+$ + H $\rightarrow$ S$^+$ + H$_2$ & 1.86e-10  & -0.41  & 27.3 $^b$ \\ 
      \hline
      SH$^+$ + $e^-$ $\rightarrow$ S + H              & 2.00e-07 &  -0.50       &  $^c$  \\       
     \hline
      H$_2$S$^+$ + H $\rightarrow$ SH$^+$ + H$_2$     & 6.15e-10  & -0.34   & 0 $^a$  \\
     \hline
      S + H$_2$\,($v$=2) $\rightarrow$ SH + H           & $\sim$8.6e-13 & $\sim$2.3 &  $\sim$2500 $^a$  \\
      S + H$_2$\,($v$=3) $\rightarrow$ SH + H           & $\sim$1.7e-12 & $\sim$2.0 &  $\sim$1500 $^a$  \\      
      \hline
      SH + H $\rightarrow$ S + H$_2$                  & 5.7e-13 & 2.48 & 1600$^{a,\dagger}$ \\
                        					          & 7.7e-14 & 0.39 & $-$1.3$^{a,\dagger}$ \\
      \hline
      S$^+$ + H$_2$\,($v$=2) $\rightarrow$ SH$^+$ + H & 2.88e-10  & -0.15  & 42.9 $^b$  \\
      S$^+$ + H$_2$\,($v$=3) $\rightarrow$ SH$^+$ + H & 9.03e-10  & -0.11  & 26.2 $^b$ \\
      S$^+$ + H$_2$\,($v$=4) $\rightarrow$ SH$^+$ + H & 1.30e-09 & -0.04  & 40.8  $^b$ \\
      S$^+$ + H$_2$\,($v$=5) $\rightarrow$ SH$^+$ + H & 1.21e-09 &  0.09  & 34.5 $^b$ \\
      \hline
      \end{tabular}
  \tablefoot{$^{(a)}$\,This work. $^{(b)}$\,From \citet{Zanchet19}. $^{(c)}$ From \citet{Prasad80}.
  ${^\dagger}$Total rate is the sum of the two expressions.}
  \end{center}
\end{table}

The H$_2$S$^+$ formation rate  through reaction~(\ref{reac-2}) with \mbox{H$_2$\,($v$\,=\,0)} is very slow.  For \mbox{H$_2$\,($v$\,=\,1)}, the rate constant increases at $\approx$\,500\,K, corresponding to the opening of the \mbox{H$_2$S$^+$ + H} threshold. 
For H$_2$\,($v$\,=\,2) and  H$_2$\,($v$\,=\,3), the reaction rate is much faster, close to the Langevin limit (see \mbox{Appendix~\ref{appendix-collisions}}). 
\mbox{However}, our estimated vibrational-state specific  rates for SH formation
through \mbox{reaction~(\ref{reac-4})} (\mbox{S + H$_2$}) are considerably smaller than for \mbox{reactions~(\ref{reac-1}) and (\ref{reac-2})}, and show an \mbox{energy barrier} even for  \mbox{H$_2$\,($v$\,=\,2)} and  \mbox{H$_2$\,($v$\,=\,3)}. We anticipate that this reaction is not a relevant formation route for SH.

In \mbox{FUV-illuminated} environments, collisions with H atoms  are very important because they compete with electron recombinations in destroying  molecular ions, and also they contribute to their excitation.
An important result of our calculations is that the 	\mbox{destruction} rate of H$_2$S$^+$ (SH$^+$) in reactions with H atoms are a factor
of \mbox{$\geq$\,3.5} (\mbox{$\geq$\,1.7}) faster (at \mbox{$T_{\rm k}$\,$\leq$\,200\,K})  than those previously used in astrochemical
models \mbox{\citep[][]{Millar86}}. 
\mbox{Conversely}, we find that  destruction of SH in reactions with H atoms \mbox{(Appendix~\ref{Apendix:S_H2})} is slower than previously assumed.

\section{PDR models of S-bearing hydrides}\label{sec:pdr-mods}

We now investigate the chemistry of S-bearing hydrides
and the effect of the new reaction rates in PDR models adapted to the Orion Bar conditions. In this analysis we used \mbox{version\,1.5.4.} of the Meudon PDR code \citep[][]{LePetit06,Bron14}. \mbox{Following} our previous studies, 
 we model  the Orion Bar  as a \mbox{stationary} PDR at constant thermal-pressure (i.e., with density and temperature gradients). When compared to time-dependent hydrodynamic PDR models \citep[e.g.,][]{Hosokawa06,Bron18,Kirsanova19},  \mbox{stationary} isobaric models seem a good description  of the most exposed and compressed
 gas layers of the PDR, from \mbox{$A_V$\,$\approx$\,0.5} to 
\mbox{$\approx$\,5\,mag} \citep[][]{Goico16,Joblin18}.

In our models, the FUV \mbox{radiation} field incident at the PDR edge is $G_0$\,=\,2$\times$10$^4$ \mbox{\citep[e.g.,][]{Marconi_1998}}.
We adopted an extinction to color-index 
ratio, $R_V$\,=\,$A_V$/$E_{B-V}$, of 5.5 \citep{Joblin18}, consistent with the flatter 
extinction curve observed  in Orion \citep[][]{Lee68,Cardelli89}. This choice implies slightly more penetration of FUV radiation into the cloud \mbox{\citep[e.g.,][]{Goico07}}.
The main input parameters and elemental abundances of these PDR models are summarized in Table~\ref{table:PDR-mods}. Figure~\ref{fig:PDR-str} shows the resulting H$_2$, H, and electron density profiles, as well as the $T_{\rm k}$ and $T_{\rm d}$  gradients.

Our chemical network is that of the  Meudon  code 
updated with the new reaction rates listed in Table~\ref{rates}.
This network includes updated photoreaction rates  from \citet{Heays17}.
To increase the accuracy of our abundance predictions, we included
the explicit integration of wavelength-dependent
SH, SH$^+$, and H$_2$S photodissociation cross sections ($\sigma_{\rm diss}$), as well as SH and H$_2$S photoionization cross sections ($\sigma_{\rm ion}$).
These cross sections are shown in Fig~\ref{fig:photo-cross-sect} of the Appendix. The integration is performed over the specific FUV radiation field at each position of the PDR. In particular,  we took   $\sigma_{\rm ion}$(SH) from \mbox{\cite{Hrodmarsson19}} and $\sigma_{\rm diss}$(H$_2$S) from \cite{Zhou20}, both  determined in laboratory experiments. 
Figure~\ref{fig:Network} summarizes the relevant chemical network that leads to the formation of  S-bearing hydrides and that we discuss in the following
sections.

\begin{table}[t]
\caption{Main parameters used in the PDR models of the Orion Bar.\label{table:PDR-mods}} 
\centering
\begin{tabular}{ccc@{\vrule height 8pt depth 5pt width 0pt}}
\hline\hline
Model parameter                                 &     Value                                     &     Note        \\ 
\hline
FUV illumination, $G_0$        				    &     2$\times$10$^4$ Habing                     &   $(a)$           \\
Total depth $A_{\rm V}$                         &     10 mag                   			        &                 \\
Thermal pressure $P_{\rm th}/k$                 &  2$\times$10$^8$\,cm$^{-3}$K                   &                 \\
Density $n_{\rm H}$\,=\,$n$(H)\,+\,2$n$(H$_2$)  &  $n_{\rm H}$\,=\,$P_{\rm th}\,/\,kT_{\rm k}$  & Varying          \\
\hline
Cosmic Ray $\zeta_{\rm CR}$                     & 10$^{-16}$\,H$_2$\,s$^{-1}$ 	                &  $(b)$           \\
$R_{\rm V}$\,=\,$A_{\rm V}$/$E_{\rm B-V}$       &          5.5                	                &  Orion$^c$      \\
$M_{\rm gas}/M_{\rm dust}$                      & 100                                           &  Local ISM      \\
Abundance O\,/\,H                               & 3.2$\times$10$^{-4}$                           &                  \\
Abundance C\,/\,H						        & 1.4$\times$10$^{-4}$			                &  Orion$^d$      \\
Abundance S\,/\,H							    & 1.4$\times$10$^{-5}$			                &  Solar$^e$	   \\
\hline                                    
\end{tabular}
\tablefoot{$^a$\cite{Marconi_1998}. $^b$\cite{Indriolo15}. $^c$\citet{Cardelli89}. $^d$\cite{Sofia04}.
$^e$\cite{Asplund09}.}
\end{table}

\begin{figure}[t]
\centering   
\includegraphics[scale=0.46, angle=0]{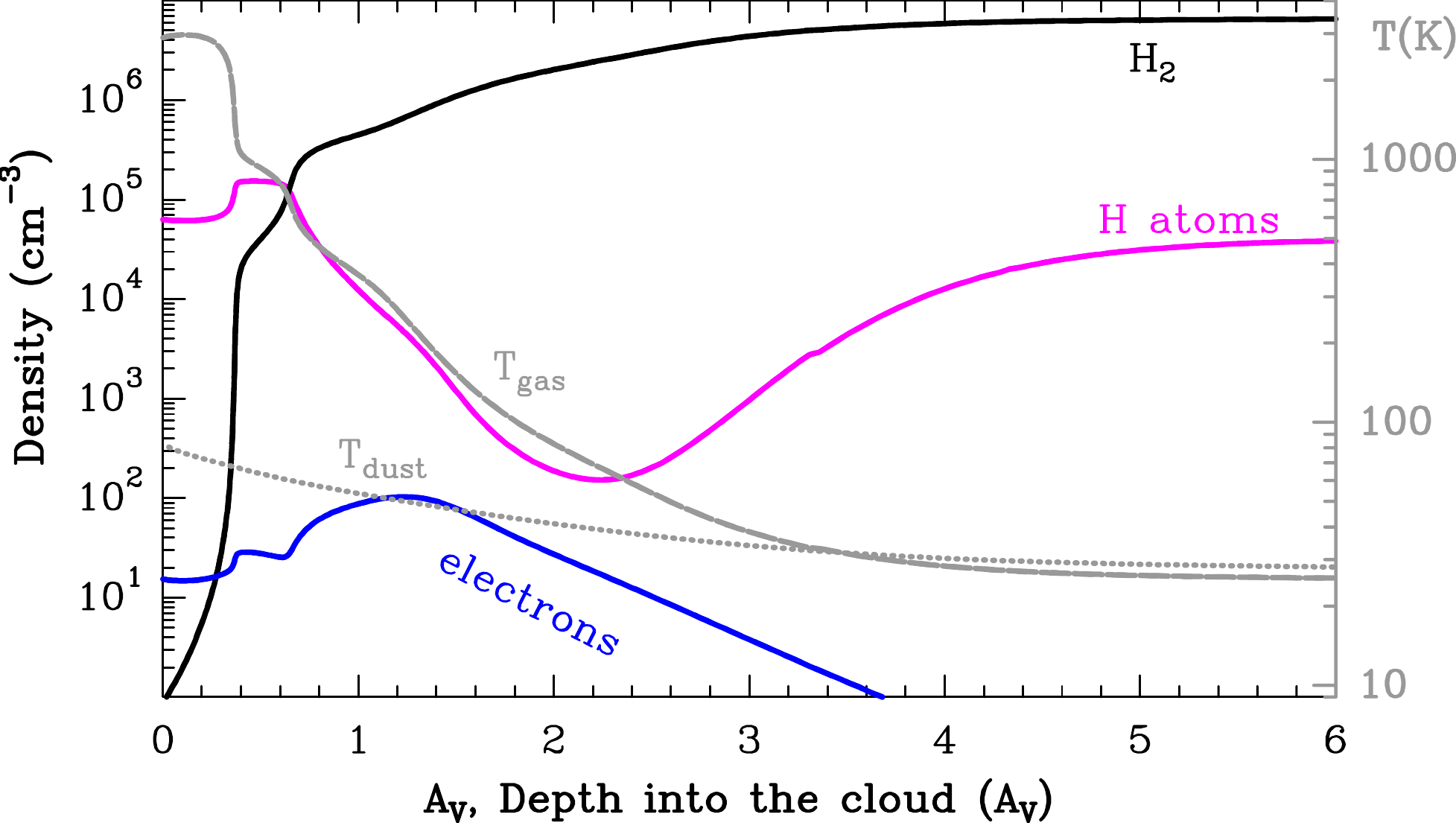}
\caption{Structure of an isobaric PDR representing the most FUV-irradiated gas layers of the Orion Bar (see Table~\ref{table:PDR-mods} for the adopted parameters). This plot shows the H$_2$, H, and electron density profiles (left axis scale), and the gas and dust temperatures  (right axis scale).}
\label{fig:PDR-str}
\end{figure}

\begin{figure}[h]
\centering   
\includegraphics[scale=0.285, angle=0]{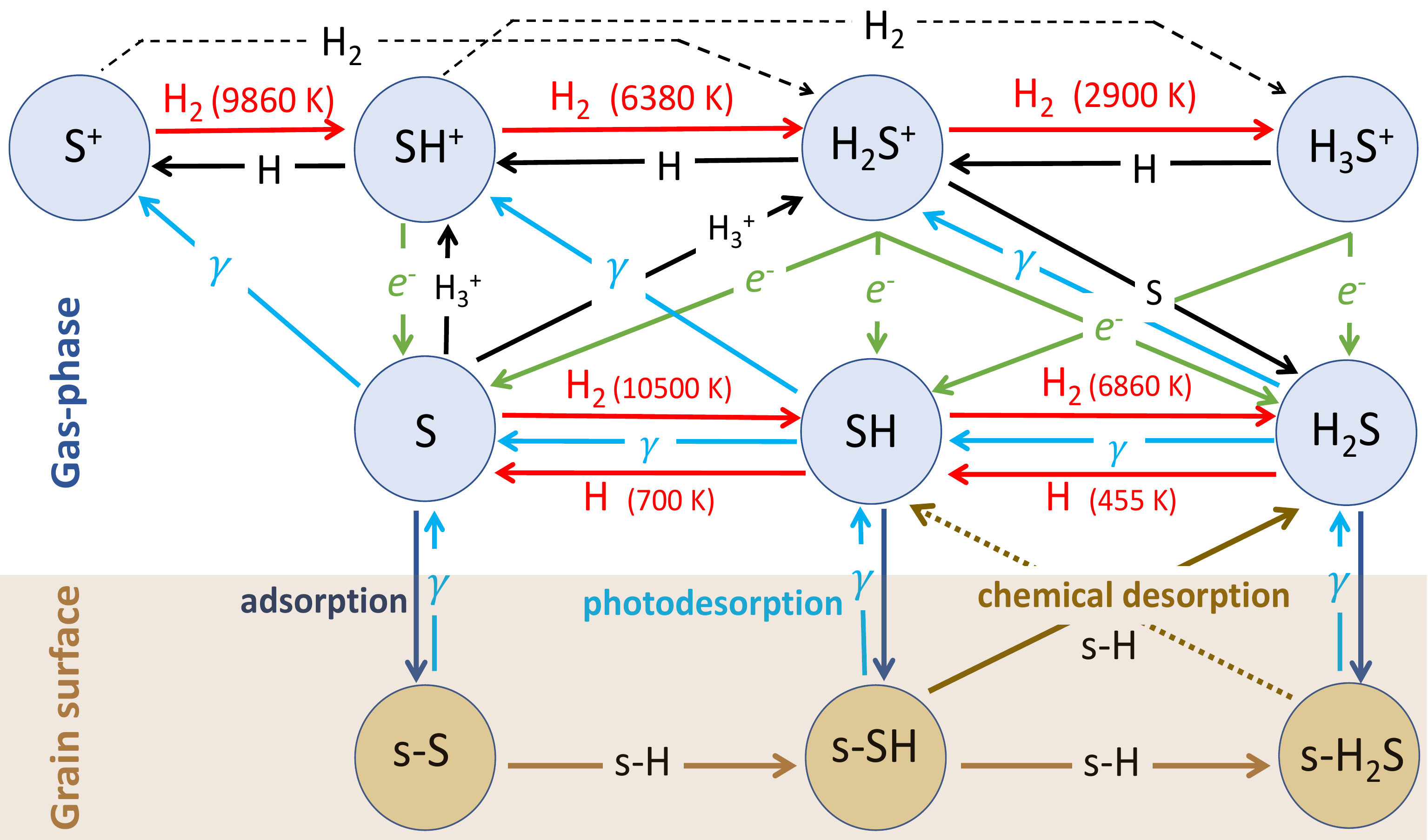}
\caption{Main gas and grain  reactions leading to the formation of sulfur hydrides. Red arrows represent endoergic reactions (endothermicity given in units of K). Dashed arrows are uncertain radiative associations (see Sect.~\ref{Apendix:destruction}), $\gamma$ stands for a FUV photon, and  ``s-'' for solid.}
\label{fig:Network}
\end{figure}

\subsection{Pure gas-phase PDR model results\label{sec:gas_models}}

Figure~\ref{fig:PDR-gas-new-old} shows results of the
\mbox{``new gas-phase''} model  using the reaction rates
in Table~\ref{rates}. The continuous curves  display the predicted fractional abundance profiles as a function of cloud depth in magnitudes of visual extinction ($A_V$). The \mbox{dashed} curves are for a model that uses the standard thermal rates previously adopted in the literature  \mbox{\citep[see, e.g.,][]{Neufeld15}}. As noted by \citet{Zanchet13a,Zanchet19}, the inclusion of \mbox{H$_2$\,($v$\,$\geq$\,2)} state-dependent quantum rates for \mbox{reaction~(\ref{reac-1})} enhances the
formation of SH$^+$ in a narrow layer at the edge of the PDR 
\mbox{($A_V$\,$\simeq$\,0 to 2\,mag)}. This agrees with the morphology of the SH$^+$ emission revealed by
ALMA images (Fig.~\ref{fig:SHp-ALMA}).
 For \mbox{H$_2$\,($v$\,$=$\,2)}, the reaction rate enhancement 
 with respect to the thermal rate \mbox{$\Delta k$\,=\,$k_2(T)/k_0(T)$}   
  \citep[see   discussion by][]{Agundez10}   
 is about 4$\times$10$^8$
  at $T_{\rm k}$\,=\,500\,K \citep[][]{Millar86}. 
\mbox{Indeed}, when the fractional abundance of  H$_2$\,($v$\,=\,2) with respect to
H$_2$\,($v$\,=\,0), defined as \mbox{$f_{\,2}$\,=\,$n$\,(H$_2$\,$v$\,=\,2)/$n$\,(H$_2$\,$v$\,=\,0)}, exceeds a few times 10$^{-9}$, meaning \mbox{$\Delta k \cdot f_{\,2} > 1$}, reaction~(\ref{reac-1}) with H$_2$\,($v$\,$\geq$\,2) dominates SH$^+$ formation.  
This reaction enhancement takes place only at the edge of the PDR, where FUV-pumped H$_2$\,($v$\,$\geq$\,2) molecules are abundant enough (gray dashed curves in Fig.~\ref{fig:PDR-gas-new-old}) and drive the formation of SH$^+$. The resulting SH$^+$ column density  increases by an order of magnitude compared to models that use the thermal rate.

In this isobaric model, the SH$^+$ abundance peak occurs at \mbox{$A_V$\,$\simeq$\,0.7\,mag}, where the gas density has increased from \mbox{$n_{\rm H}$\,$\simeq$\,6$\times$10$^4$\,cm$^{-3}$} at the PDR edge (the IF) to $\sim$5$\times$10$^5$\,cm$^{-3}$ (at the DF). At this point, SH$^+$ destruction is dominated by recombination with electrons and  by reactive collisions with H atoms. This implies 
 \mbox{$D$(SH$^+$) $[$s$^{-1}$$]$\,$\sim$\,$n_e$\,$k_e$\,$\simeq$\,$n_{\rm H}$\,$k_{\rm H}$\,$\gg$\,$n$(H$_2$\,$v$$\geq$\,2)\,$k_2$}, as we assumed in the single-slab SH$^+$  excitation models  (\mbox{Sec.~\ref{sec-MTC-SHp}}).
Therefore, only a small fraction of SH$^+$ molecules further react with H$_2$\,($v$\,$\geq$\,2) to form H$_2$S$^+$. 
The resulting low H$_2$S$^+$ abundances limit the formation of abundant SH from  dissociative recombinations of H$_2$S$^+$ (recall that we estimated that reaction S\,+\,H$_2$\,($v$\,$\geq$2)\,$\rightarrow$\,SH\,+\,H is very slow). The SH abundance peak is shifted deeper inside the cloud, at about $A_V$\,$\simeq$\,1.8\,mag, where SH forms by dissociative recombination of H$_2$S$^+$ and it is destroyed by FUV photons and reactions with H atoms. 
In these  gas-phase models  the H$_2$S abundance peaks even deeper inside the PDR, at \mbox{A$_V$\,$\simeq$\,5\,mag}, where it forms by  recombinations of H$_2$S$^+$ and  H$_3$S$^+$ with electrons as well as by charge exchange  \mbox{S\,+\,H$_2$S$^+$}.  However, the new rate of reaction \mbox{H$_2$S$^+$\,+\,H} is higher than assumed in the past, so the new models predict lower H$_2$S$^+$ abundances at intermediate PDR depths 
(thus, less H$_3$S$^+$ and H$_2$S; see Fig.~\ref{fig:PDR-gas-new-old}).

\begin{table*}[ht]
\caption{Column density predictions from different PDR models (up to $A_V$\,=\,10\,mag) and estimated values from observations (single-slab approach).\label{table:PDR-columns}} 
\centering
\begin{tabular}{cccccc@{\vrule height 8pt depth 5pt width 0pt}}
\hline\hline
                                             &  log $N$\,(cm$^{-2}$) &                      &              &             &           \\ 
\textit{Type of PDR model$^a$}               &        SH$^+$              &   SH                 &  H$_2$S         &     H$_2$S$^+$   &  H$_3$S$^+$ \\
\hline
Standard gas-phase                           &	11.0$^a$--12.2$^b$  &   11.4$^a$--12.5$^b$ &  11.3$^a$--12.4$^b$  &   9.9$^a$--11.1$^b$  &  7.8$^a$--9.0$^b$   \\
New gas-phase (Table~\ref{rates})            &	12.1$^a$--13.2$^b$	&   11.4$^a$--12.5$^b$ &  10.6$^a$--11.7$^b$  &   9.9$^a$--11.0$^b$  &  7.7$^a$--8.9$^b$   \\
Gas-grain (low $E_{\rm b}$, $\epsilon$=1\%)                  &	12.0$^a$--13.2$^b$	&   13.2$^a$--14.4$^b$ &  12.9$^a$--14.1$^b$  &   9.6$^a$--10.7$^b$  & 10.1$^a$--11.2$^b$  \\
Gas-grain (high $E_{\rm b}$, $\epsilon$=1\%)                 &	12.0$^a$--13.1$^b$  &   13.6$^a$--14.8$^b$ &  13.7$^b$--14.8$^b$  &   9.9$^b$--11.0$^b$  & 10.8$^b$--12.0$^b$   \\
\hline
Estimated from observations                  &         $\sim$13.1            &   $<$13.8         &   $\sim$14.4                &     --               &  $<$\,10.7     \\
\hline
\end{tabular}  
\tablefoot{$^a$Column densities for a face-on PDR.  $^b$Edge-on PDR with a tilt angle \mbox{$\alpha$\,=\,4$^o$}, leading to the maximum expected geometrical enhancement.}
\end{table*}

\begin{figure}[ht]
\centering   
\includegraphics[scale=0.42, angle=0]{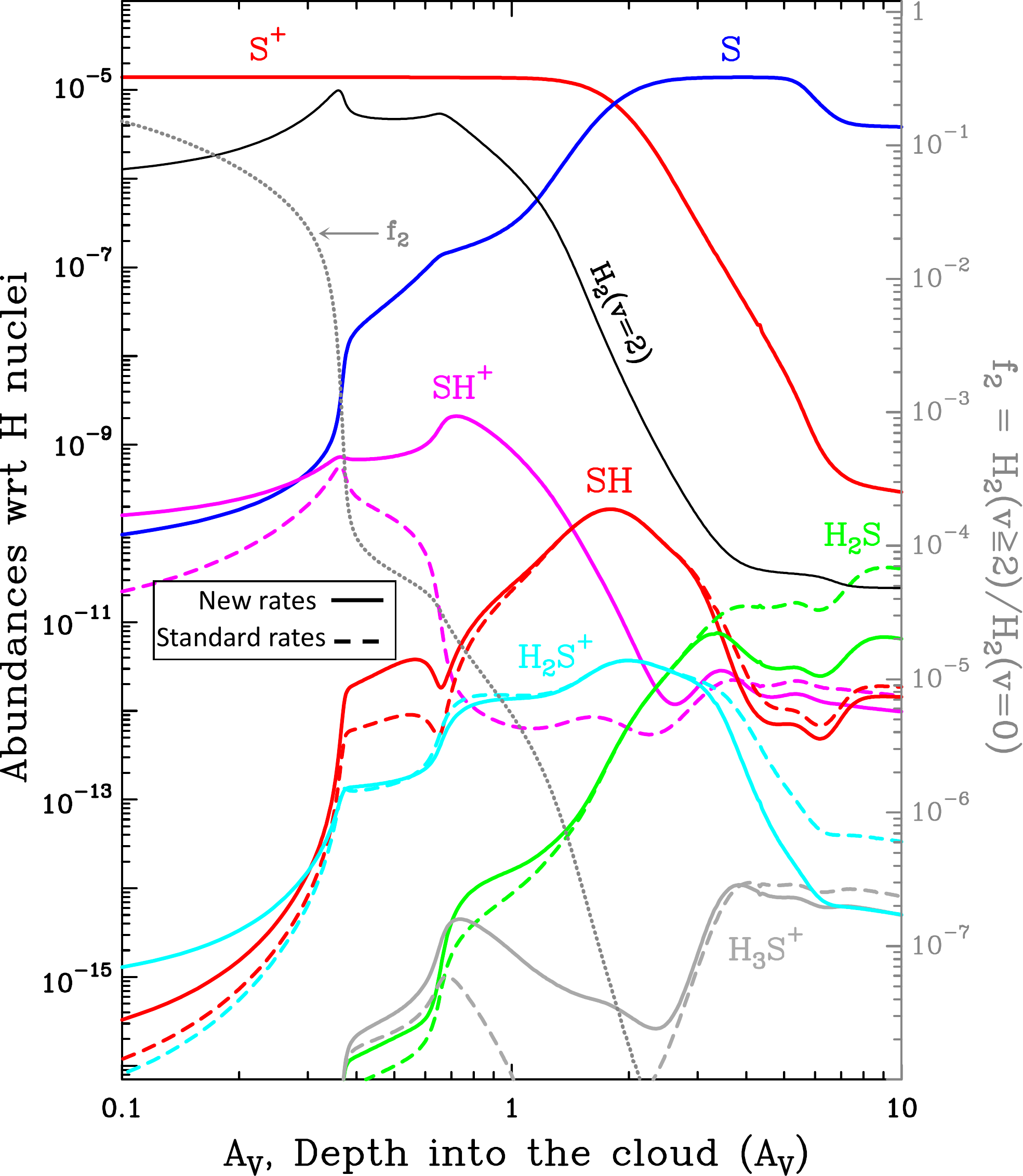}
\caption{Pure gas-phase  PDR models of the Orion Bar. Continuous curves show fractional abundances as a function of cloud depth, in logarithm scale to better display 
the irradiated edge of the PDR, using the new reaction rates listed in \mbox{Table~\ref{rates}}.
The gray dotted curve shows $f_2$, the fraction of H$_2$ that is in \mbox{vibrationally} excited levels \mbox{$v$\,$\geq$\,2} (right axis scale). 
Dashed curves are for a model using standard reaction rates.}
\label{fig:PDR-gas-new-old}
\end{figure}  

The SH column density  predicted by the 
 \mbox{new gas-phase} model is below the upper limit determined from SOFIA. 
However, the predicted  H$_2$S column density is much lower than the value we derive from observations (Table~\ref{table:PDR-columns}) and the predicted H$_2$S line intensities are too faint (see Sect.~\ref{line-models}).
 
Because the cross sections of the different H$_2$S photodissociation channels have
different wavelength dependences \citep{Zhou20}, the H$_2$S and SH abundances between $A_V$\,$\approx$\,2 and 6\,mag are sensitive to the specific shape of the FUV radiation field \citep[determined by  line blanketing, dust absorption, and grain 
\mbox{scattering}; e.g.,][]{Goico07}. Still, we checked that using steeper extinction curves does not increase H$_2$S column density any closer to the observed levels.
This disagreement between the observationally inferred $N$(H$_2$S)  
column density and the predictions of  gas-phase PDR models is even worse\footnote{Older  gas-phase PDR models previously  predicted
low H$_2$S \mbox{column densities} \citep[][]{Jansen95,Sternberg95}.}   if one considers the uncertain rates of radiative association
reactions \mbox{S$^+$\,+\,H$_2$\,$\rightarrow$\,H$_2$S$^+$\,+\,$h\nu$} and \mbox{SH$^+$\,+\,H$_2$\,$\rightarrow$\,H$_3$S$^+$\,+\,$h\nu$}  included in the new gas-phase model. For the latter reaction, the main problem is that the
 electronic states of the reactants do not correlate with the $^1A_1$ ground electronic
state of the  activated complex H$_3$S$^+$$^*$ (denoted by $*$). Instead,  H$_3$S$^+$$^*$  forms in an excited triplet state ($^3A$).
\mbox{\cite{Herbst-etal:89}} proposed that a spin-flip followed by a radiative association can occur in interstellar conditions  and form H$_3$S$^+$$^*$($X^1A_1$)  
 \mbox{\citep[][]{Millar-Herbst:90}}.
In Appendix~\ref{Apendix:destruction}, we give arguments against this mechanism. 
For similar reasons, \mbox{\citet{Prasad82}}  avoided to include the \mbox{S$^+$\,+\,H$_2$} radiative association in their models.
Removing these reactions in pure gas-phase models  drastically decreases the H$_2$S$^+$ and H$_3$S$^+$
abundances, and thus those of SH and H$_2$S  (by a factor of $\sim$100 in these models). 
The alternative H$_2$S$^+$ formation route
through reaction \mbox{SH$^+$\,+\,H$_2$($v$\,=\,2)} is only efficient 
at the PDR surface (\mbox{$A_V$\,$<$\,1\,mag}). This is due to the large
H$_2$($v$\,=\,2) fractional abundances, \mbox{$f_2$\,$>$\,10$^{-6}$} at $T_k$\,$>$\,500\,K,  required to enhance the H$_2$S$^+$ production.
Therefore, and contrary to S$^+$ destruction, reaction of SH$^+$ with H$_2$
is not the dominant destruction pathway for SH$^+$. 
Only deeper inside the PDR,  reactions of S  with H$_3^+$ produce  small abundances of SH$^+$ and H$_2$S$^+$, but the \mbox{hydrogenation} of H$_n$S$^+$ ions is not efficient and  limits the gas-phase production  H$_2$S.

\subsection{Grain surface formation of solid H$_2$S \label{sec:surface_formation}}


Similarly to the formation of \mbox{water} ice (s-H$_2$O) on  grains  \mbox{\citep[e.g.,][]{Hollenbach09,Hollenbach12}}, the formation of H$_2$S may be dominated by
 grain  surface reactions followed by desorption back to the gas \mbox{\citep[e.g.,][]{Charnley87}}. Indeed,  water vapor is relatively abundant in the Bar  
 \citep[\mbox{$N$(H$_2$O)\,$\approx$\,10$^{15}$\,cm$^{-2}$};][]{Choi14,Putaud19} and
large-scale maps show  that the H$_2$O abundance peaks close to  cloud surfaces  \citep[][]{Melnick20}.

To investigate the \mbox{s-H$_2$S} formation on grains, we  updated the chemical model  by allowing S atoms to deplete onto grains as the gas temperature drops inside the molecular cloud \citep[for the basic grain chemistry formalism, see,][]{Hollenbach09}. The timescale  of this process ($\tau_{\rm gr,\,S}$) goes as
\mbox{$x({\rm S})^{-1}$\,$n_{\rm H}^{-1}\,T_{\rm k}^{-1/2}$}, where $x$(S) is the abundance of neutral \mbox{sulfur} atoms with respect to H nuclei. In a PDR, the abundance of H atoms is typically higher than that of S atoms\footnote{We only consider the
depletion of neutral S atoms. \mbox{S$^+$ ions} are expected to be more abundant than S atoms at the edge of the Orion Bar ($A_V$\,$\lesssim$\,2\,mag) where $T_{\rm k}$ and $T_{\rm d}$ are too high, and the FUV radiation 
field  too strong, to allow the formation of abundant grain mantles.} and H atoms 
stick on grains more frequently than S atoms unless \mbox{$x$(H)\,$<$\,$x$(S)$\cdot$0.18}. An adsorbed H atom 
(\mbox{s-H}) is weakly bound, mobile, and can diffuse throughout the grain surface until it finds an adsorbed S atom (\mbox{s-S}). 
If the timescale for a grain to be hit by a H atom ($\tau_{\rm gr,\,H}$) is shorter that the timescale for a \mbox{s-S} atom to \mbox{photodesorb} (\mbox{$\tau_{\rm photdes,\,S}$}) or sublimate  ($\tau_{\rm subl,\,S}$) then reaction of \mbox{s-H} with \mbox{s-S} will  proceed and form a \mbox{s-SH} radical roughly upon ``collision'' and without energy barriers \mbox{\citep[e.g.,][]{Tielens82,Tielens10}}. 
\mbox{Likewise}, if  \mbox{$\tau_{\rm gr,\,H}$\,$<$\,$\tau_{\rm photdes,\,SH}$} and \mbox{$\tau_{\rm gr,\,H}$\,$<$\,$\tau_{\rm subl,\,SH}$}, a newly adsorbed \mbox{s-H} atom can diffuse, find a grain site with an \mbox{s-SH} radical and react without barriers to form \mbox{s-H$_2$S}.  In these  surface processes,
a significant amount of S is ultimately transferred to \mbox{s-H$_2$S} \citep[e.g.,][]{Vidal17},
which can subsequently desorb: thermally, by FUV photons, or by cosmic rays. 
 In addition, laboratory experiments show
that the excess energy of certain exothermic surface reactions can
 promote the direct desorption of the product \mbox{\citep[][]{Minissale16}}.
 In particular, reaction \mbox{s-H\,+\,s-SH} directly desorbs
H$_2$S  with a maximum efficiency of $\sim$\,60\,\%~\citep[as observed in experiments,][]{Oba18}.
Due to the high flux of FUV photons in PDRs, chemical desorption may not always compete with photodesorption. However, it can be a dominant process inside molecular clouds \cite[][]{Garrod07,Esplugues16,Vidal17,Almaida20}.

The photodesorption timescale of an ice mantle is proportional to \mbox{$Y^{-1}$\,$G_0^{-1}$\,exp\,(+$b$\,$A_V$)}, where $Y$ is the photodesorption yield (the number of desorbed atoms or molecules per incident photon) and $b$ is a dust-related FUV field absorption factor. The timescale for mantle sublimation (thermal desorption) goes as \mbox{$\nu_{\rm ice}^{-1}$\,exp\,(+$E_{\rm b}$\,/\,$k\,T_{\rm d}$)}, where 
$\nu_{\rm ice}$ is the characteristic vibrational frequency of the solid lattice, $T_{\rm d}$ is the dust grain temperature,
and $E_{\rm b}/k$ is the adsorption binding energy of the species (in K). 
Binding energies play a crucial role in model predictions because they determine the freezing temperatures and sublimation timescales. 
Table~\ref{table:surf-param} lists the $E_{\rm b}/k$ and $Y$ values  considered here.

\begin{figure}[t]
\centering   
\includegraphics[scale=0.56, angle=0]{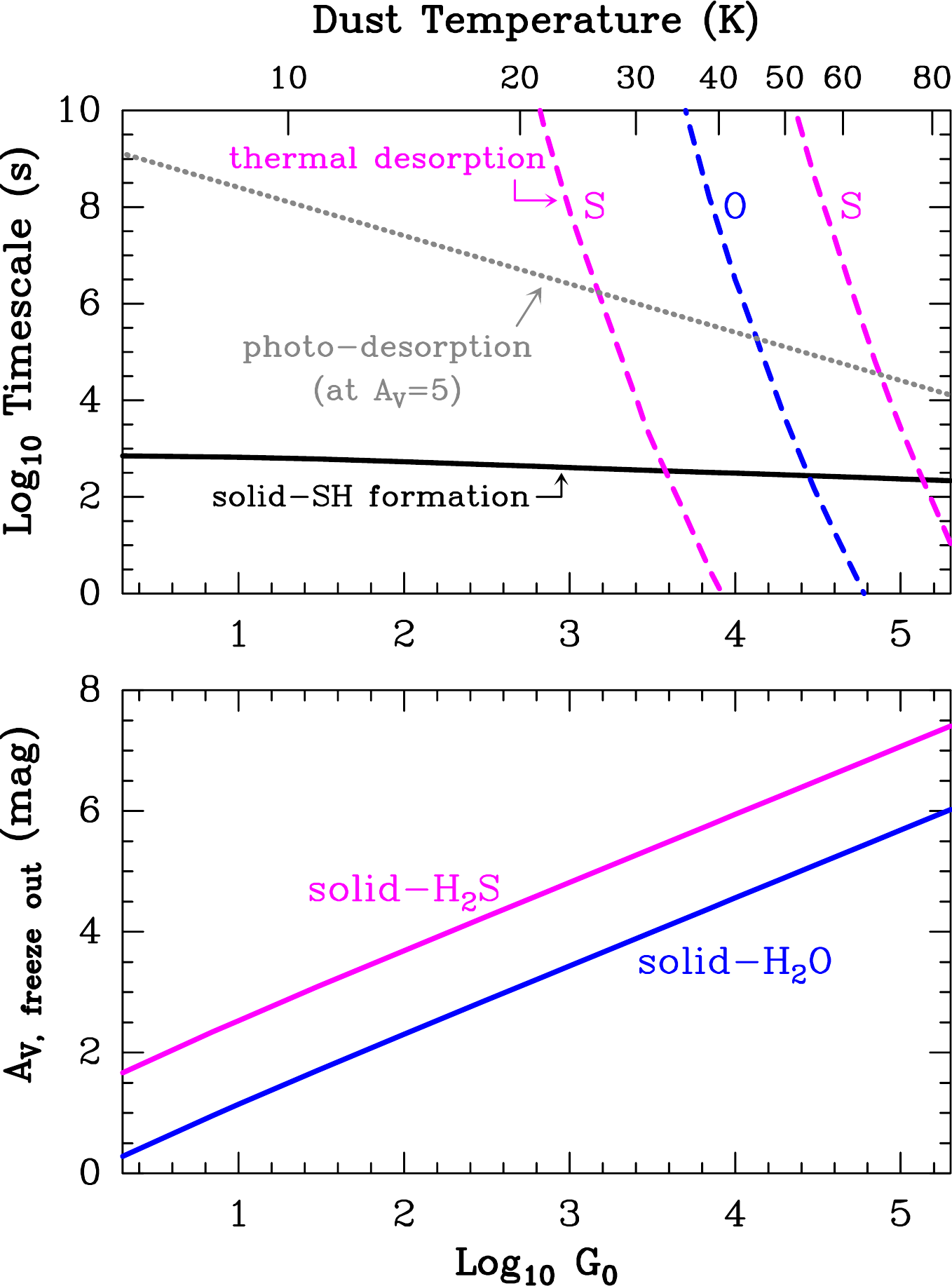}
\caption{Representative timescales relevant to the formation of \mbox{s-H$_2$S} and \mbox{s-H$_2$O} as well as their freeze-out depths.
$\textit{Upper panel}$: 
The continuous black curve is the timescale for a grain to be hit by an H atom.
Once in the grain surface, the H atom  \mbox{diffuses} and can react  with an adsorbed  S atom to form  \mbox{s-SH}.
The dashed magenta curves show the timescale for thermal \mbox{desorption} of an \mbox{s-S} atom 
(\mbox{$E_{\rm b}/k$\,(S)\,=\,1100\,K} left curve, and 2600\,K right curve) and of an \mbox{s-O} atom (blue curve; \mbox{$E_{\rm b}/k$\,(O)\,=\,1800\,K}).  The gray dotted curve is the  photodesorption timescale  of \mbox{s-S}. 
At $G_0$ values where the 
continuous line is below the dashed and dotted lines, \mbox{s-O} and \mbox{s-S} atoms remain on grain surfaces sufficiently long to combine with an adsorbed H atom and form \mbox{s-OH} and \mbox{s-SH} (and then \mbox{s-H$_2$O} and \mbox{s-H$_2$S}). These timescales are for \mbox{$n_{\rm H}$\,=\,10$^5$\,cm$^{-3}$} and \mbox{$n$(H)\,=\,100\,cm$^{-3}$}.
$\textit{Bottom panel}$: Freeze-out depth at which most O and S are incorporated as 
\mbox{s-H$_2$O} and \mbox{s-H$_2$S} (assuming no chemical desorption and $T_{\rm k}$\,=\,$T_{\rm d}$).}
\label{fig:ice-time-scales}
\end{figure}

Representative timescales of the basic grain  processes described above
are summarized in the upper panel of Fig.~\ref{fig:ice-time-scales}.
In this plot, $T_d$ is a characteristic dust temperature inside the PDR, \mbox{$T_d$\,=\,(3$\cdot$10$^4$\,+\,2$\cdot$10$^3$\,$G_{0}^{1.2}$)$^{0.2}$}, taken  from
\citet{Hollenbach09}. In~the upper panel, the continuous black curve is the timescale for a grain to be hit by an H atom ($\tau_{\rm gr,\,H}$).
The dashed magenta curves show the timescale for thermal desorption of an \mbox{s-S} atom 
($\tau_{\rm subl,\,S}$) (left curve for \mbox{$E_{\rm b}/k$\,(S)\,=\,1100\,K}  and right curve for \mbox{$E_{\rm b}/k$\,(S)\,=\,2600\,K}), and the same for an \mbox{s-O} atom (blue curve).  The gray dotted curve is the timescale for \mbox{s-S}
atom photodesorption  ($\tau_{\rm photodes,\,S}$)  at $A_V$\,=\,5\,mag. 
At $G_0$ strengths where the continuous line is below the dashed and dotted lines, an adsorbed \mbox{s-S} atom remains on the grain surface sufficiently long to react with 
a diffusing \mbox{s-H atom}, form \mbox{s-SH}, and ultimately   \mbox{s-H$_2$S}.

Figure~\ref{fig:ice-time-scales} shows that, if one takes
\mbox{$E_{\rm b}/k$\,(S)\,=\,1100\,K} \citep[the most common value  in the literature;][] {Hasegawa93}, the formation of \mbox{s-H$_2$S} is possible inside clouds illuminated by modest FUV fields, when grains are sufficiently cold (\mbox{$T_{\rm d}$\,$<$\,22\,K}). However, recent calculations of \mbox{s-S} atoms adsorbed on water ice surfaces suggest higher binding energies \citep[\mbox{$\sim$\,2600\,K};][]{Wakelam17}. This would imply that S atoms freeze at higher $T_{\rm d}$ ($\lesssim$\,50\,K) and that \mbox{s-H$_2$S} mantles form in more strongly illuminated PDRs 
\citep[the observed $T_{\rm d}$ at the edge of the Bar is \mbox{$\simeq$\,50\,K} and decreases to $\simeq$35\,K behind the PDR; see,][]{Arab_2012}.

The freeze-out depth for sulfur in a PDR, the $A_V$ at which most sulfur is incorporated as \mbox{S-bearing} solids (\mbox{s-H$_2$S} in our simple model) can be estimated by equating  \mbox{$\tau_{\rm gr,\,S}$} and \mbox{$\tau_{\rm photdes,\,H_2S}$}. This implicitly assumes the H$_2$S chemical desorption does not dominate in \mbox{FUV-irradiated} regions,
which is in line with the particularly large FUV absorption cross section of 
\mbox{s-H$_2$S} measured in laboratory experiments \citep{Cruz14}.
With these assumptions, the lower panel of Fig.~\ref{fig:ice-time-scales} shows the
predicted \mbox{s-H$_2$S} and \mbox{s-H$_2$O} freeze-out depths.
Owing to the lower abundance and higher  atomic mass
of sulfur atoms  (i.e., grains are hit slower by S atoms than by O atoms), the H$_2$S  freeze-out depth appears slightly deeper than that of \mbox{water ice}.  For the FUV-illumination conditions in the  Bar, the freeze-out depth of sulfur is expected at \mbox{$A_V$\,$\gtrsim$\,6\,mag}. This implies that photodesorption of \mbox{s-H$_2$S}  can produce enhanced abundances of gaseous \mbox{H$_2$S at $A_V$\,$<$\,6\,mag}.

\begin{table}[t]
\caption{Adopted binding energies and photodesorption yields. \label{table:surf-param}} 
\centering
\begin{tabular}{ccc@{\vrule height 8pt depth 5pt width 0pt}}
\hline\hline
Species        &   $E_{\rm b}/k$       &       Yield  \\
 		       &       (K)           &   (FUV photon)$^{-1}$ \\
\hline
S              & 1100\,$^a$/2600\,$^b$ &   10$^{-4}$         	\\
SH             & 1500\,$^a$/2700\,$^b$ &   10$^{-4}$         	\\
H$_2$S         & 2700\,$^{b,c}$        &  1.2$\times$10$^{-3}$\,$^g$ (as H$_2$S)  \\
CO             &      1300\,$^d$       &  3$\times$10$^{-3}$\,$^h$    \\
O              &      1800\,$^e$       &  10$^{-4}$\,$^h$         	\\
O$_2$          &      1200\,$^d$       &  10$^{-3}$\,$^h$         	 \\
OH             &      4600\,$^a$       &  10$^{-3}$\,$^h$         	 \\
H$_2$O         &      4800\,$^f$       &  10$^{-3}$\,$^h$ (as H$_2$O)\\
		       &                       &  2$\times$10$^{-3}$\,$^h$ (as OH)\\
\hline                                    
\end{tabular}
\tablefoot{$^a$\citet{Hasegawa93}. $^b$\citet{Wakelam17}. $^c$\citet{Collings04}. $^d$\citet{Minissale16}.
$^e$\citet{He15}.$^f$\citet{Sandford88}. $^g$\citet{Fuente17} $^h$See, \citet{Hollenbach09}.}
\end{table}

\begin{figure*}[t]
\centering   
\includegraphics[scale=0.405, angle=0]{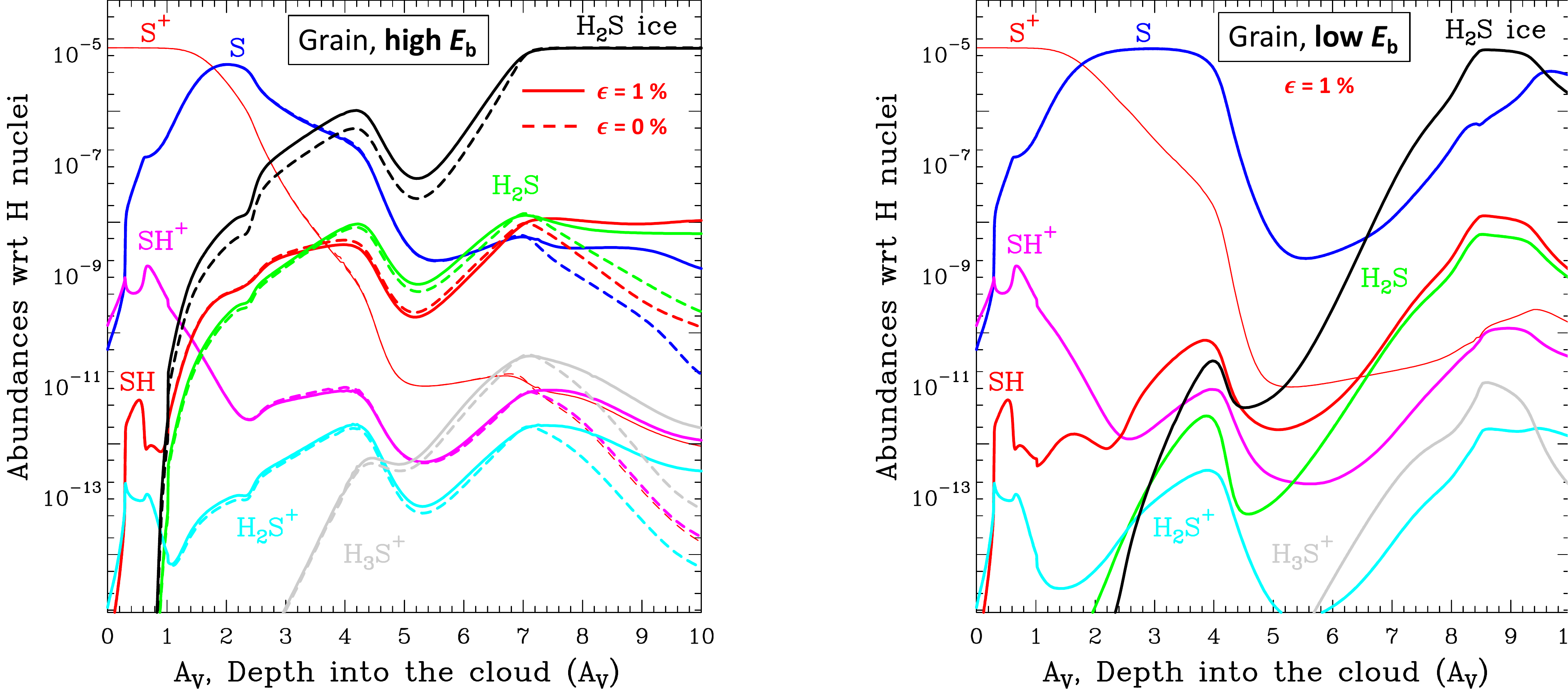}
\caption{Gas-grain  PDR models leading to the formation of \mbox{s-H$_2$S} (shown as black curves).  \mbox{Continuous} colored curves show gas-phase fractional abundances as a function of depth into the cloud. $\epsilon$ refers to the efficiency of the 
 chemical desorption reaction  \mbox{s-H\,+\,s-H$_2$S\,$\rightarrow$\,SH\,+\,H$_2$} (see text). \textit{Left panel}: Gas-grain \mbox{high $E_{\rm b}$} model (high adsorption binding energies for S and SH, see Table~\ref{table:surf-param}). \textit{Right panel}:  \mbox{Low $E_{\rm b}$} model.}
\label{fig:PDR_grains}
\end{figure*}

FUV-irradiation and thermal desorption of H$_2$S ice mantles have been studied in the laboratory \citep[e.g.,][]{Cruz14,Escobar11}. These experiments show that pure \mbox{s-H$_2$S} ices thermally desorb around 82\,K, and at higher  temperatures for \mbox{H$_2$S--H$_2$O} ice mixtures. 
 These experiments determine a photodesorption yield of 
\mbox{$Y_{\rm H_2S}$\,$\sim\,$1.2$\times$10$^{-3}$} molecules per FUV photon 
\citep[see also][]{Fuente17}. Regarding surface grain chemistry, experiments show that  
reaction \mbox{s-H\,+\,s-SH\,$\rightarrow$\,s-H$_2$S} is exothermic \mbox{\citep{Oba18},}
whereas  reaction \mbox{s-H\,+\,s-H$_2$S}, although it has an activation energy barrier of $\sim$1500\,K, it may directly desorb gaseous SH. Finally, reaction  \mbox{s-SH\,+\,s-SH\,$\rightarrow$\,s-H$_2$S$_2$} 
may trigger the formation of doubly sulfuretted species, but it requires mobile 
\mbox{s-SH} radicals \mbox{\citep[e.g.,][]{Escobar11,Fuente17}}. Here we will only
consider surface reactions with mobile \mbox{s-H}.

\subsection{Gas-grain  PDR model results}\label{sec:grain_models}

Here we show PDR model results in which we add a simple  network of gas-grain reactions for a small number of  \mbox{S-bearing} \mbox{(S, SH, and H$_2$S)} and \mbox{O-bearing}
\mbox{(O, OH, H$_2$O, O$_2$, and CO)} species. These species can 
adsorb on grains as temperatures drop, photodesorb by FUV photons (stellar and secondary), desorb by direct impact of cosmic-rays, or sublimate at a given PDR depth (depending on $T_{\rm d}$ and on their $E_{\rm b}$). Grain size distributions 
(\mbox{$n_{\rm gr}$\,$\propto$\,$a^{-3.5}$}, where $a$ is the grain radius) and  gas-grain reactions are treated within the  Meudon code formalism 
\citep[see,][]{LePetit06,Goico07,LeBourlot12,Bron14}.
As grain surface chemistry reactions we include  
\mbox{s-H\,+\,s-X\,$\rightarrow$\,s-XH} and \mbox{s-H\,+\,s-XH\,$\rightarrow$\,s-H$_2$X}, where \mbox{s-X} refers to \mbox{s-S} and \mbox{s-O}. In addition, we add the direct chemical desorption reaction \mbox{s-H\,+\,s-SH\,$\rightarrow$\,H$_2$S}
with an efficiency of 50\,$\%$ per reactive event, and also tested different efficiencies ($\epsilon$) for the chemical desorption process
\mbox{s-H\,+\,s-H$_2$S\,$\rightarrow$\,SH\,+\,H$_2$}.

In our models we compute the relevant gas-grain timescales and atomic abundances at every depth $A_V$ of the PDR. If the timescale for a grain to be struck by an H atom ($\tau_{\rm gr,\,H}$) is shorter than the timescales to sublimate or to \mbox{photodesorb}  an  \mbox{s-X} atom or a \mbox{s-XH} molecule; and if H atoms stick on grains  more frequently than X atoms, we simply assume
these surface  reactions proceed instantaneously.  
At large $A_V$, larger than the freeze-out depth, this grain chemistry builds
abundant \mbox{s-H$_2$O} and \mbox{s-H$_2$S} ice mantles.

Figure~\ref{fig:PDR_grains} shows results of two types of gas-grain models. The only difference
between them is the adopted adsorption binding energies for \mbox{s-S} and \mbox{s-SH}. Left panel is for a \mbox{``high $E_{\rm b}$''} model  and right panel is for a \mbox{``low $E_{\rm b}$''} model (see Table~\ref{table:surf-param}). 
We note that these models do not include the gas-phase radiative association reactions 
\mbox{S$^+$\,+\,H$_2$\,$\rightarrow$\,H$_2$S$^+$\,+\,$h\nu$} and \mbox{SH$^+$\,+\,H$_2$\,$\rightarrow$\,H$_3$S$^+$\,+\,$h\nu$}; although their effect
is smaller than in pure gas-phase models.

The chemistry of the most exposed PDR surface layers ($A_V$\,$\lesssim$\,2\,mag) is the same to that of the gas-phase models discussed in Sect.~\ref{sec:gas_models}. Photodesorption keeps dust grains free of ice mantles, and fast gas-phase ion-neutral reactions, photoreactions, and reactions with \mbox{FUV-pumped} H$_2$ drive the chemistry.
The resulting SH$^+$ abundance profile is nearly identical and 
 there is no need to invoke  depletion of elemental sulfur from the gas-phase to explain the observed SH$^+$ emission (see Fig.~\ref{fig:PDR2MTC_gas}).
Beyond these first PDR  irradiated layers, the chemistry does change because the  
formation of \mbox{s-H$_2$S} on grains and subsequent desorption alters the chemistry of the other \mbox{S-bearing} hydrides.

\begin{figure*}[ht]
\centering   
\includegraphics[scale=0.455, angle=0]{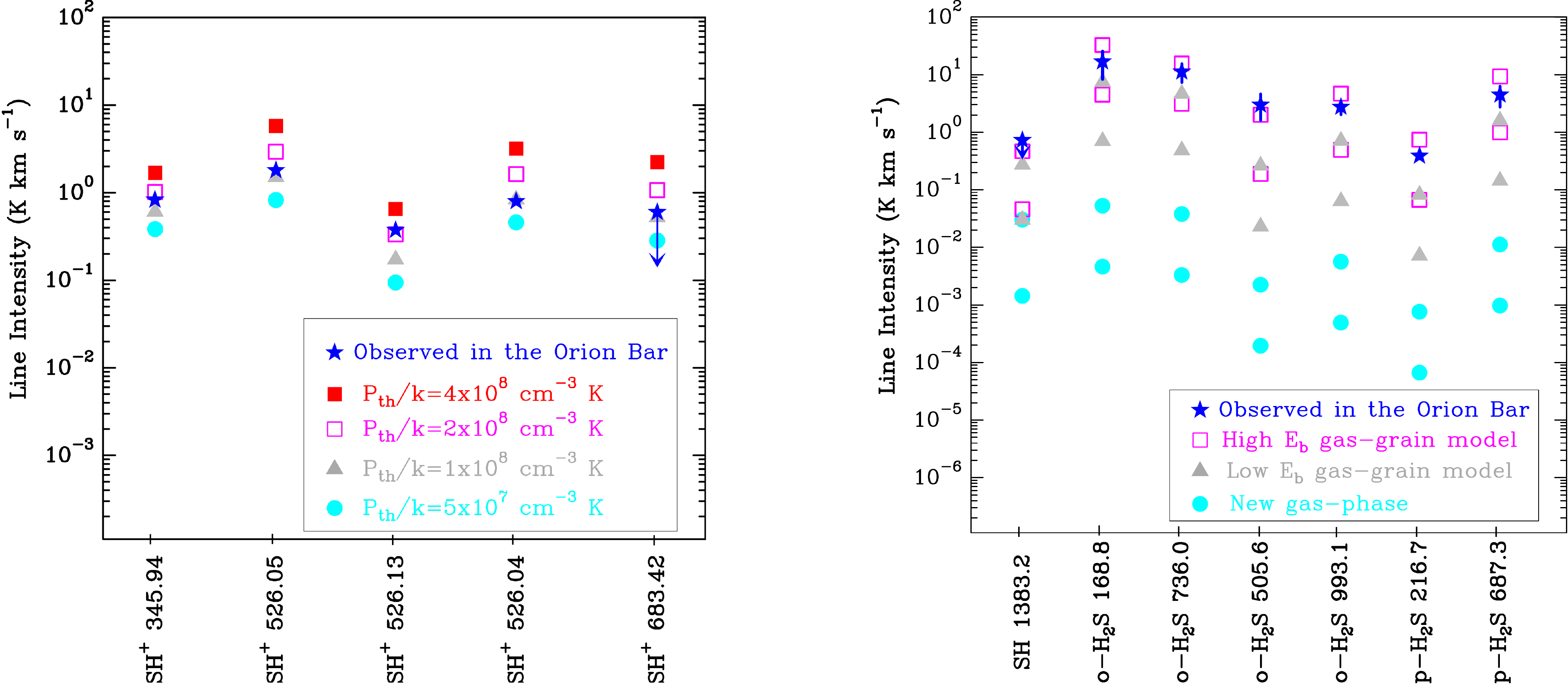}
\caption{Line intensity predictions for different isobaric PDR models.
Calculations were carried out in a multi-slab  Monte Carlo code \mbox{(Sect.~\ref{sec:MTC_mods})} that uses the output of the PDR model. Blue stars show the line intensities  observed toward the Bar (corrected by beam dilution).
\textit{Left panel}: SH$^+$ emission models for  PDRs of different 
\mbox{$P_{\rm th}$} values and $\alpha$\,=\,5$^o$. 
\textit{Right panel}: SH and H$_2$S (adopting an OTP ratio of~3) emission from:
\mbox{high $E_{\rm b}$}  (magenta squares), \mbox{low $E_{\rm b}$} (gray triangles), and gas-phase  (cyan circles) PDR models, all with 
\mbox{$P_{\rm th}$\,/\,$k$}\,=\,2$\times$10$^{8}$\,K\,cm$^{-3}$.  
Upper limit intensity predictions are for a PDR with an inclination angle of $\alpha$\,=\,5$^o$ with respect to a edge-on geometry. Lower limit intensities refer to a face-on PDR model.
\label{fig:PDR2MTC_gas}}
\end{figure*}

In model \mbox{high $E_{\rm b}$},   S atoms start to freeze out  closer to the PDR edge 
\mbox{($T_{\rm d}$\,$<$\,50\,K)}. Because of the increasing densities and decreasing temperatures,  the \mbox{s-H$_2$S} abundance with respect to H nuclei reaches $\sim$10$^{-6}$ at $A_V$\,$\simeq$\,4\,mag. In model \mbox{low $E_{\rm b}$}, this level of \mbox{s-H$_2$S}  abundance is only reached  beyond an $A_V$ of 7\,mag. At lower $A_V$, the formation of \mbox{s-H$_2$S} on bare grains and subsequent photodesorption produces more \mbox{H$_2$S} than pure-gas phase models independently of whether H$_2$S chemical desorption  is included or not. In these intermediate PDR  layers,  at $A_V$\,$\simeq$\,2-7\,mag for the strong irradiation conditions in the Bar, the flux of  FUV photons drives much of the chemistry, desorbing grain mantles, preventing complete freeze out, and dissociating the gas-phase products.

There are two H$_2$S abundance peaks at \mbox{$A_V$\,$\simeq$\,4 and 7\,mag}.  The H$_2$S abundance in these  ``photodesorption peaks'' depends on the amount of  \mbox{s-H$_2$S} mantles formed on grains  and on the balance between \mbox{s-H$_2$S} photodesorption and \mbox{H$_2$S} photodissociation (which now becomes the major source of SH).
 The enhanced H$_2$S abundance modifies the chemistry of H$_2$S$^+$ and H$_3$S$^+$ as well: H$_2$S photoionization (with a threshold at $\sim$10.4\,eV) becomes the dominant source of H$_2$S$^+$  at \mbox{$A_V$\,$\simeq$\,4\,mag}  because the H$_2$\,($v$\,$\geq$2) abundance is too low to make reaction~(\ref{reac-2}) competitive. Besides, reactions of  H$_2$S with abundant molecular ions such as HCO$^+$, H$_{3}^{+}$, and H$_3$O$^+$ dominate the H$_3$S$^+$ production.
  
Our \mbox{gas-grain} models predict that other \mbox{S-bearing}  molecules, such as SO$_2$ and SO, can be the major sulfur reservoirs at these intermediate PDR depths. However, their abundances strongly depend on those of O$_2$ and OH through reactions \mbox{S\,+\,O$_2$\,$\rightarrow$\,SO\,+\,O} and \mbox{SO\,+\,OH\,$\rightarrow$\,SO$_2$\,+\,H} \citep[see e.g.,][]{Sternberg95,Fuente16,Fuente19}. These reactions link the chemistry of  \mbox{S-} and \mbox{O-bearing} neutral molecules
\mbox{\citep{Prasad82}} and are an
important sink of S atoms  at \mbox{$A_V$\,$\gtrsim$\,5\,mag}. \mbox{However}, while large column densities of OH have been detected in the Orion Bar \citep[$\gtrsim$\,10$^{15}$\,cm$^{-2}$;][]{Goicoechea11}, O$_2$ remains undetected despite deep searches  \citep{Melnick12}. \mbox{Furthermore}, the inferred upper limit $N$(O$_2$) columns   are below the expectations of PDR models \citep{Hollenbach09}. This discrepancy likely implies that these \mbox{gas-grain}  models miss details of the  grain surface chemistry leading to \mbox{O$_2$} \citep[for other environments and modeling approaches see, e.g.,][]{Ioppolo08,Taquet16}. 
Here we will not discuss SO$_2$, SO, or O$_2$ further.

At large cloud depths, \mbox{$A_V$\,$\gtrsim$\,8\,mag}, the FUV flux is largely attenuated, temperatures drop, the chemistry becomes slower,
 and other chemical processes dominate.
The H$_2$S abundance is controlled by the chemical desorption reaction 
\mbox{s-H\,+\,s-SH\,$\rightarrow$\,H$_2$S}. This process keeps a floor of detectable
H$_2$S  abundances ($>$10$^{-9}$) in regions shielded from stellar FUV radiation.
In addition, and although not energetically favorable, the chemical desorption \mbox{s-H\,+\,s-H$_2$S\,$\rightarrow$\,SH\,+\,H$_2$} enhances the SH production 
at large $A_V$ (the enhancement depends on the desorption \mbox{efficiency $\epsilon$}), which in turn boosts the abundances of other \mbox{S-bearing} species, including that of neutral S atoms. 

The H$_2$S abundances predicted by the \mbox{high $E_{\rm b}$} model reproduce the 
H$_2$S line intensities observed in the Bar \mbox{(Sect.~\ref{line-models})}. In this model   \mbox{s-H$_2$S}  becomes the main sulfur reservoir. However, we stress that here we do not consider the formation of more complex \mbox{S-bearing} ices such as
\mbox{s-OCS}, \mbox{s-H$_2$S$_2$}, \mbox{s-S$_n$}, \mbox{s-SO$_2$} or \mbox{s-HSO} 
\citep[][]{Escobar11,Vidal17,Laas19}. 
Together with our steady-state solution of the chemistry, this implies that our predictions are not precise deep inside the PDR. However, we recall that our observations  refer to the edge of the Bar, so it is not plausible that the model conditions at \mbox{$A_V$\,$\gtrsim$\,8\,mag}  represent the line of sight we observe. 

Model \mbox{low $E_{\rm b}$} produces less H$_2$S in the PDR layers below $A_V$\,$\lesssim$\,8\,mag
because S atoms  do not freeze until the dust temperature drops deep inside the PDR. 
Even beyond these layers, thermal desorption of \mbox{s-S} maintains higher abundances
of S atoms at large depths.
Indeed, model \mbox{low $E_{\rm b}$} predicts that 
the major sulfur reservoir deep inside the cloud are
gas-phase S atoms. This agrees with recent chemical models of cold dark clouds \mbox{\citep{Vidal17,Almaida20}}. 

\begin{figure*}[ht]
\centering   
\includegraphics[scale=0.43, angle=0]{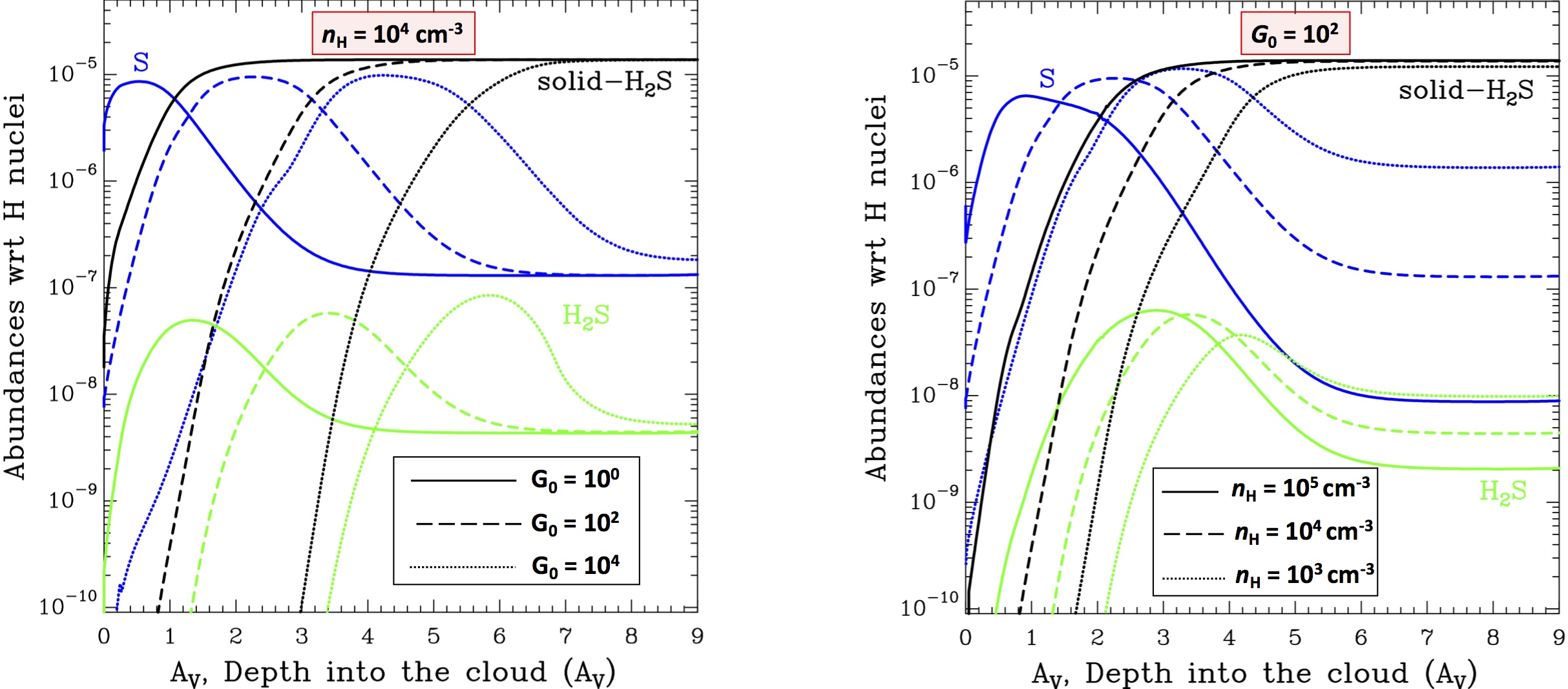}
\caption{Constant density gas-grain PDR models using the \mbox{high $E_{\rm b}$} chemical network and undepleted sulfur elemental abundances. \mbox{\textit{Left panel}: Effects of changing} the FUV radiation field. \textit{Right panel}: Effects of varying the gas density.}
\label{fig:PDR-grid}
\end{figure*}

\subsection{Line intensity comparison and H$_2$S ortho-to-para ratio}\label{line-models}

We now specifically compare the SH$^+$, SH, and H$_2$S line intensities implied by the different PDR models with the intensities observed toward the DF position of the Bar. 
We used the output of the PDR models --  $T_{\rm k}$, $T_{\rm d}$, $n$(H$_2$), $n$(H), $n_e$, $n$(SH$^+$), $n$(SH), and $n$(H$_2$S)  profiles from $A_V$\,=\,0 to 10\,mag --  as input for a \mbox{multi-slab} Monte Carlo model of their line excitation, including formation pumping \mbox{(formalism presented in Sect.~\ref{sec:MTC_mods})} and radiative transfer. 
  As the Orion Bar is not a perfectly \mbox{edge-on}, this comparison requires a knowledge of the tilt angle ($\alpha$) with respect to a pure \mbox{edge-on} PDR. 
Different studies suggest $\alpha$ of  $\approx$\,5$^o$ 
\citep[e.g.,][]{Jansen95,Melnick12,Andree17}. This inclination implies
an increase in line-of-sight column density, compared to a \mbox{face-on} PDR, by a geometrical factor \mbox{(sin\,$\alpha$)$^{-1}$}.  It also means that \mbox{optically thin} lines are limb-brightened. 

The left panel of \mbox{Fig.~\ref{fig:PDR2MTC_gas}} shows SH$^+$ line intensity predictions for isobaric PDR models of different $P_{\rm th}$ values (leading to different $T_{\rm k}$ and $n_{\rm H}$ profiles).
Since the bulk of the SH$^+$ emission arises from the PDR edge ($A_V$\,$\simeq$\,0 to 2\,mag) all models (gas-phase or gas-grain) give similar results.
 The best fit is  for 
\mbox{$P_{\rm th}$\,$\simeq$\,(1--2)$\times$10$^8$\,cm$^{-3}$\,K}  and   $\alpha$\,$\simeq$5$^o$. 
These high pressures, at least close to the DF, agree with those inferred from ALMA images of \mbox{HCO$^+$\,($J$\,=\,4-3)} emission \citep{Goico16}, Herschel observations of high-$J$ CO lines \citep{Joblin18}, and IRAM\,30\,m  detections of carbon recombination lines \citep{Cuadrado19}.

Right panel of \mbox{Fig.~\ref{fig:PDR2MTC_gas}} shows SH and H$_2$S line emission predictions for the \mbox{high $E_{\rm b}$} gas-grain model (magenta squares), \mbox{low $E_{\rm b}$} gas-grain model
(gray triangles), and a pure gas-phase model (cyan circles).
For each  model, the upper limit intensities refer to radiative transfer calculations with an inclination angle $\alpha$\,=\,5$^o$.  The lower intensity limits refer to a face-on PDR.
 Gas-phase models largely underestimate the observed H$_2$S intensities.
Model  \mbox{low $E_{\rm b}$} produces higher H$_2$S columns and brighter H$_2$S lines, but still below the observed levels (by up to a factor of ten).
Model \mbox{high $E_{\rm b}$}  provides a good agreement with observations; the two possible inclinations bracket the observed intensities, and it should be considered as the reference model of the  Bar. It is also consistent with the observational SH upper limits.

Our observations and models provide a \mbox{(line-of-sight)} \mbox{$N$($o$-H$_2$S)/$N$($p$-H$_2$S)} OTP ratio of 2.9\,$\pm$\,0.3, consistent with the  \mbox{(gas-phase) high-temperature} statistical equilibrium value. \mbox{However}, the cold \mbox{``nuclear-spin-temperatures''} ($T_{\rm spin}$\,$\ll$\,$T_{\rm k}$; see definition in \mbox{eq.~\ref{eq-OTP}}) implied by  the low  water vapor OTP ratios 
 observed in some sources (\mbox{$<$\,2.5}) have been associated with the temperature of the ice \mbox{mantles} 
 where H$_2$O  molecules might have formed 
 \mbox{\citep[i.e., \mbox{$T_{\rm spin}$\,$\simeq$\,$T_{\rm d}$};][]{Mumma87,Lis13}}. In the case of H$_2$S, our derived OTP ratio toward the DF position implies any $T_{\rm spin}$ above \mbox{30\,$\pm$10\,K} (see Fig.~\ref{fig:OTP}). Hence, this temperature  might be also compatible with \mbox{s-H$_2$S} formation\footnote{\cite{Crockett14} inferred 
\mbox{$N$($o$-H$_2$S)/$N$($p$-H$_2$S)\,=\,2.5\,$\pm$\,0.8}  in the hot core of Orion~KL using LTE rotational digrams. However, they \mbox{favored} an OTP ratio of \,1.7\,$\pm$\,0.8 based on the column density ratio of selected pairs of rotational levels with similar energies. This latter OTP ratio  implies  \mbox{$T_{\rm spin}$(H$_2$S)\,$\simeq$\,12\,K}  (Fig.~\ref{fig:OTP}), \mbox{perhaps} related to much colder dust grains than in PDRs or to colder gas conditions  
just before the hot core phase; so that reactive collisions did not have time to establish the statistical equilibrium value. We note that the observed OTP ratios
of H$_2$CO, H$_2$CS, and H$_2$CCO in the  Bar are also $\sim$3 \citep{Cuadrado17}.} in warm grains
if $T_{\rm spin}$\,$\simeq$\,$T_{\rm d}$ upon formation is preserved in the gas-phase after \mbox{photodesorption} \citep[e.g.,][]{Guzman13}. Interestingly, the H$_2$O  OTP ratio derived from observations of the Orion Bar  is \mbox{2.8\,$\pm$\,0.1} \citep{Putaud19} and implies
 \mbox{$T_{\rm spin}$(H$_2$O)\,=\,35\,$\pm$\,2\,K}. This value is compatible with
  \mbox{$T_{\rm spin}$(H$_2$S)} and might reflect the similar  $T_{\rm d}$ of the PDR layers where most \mbox{s-H$_2$O}
 and \mbox{s-H$_2$S} form and photodesorb.
\mbox{Nevertheless}, laboratory experiments have challenged this 
\mbox{$T_{\rm spin}$\,$\simeq$\,$T_{\rm d}$}  association, at least for \mbox{s-H$_2$O}: cold water ice surfaces, at 10\,K, photodesorb H$_2$O molecules  with an OTP ratio of $\sim$3 \citep{Hama16}. Follow up observations of \mbox{$p$-H$_2$S} lines across the Bar will allow us to study possible variations of the  OTP ratio as $G_0$ diminishes and grains get colder.

\subsection{Generalization to different G$_0$ and n$_{\rm H}$ conditions} 

In this section we generalize our results to a broader range of gas densities and FUV illumination conditions
(i.e., to clouds with different \mbox{$G_0$\,/\,$n_{\rm H}$} ratios). We run several PDR models using the \mbox{high $E_{\rm b}$} gas-grain chemistry. 
 The main difference compared to the Orion Bar models is that here we model constant density clouds with standard interstellar grain properties  ($R_V$\,=\,3.1).
\mbox{Figure~\ref{fig:PDR-grid}} (left panel) shows models of clouds  with constant \mbox{$n_{\rm H}$\,=\,10$^4$\,cm$^{-3}$} and varying
FUV radiation fields, while \mbox{Fig.~\ref{fig:PDR-grid} (right panel)}  show models of constant FUV illumination ($G_0$\,=\,100) and varying gas densities\footnote{In these models we consider undepleted [S/H] abundances
and only  the chemical \mbox{desorption}   \mbox{s-H\,+\,s-SH\,$\rightarrow$\,H$_2$S} 
(with a 50\,\% efficiency).}.
The main result of this study is the similar gas-phase H$_2$S column density (a few 10$^{14}$\,cm$^{-2}$ up to \mbox{$A_V$\,=\,10}) and H$_2$S abundance peak (a few 10$^{-8}$ close to the \mbox{FUV-irradiated cloud} edge) predicted by these models nearly irrespective of $G_0$\ and $n_{\rm H}$. A
similar conclusion was  reached previously for water vapor in \mbox{FUV-illuminated clouds}  \mbox{\citep[][]{Hollenbach09,Hollenbach12}}.
\mbox{Increasing} $G_0$ shifts the position of the 
H$_2$S abundance peak to larger $A_V$  until the rate of S atoms sticking on grains balances
the  H$_2$S photodissociation rate  (the dominant H$_2$S destruction mechanism except in shielded gas;
see also Fig.\ref{fig:ice-time-scales}). Since \mbox{s-H$_2$S} photodesorption and H$_2$S photodissociation rates depend on $G_0$, the peak H$_2$S abundance in the PDR
is roughly the same independently of  \mbox{$G_0$}. 
On the other hand, the formation rate of \mbox{s-H$_2$S} mantles depends on the product 
\mbox{$n$(S)\,$n_{\rm gr}$\,$\propto$\,$n_{\rm H}^{2}$}, whereas the H$_2$S photodesorption rate depends on \mbox{$n_{\rm gr}$\,$\propto$\,$n_{\rm H}$}. Hence, the H$_2$S abundance peak moves toward the cloud surface for denser PDRs (like the Orion Bar).
The exact abundance value depends on the adopted grain-size distribution and on the H$_2$S photodesorption yield \citep[which is well constrained by  experiments; see,][]{Cruz14,Fuente17}. 

The role of chemical desorption increases and can dominate beyond the photodesorption peak  as the flux of stellar FUV photons is attenuated. 
Here we do not carry out an exhaustive study of this mechanism, which is hard to model in full detail because its efficiency decreases considerably  with the properties of grain surfaces \citep[bare vs. icy; see e.g.,][]{Minissale14}.
In our models, and depending on $\zeta_{\rm CR}$, photodesorption by secondary FUV photons can also be important in cloud interiors.
These processes limit the conversion of most of the sulfur
reservoir into \mbox{S-bearing} ices  and increase the  abundance of other gas-phase species  deep inside clouds, notably S atoms and H$_2$S molecules. 

The H$_2$S abundance in shielded  gas depends on the \mbox{destruction} rate by gas-phase reactions different than photodissociation, in particular H$_2$S reactions with H$_{3}^+$. The H$_{3}^+$ abundance increases with $\zeta_{\rm CR}$ and decreases with the electron density.   Figure~\ref{fig:PDR-grid} (\textit{right}) shows models of constant  $G_0$ 
and constant $\zeta_{\rm CR}$ in which the H$_2$S abundance at large depths increases
with decreasing density (more penetration of FUV photons, more ionization, more electrons, less H$_{3}^+$). The lowest gas density model, \mbox{$n_{\rm H}$\,=\,10$^3$\,cm$^{-3}$}, shows the highest  H$_2$S abundance at large $A_V$.
\mbox{Because} S freeze-out is less efficient at low densities, the low-density model  shows  higher gas-phase S  abundances  at large depths, making
atomic S a dominant gas-phase sulfur reservoir. 
Unfortunately, direct observation of atomic S in cold gas is complicated, which makes
it difficult to benchmark this prediction. 

In warm PDRs, in addition to S radio \mbox{recombination} lines \cite[e.g.,][]{Smirnov95}, the $^3P$  fine-structure  lines of atomic sulfur, \mbox{the [\SI]\,25, 56\,$\upmu$m lines}, can be interesting diagnostics of gas physical conditions and of [S/H] abundances. 
Unfortunately, the low sensitivity of previous infrared telescopes was not sufficient to detect the [\SI]\,25\,$\upmu$m line (\mbox{$\Delta E_{12}$\,=\,570\,K}) in the Orion  Bar \mbox{\citep{Rosenthal00}}; although it is detected in  protostellar outflows  
\citep[e.g.,][]{Neufled09,Goicoechea12}.
Moreover, the \mbox{$^3P_2$-$^1D_2$} \mbox{forbidden} line of atomic sulfur at 1.082\,$\upmu$m can be an interesting tracer of the ionization and dissociation fronts in PDRs.  Some of these lines will be accesible to high-angular-resolution and high sensitivity observations with \textit{JWST}.

\subsubsection{The origin of H$_2$S emission in other  environments}

Irrespective of $n_{\rm H}$ and $G_0$, grain surface formation of  \mbox{s-H$_2$S} and \mbox{photodesorption} back to the gas-phase  lead to H$_2$S column densities of a few 10$^{14}$\,cm$^{-2}$ in PDRs. This is in agrement with the observed column in the   Bar ($G_0$\,$\approx$\,10$^4$) as well as at the mildly illuminated rims of  \mbox{TMC-1} and \mbox{Barnard 1b} clouds \citep[$G_0$\,$\approx$\,10;][]{Almaida20}.
The inferred H$_2$S abundance in the shielded interior 
 of these  dark clouds \mbox{($A_V$\,$>$\,10\,mag})  drops to a few 10$^{-9}$, but the species clearly does not disappear from the gas  \citep[$N$(H$_2$S) of a few 10$^{13}$\,cm$^{-2}$;][]{Almaida20}. Interestingly,  neither in the Bar the H$_2$S line emission at $\sim$168\,GHz   decreases much  behind the PDR  
 (Fig.~\ref{fig:IRAM-map}) even if the flux of FUV photons is largely attenuated compared to the irradiated PDR edge.
 
Despite oxygen is $\sim$25 times more abundant than sulfur, the  H$_2$O to H$_2$S column density ratio in the Orion Bar PDR  is only about $\sim$\,5. This similarity must also reflect the higher abundances of CO compared to CS.
Furthermore, the H$_2$S column density in cold cores is strikingly  similar to that of \mbox{water vapor}  \citep[][]{Caselli10,Caselli12}.
 This coincidence  points to a more efficient desorption mechanism of \mbox{s-H$_2$S} compared to \mbox{s-H$_2$O} in gas shielded from  stellar FUV photons. \cite{Almaida20} argues that
 chemical desorption is able to reproduce the observed H$_2$S abundance floor if the efficiency of this process diminishes as ice grain mantles get thicker inside cold  dense cores. 

Turning  back to warmer star-forming environments, 
our predicted H$_2$S abundance in  \mbox{FUV-illuminated} gas is comparable to  that observed toward many hot cores \citep[\mbox{$\sim$10$^{-9}$-10$^{-8}$};][]{vdT03,Herpin09}. In these massive protostellar environments, thermal \mbox{desorption}  of icy mantles, suddenly heated
to \mbox{$T_{\rm d}$\,$\gtrsim$\,100\,K} by the luminosity of the embedded massive protostar, drives the H$_2$S production. Early in their evolution,  young hot cores ($\lesssim$\,10$^4$\,yr) can show even higher abundances of recently desorbed H$_2$S  \citep[before
 further chemical processing takes place in the gas-phase; e.g.,][]{Charnley87,Hatchell98,Jimenez-Serra12,Esplugues14}. 
 \mbox{Indeed}, \mbox{\cite{Crockett14}} reports a gas-phase H$_2$S abundance of several 10$^{-6}$ toward the hot core in Orion~KL. This high value likely reflects the
 minimum  \mbox{s-H$_2$S} abundance locked as \mbox{s-H$_2$S}  mantles just before thermal desorption. In addition,  the H$_2$S abundance in the Orion Bar  is  only slightly 
 lower than that inferred in protostellar outflows \mbox{(several 10$^{-8}$)}. In these
 regions, fast shocks erode and sputter the  grain mantles, releasing a large fraction  of their molecular content and activating a 
 high-temperature gas-phase chemistry that quickly reprocesses the gas  \mbox{\citep[e.g.,][]{Holdship19}}.
All in all, it seems reasonable to conclude that everywhere \mbox{s-H$_2$S} grain mantles form, or already formed in a previous evolutionary stage,  emission lines from
\mbox{gas-phase} H$_2$S will be detectable.

 In terms of its detectability with single-dish telescopes, H$_2$S rotational lines are  bright in hot cores  \citep[\mbox{$T_{\rm peak,\,168\,GHz}$\,$\simeq$\,30\,K} in Orion KL but \mbox{$\simeq$\,1-3\,K} toward most hot cores;][]{Tercero10,vdT03,Herpin09}, in strongly irradiated PDRs (\mbox{$\simeq$\,6\,K}, this work), and  in lower-illumination PDRs such as the Horsehead  \citep[\mbox{$\simeq$\,1\,K};][]{Marichalar19}. \mbox{The H$_2$S} emission
is fainter toward  cold dark clouds 
\citep[\mbox{$\simeq$\,0.2\,K} in \mbox{TMC-1};][]{Almaida20} and 
protostellar outflows \citep[\mbox{$\simeq$\,0.6\,K} in  L1157;][]{Holdship19}.
These line intensity  differences are mostly produced by  different gas physical conditions
and not by enormous changes of the H$_2$S abundance.

Finally, H$_2$S is also detected  outside the Milky Way \citep[firstly by][]{Heikkila99}. Lacking enough spatial-resolution  it is more difficult to determine the origin of the  extragalactic H$_2$S emission. The derived abundances in starburst galaxies such as NGC\,253  \citep[$\sim$10$^{-9}$;][]{Martin06} might be interpreted as arising from a collection of spatially unresolved hot cores \mbox{\citep{Martin11}}. However, hot cores have low filling factors at star-forming cloud  scales. Our study suggests that much of this emission can arise from (the most common) extended  molecular gas illuminated by stellar \mbox{FUV radiation} \citep[e.g.,][]{Goico19}.


\section{Summary and conclusions}

We carried out a self-consistent observational and modeling study of the chemistry of S-bearing hydrides in \mbox{FUV-illuminated gas}. We obtained the following results:\\
-- ALMA images of the Orion Bar  show that SH$^+$  is confined to narrow gas layers
of the PDR edge, close to the H$_2$ \mbox{dissociation} front. Pointed observations
carried out with the IRAM\,30m telescope show bright H$_{2}^{32}$S, H$_{2}^{34}$S, H$_{2}^{33}$S  emission toward the PDR (but no H$_3$S$^+$, a key gas precursor of H$_2$S) as well as behind the Bar, where the flux of FUV photons is largely attenuated. 
SOFIA observations provide tight limits to the SH emission.\\
-- The SH$^+$ line emission arises from a high-pressure gas component,
\mbox{$P_{\rm th}$\,$\simeq$\,(1--2)$\times$10$^8$\,cm$^{-3}$\,K}, where SH$^+$ ions are destroyed by reactive collisions with H atoms and electrons (as most H$_n$S$^+$ ions do). 
We derive $N$(SH$^+$)\,$\simeq$\,10$^{13}$\,cm$^{-2}$ and an abundance peak of several $\sim$10$^{-9}$. 
 H$_2$S shows larger column densities toward the PDR,  \mbox{$N$(H$_2$S)\,=\,$N$($o$-H$_2$S)\,+\,$N$($p$-H$_2$S)\,$\simeq$\,2.5$\times$10$^{14}$\,cm$^{-2}$}. 
\mbox{Our tentative} detection of SH translates into an upper limit  
column density ratio \mbox{$N$(SH)/$N$(H$_2$S)} of \mbox{$<$\,0.2-0.6},  already lower than the ratio of \mbox{1.1-3.0} observed in
low-density diffuse molecular clouds \mbox{\citep[][]{Neufeld15}}. 
This implies an enhanced H$_2$S production mechanism in \mbox{FUV-illuminated} dense gas.\\
-- All gas-phase reactions \mbox{X + H$_2$($v$=0)\,$\rightarrow$\,XH + H} (with X\,=\,S$^+$, S, SH$^+$, or H$_2$S$^+$) are highly endoergic. While  reaction of \mbox{FUV-pumped} H$_2$($v$\,$\geq$\,2) molecules with S$^+$ ions becomes exoergic and explains the observed levels of SH$^+$, further reactions of H$_2$($v$\,$\geq$\,2) with SH$^+$ or with neutral S atoms, both reactions studied here through  \mbox{\textit{ab initio}} quantum  calculations, do not form enough H$_2$S$^+$ or H$_3$S$^+$ to ultimately  produce abundant H$_2$S.
In particular,  pure gas-phase models underestimate the  H$_2$S column density
observed in the Orion Bar by more than two orders of magnitude. This implies that 
these models miss the main H$_2$S formation route.
 The disagreement is even worse as  we favor, after considering the potential energy surfaces of  the  H$_2$S$^{+*}$ and H$_3$S$^{+*}$ \mbox{complexes}, that the radiative associations 
\mbox{S$^+$\,+\,H$_2$\,$\rightarrow$\,H$_2$S$^+$\,+\,$h\nu$} and \mbox{SH$^+$\,+\,H$_2$\,$\rightarrow$\,H$_3$S$^+$\,+\,$h\nu$} may
actually not occur or possess slower rates than  considered in the literature.\\
-- To overcome these bottlenecks, we built  PDR models that include a simple network of gas-grain and grain surface reactions. 
 The higher binding energies of S and SH suggested by recent studies imply that  bare grains start to grow \mbox{s-H$_2$S} mantles 
not far from the illuminated edges of molecular clouds.
Indeed, the observed $N$(H$_2$S)  in the Orion Bar can only be explained by the freeze-out of S atoms, grain surface  formation of \mbox{s-H$_2$S} mantles, and subsequent photodesorption back to the gas phase.
The inferred  H$_2$S OTP ratio of \mbox{2.9\,$\pm$\,0.3} (equivalent to $T_{\rm spin}$\,$\geq$\,30\,K) is compatible with the high-temperature statistical ratio as well as with warm grain surface formation   if \mbox{$T_{\rm spin}$\,$\simeq$\,$T_{\rm d}$} and if $T_{\rm spin}$ is preserved in the gas-phase
after desorption.\\
-- Comparing observations with chemical and excitation models, we conclude that the \mbox{SH$^+$-emitting} layers at the edge 
of the Orion  Bar \mbox{($A_V$\,$<$\,2\,mag)} are charaterized by no or very little depletion of  sulfur from the gas-phase. At intermediate PDR depths \mbox{}($A_V$\,$<$\,8\,mag) the observed H$_2$S column densities do not require depletion of elemental (cosmic) sulfur abundances either.\\
-- We conclude that everywhere \mbox{s-H$_2$S} grain mantles form (or formed) gas-phase H$_2$S will be present in detectable amounts.
Independently of $n_{\rm H}$ and $G_0$,  \mbox{FUV-illuminated} clouds
produce roughly the same H$_2$S column density (a few 10$^{14}$\,cm$^{-2}$) and H$_2$S peak abundances (a few 10$^{-8}$). This agrees with the H$_2$S column densities
derived in the Orion Bar and at the edges of mildly  illuminated clouds. Deep inside 
 molecular clouds ($A_V$\,$>$\,8\,mag), H$_2$S still forms by direct chemical desorption 
and photodesorption by secondary FUV photons. These processes alter the abundances of other \mbox{S-bearing} species and makes difficult to predict the dominant sulfur
 reservoir in cloud interiors.

In this study we focused on \mbox{S-bearing} hydrides. Still, many subtle details  remain to be fully understood: radiative associations, electron recombinations, and \mbox{formation} of multiply sulfuretted molecules. \mbox{For example}, the low-temperature (T$_{\rm k}$\,$<$\,1000\,K) rates  of the \mbox{radiative} and \mbox{dielectronic} recombination of S$^+$  used in PDR models may still be not accurate enough \citep{Badnell91}.
In addition, the main ice-mantle  sulfur reservoirs are not fully constrained observationally. Thus, some of the narrative may be subject to speculation. \mbox{Similarly}, reactions of S$^+$ with abundant organic molecules  desorbed from grains  (such as \mbox{s-H$_2$CO}, not considered in our study) may contribute to enhance the H$_2$S$^+$ abundance through gas-phase reactions   \mbox{\citep[e.g., \mbox{S$^+$\,+\,H$_2$CO\,$\rightarrow$\,H$_2$S$^+$\,+\,CO};][]{Prasad82}}. 
Future observations of the abundance and freeze out depths of the key ice carriers with \textit{JWST}  will clearly help in these fronts.

\begin{acknowledgements}  
We warmly thank Prof. Gy\"orgy Lendvay for interesting discussions and for
sharing the codes related to  their \mbox{S($^3P$) + H$_2$($^1\Sigma_g^+$,v)} PES. We  thank Paul Dagdigian, Fran\c{c}ois Lique, and Alexandre Faure  for sharing their \mbox{H$_2$S--H$_2$}, \mbox{SH$^+$--H}, and \mbox{SH$^+$--e$^-$} inelastic collisional rate coefficients
and for interesting discussions in Grenoble and Salamanca.
We  thank Helgi Hrodmarsson for sending his experimental SH photoionization cross section
in tabulated format. We finally thank our referee, John H. Black, for encouraging and  insightful suggestions.
This paper makes use of the ALMA data ADS/JAO.ALMA\#2012.1.00352.S.  ALMA is a partnership of ESO (representing its member states), NSF (USA), and NINS (Japan), together with NRC (Canada), and NSC and ASIAA (Taiwan), in cooperation with the Republic of Chile. The Joint ALMA Observatory is operated by ESO, AUI/NRAO, and NAOJ. It also includes IRAM\,30\,m telescope observations. IRAM is supported by INSU/CNRS (France), MPG (Germany), and IGN (Spain).
We thank the staff at  the IRAM\,30m telescope
and the work of the USRA and NASA staff of the Armstrong Flight Research Center in Palmdale and of the Ames Research Center in Mountain View (California), and the Deutsches SOFIA Institut.
 We thank the Spanish MICIU for funding support under grants 
 \mbox{AYA2016-75066-C2-2-P}, \mbox{AYA2017-85111-P}, \mbox{FIS2017-83473-C2} 
\mbox{PID2019-106110GB-I00}, and \mbox{PID2019-106235GB-I00} and the
French-Spanish collaborative project PICS (PIC2017FR). We finally acknowledge computing time at Finisterrae (CESGA) under RES grant \mbox{ACCT-2019-3-0004}.
  
\end{acknowledgements}

%
%

\bibliographystyle{aa}
\bibliography{references}

\begin{appendix}\label{Sect:Appendix}

\section{H$_2$S$^+$ formation and destruction}\label{Apendix:ab-initio} 

In this Appendix we give details about how we calculated the H$_2$ 
vibrational-state-dependent rates of reaction~(\ref{reac-2}) and of the reverse reaction, the \mbox{destruction} of H$_2$S$^+$\,($^2A'$) by reactive collisons with H\,($^2S$) atoms (summarized in Fig.~\ref{fig-rates}). 

 We first built a full dimensional potential energy surface (PES) of the triplet H$_3$S$^+$\,($^3A$) system
by fitting more than 150,000 \mbox{\textit{ab initio}} points, including the long range interactions in the reactants and products channels. 
The main topological features of the PES are summarized in the minimum energy path between reactants and products (see middle panel of Fig.~\ref{fig:PES}).
 These \mbox{\textit{ab initio}} points were calculated with an explicitly correlated restricted coupled cluster including a single, double, and (perturbatively) triple excitations (RCCSD(T)-F12a) method \citep{Knizia-etal:09}.
The analytical fit has a overall rms error of $\simeq$\,0.01\,eV
(Fig.~\ref{fig:rms-reaction}). Appendix~\ref{Apendix:pes} provides more details.

Reaction~(\ref{reac-2}) is endothermic by 0.672\,eV, and the PES of the triplet state 
shows two shallow wells in the \mbox{H$_2$ + SH$^+$} entrance channel (named $^3W_{1a}$ and $^3W_{1b}$, with a depth  of $\simeq$\,0.118\,eV) and another one near the 
\mbox{H + H$_2$S$^+$} products (named $^3W_2$, with a  depth of  0.08\,eV).
Between the reactants and products wells there is a saddle point, with an energy of 0.601\,eV. This saddle point, slightly below the products, has a geometry similar to $^3W_2$ in which the 
\mbox{H--H} distance is strongly elongated compared to that of H$_2$. 
These features are also present in the maximum multiplicity PES of reactions 
\mbox{H$_2$ + S$^+(^4S)$} and \mbox{H$_2$ + H$_2$S$^+$($^2A$)}
(see Fig.~\ref{fig:PES}).
We determine the state-dependent rates of \mbox{reaction~(\ref{reac-2})} and  of the  reverse reaction using a quasi-classical trajectory (QCT) method on our ground triplet PES.
We provide more details on how the reactive cross sections for fixed collision energies were calculated in
Appendix.~\ref{appendix-collisions}.

The formation rate of H$_2$S$^+$  from H$_2$\,($v$\,=\,0) is very slow.  For H$_2$\,($v$\,=\,1), the rate constant significantly increases at $\approx$\,500\,K, corresponding with the opening of the 
\mbox{H$_2$S$^+$ + H} threshold. At this point,  it is important
to consider the zero-point energy (ZPE) of the products (see next section for details).
For H$_2$\,($v$\,=\,2) and  H$_2$\,($v$\,=\,3),  reaction rates are faster, close to the Langevin limit.
\mbox{Finally}, the H$_2$S$^+$ destruction rate constant is very similar to that of its formation from H$_2$\,($v$\,=\,2). In \mbox{Appendix~\ref{Apendix:destruction}} we provide more information about the destruction 
of H$_n$S$^+$ ions through radiative association and spin flip mechanisms.

\begin{figure}[t]
\centering
\includegraphics[scale=0.6]{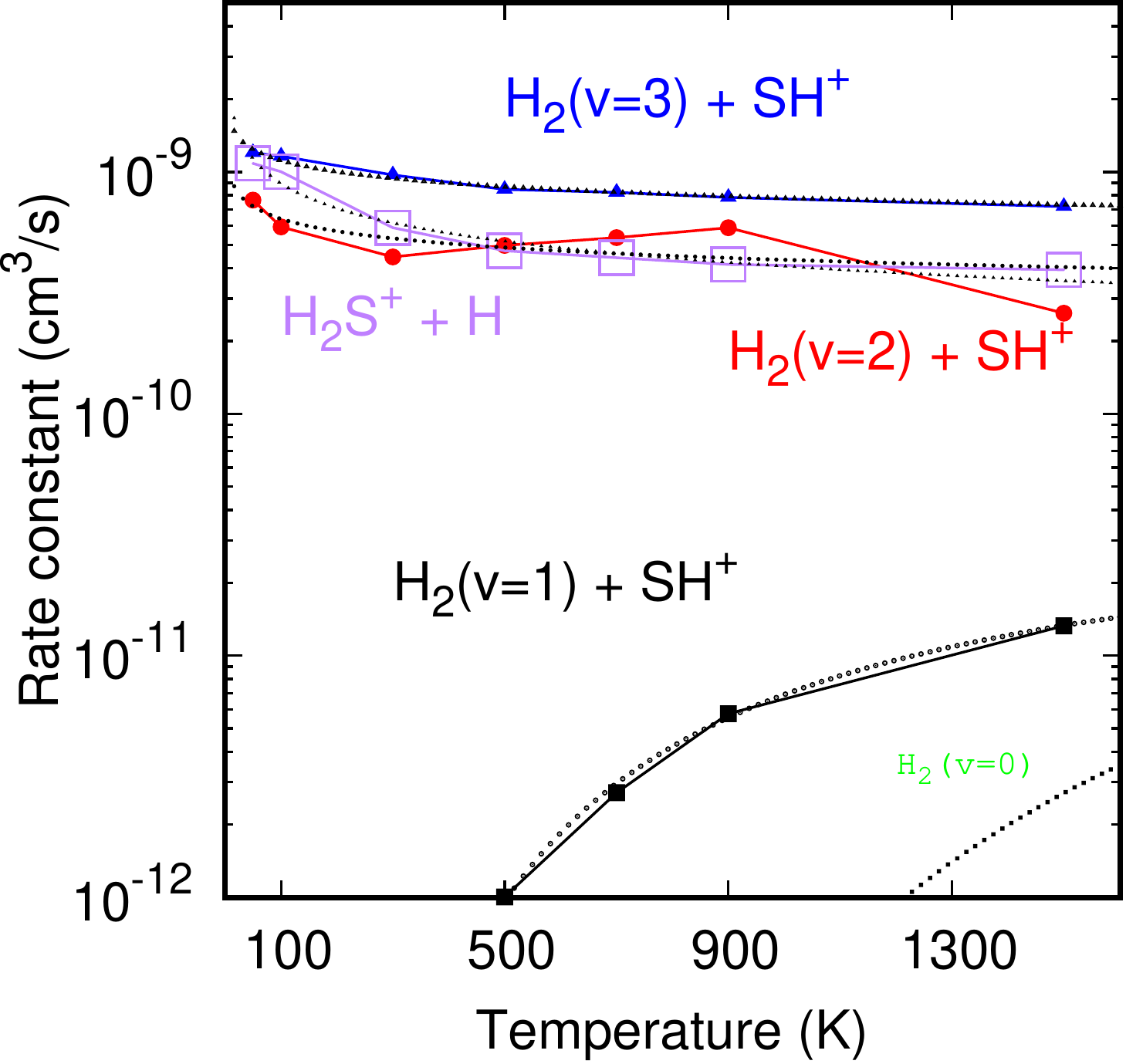}
\caption{Calculated rate constants as a function of temperature (for translation and rotation) for \mbox{SH$^+$\,($v$\,=\,0,~$j$\,=\,0)\,+\,H$_2$\,($v$\,=\,1, 2, 3,~$j$\,=\,0)} and  \mbox{H$_2$S$^+$\,($v$\,=\,0,\,$j$\,=\,0)\,+\,H} reactions (lavender) using ZPE corrected QCT method. 
\mbox{Dotted curves} are fits of the form $k(T)$\,=\,$\alpha$\,($T$/300)$^{\beta}$\,exp$(-\gamma/T)$.
Rate coefficients are listed in Table~\ref{rates}.}
\label{fig-rates}
\end{figure}

\begin{figure}[h]
\centering
\includegraphics[scale=0.38]{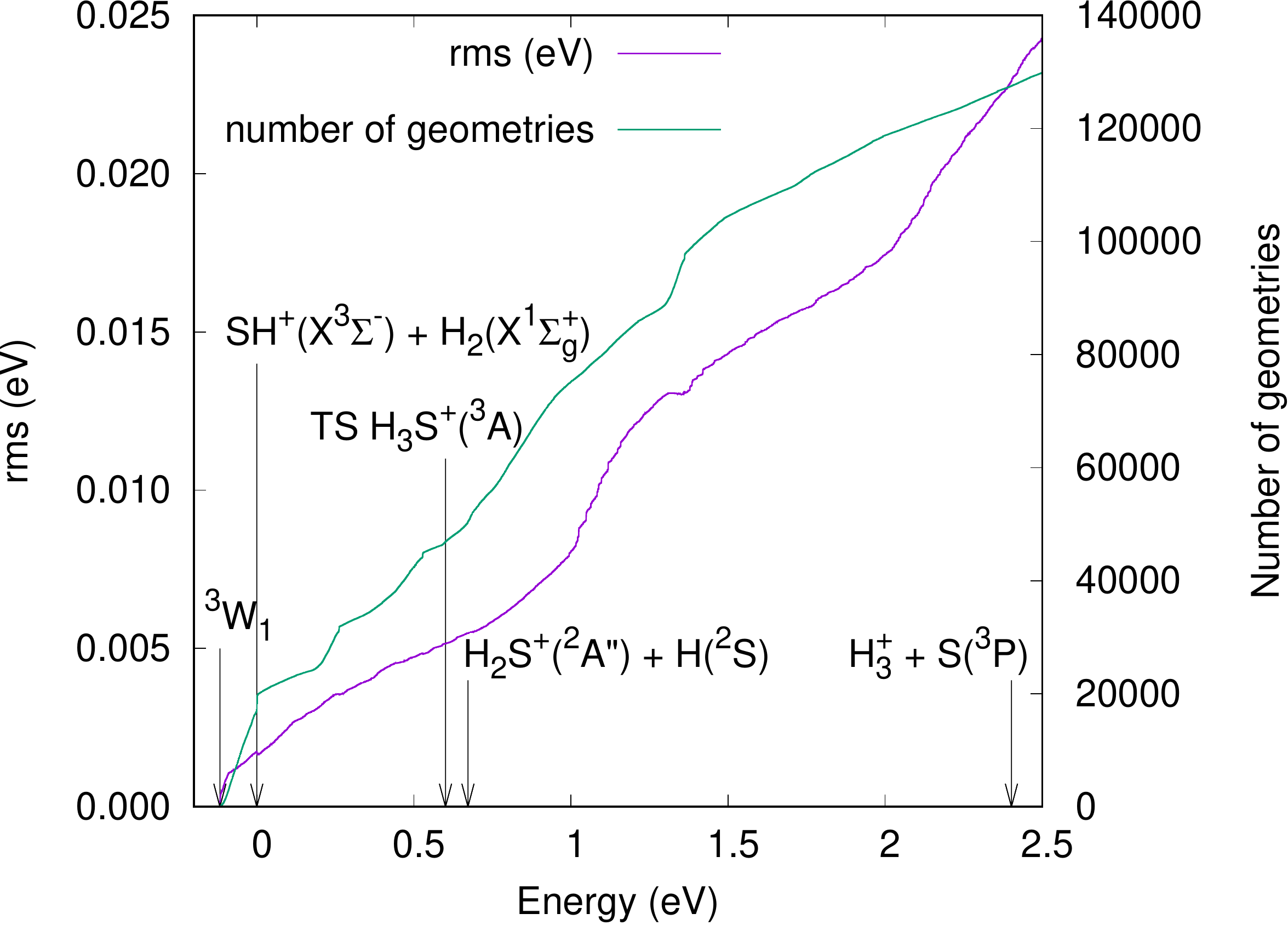}
\caption{Rms error as a function of total energy, showing the number of
\textit{ab initio} points used to evaluate the error in the PES calculation.
Arrows indicate selected critical points in the PES and provide an estimate
of the error in each region. TS means transition state.}
\label{fig:rms-reaction}
\end{figure}

\subsection{Ab initio calculations and PES}\label{Apendix:pes}

\citet{Dagdigian:19} presented a PES for the \mbox{SH$^+$-H$_2$} system that includes 4-dimensions and is based on RCCSD(T)-F12a \mbox{\textit{ab initio}} calculations.
This PES was used to study \mbox{SH$^+$--H$_2$} inelastic collisions
using a rigid rotor approach in which
the two diatomic molecules are kept fixed at their equilibrium distances.
However, in order to study the reactivity of the collision, the two diatomic distances have to be included to account for the breaking and formation of new bonds.
 
Reaction~(\ref{reac-2}) corresponds to a triplet state \mbox{H$_3$S$^+$\,($^3A$)}. The \mbox{H$_2$S$^+$\,($^2A'$)\,+\,H\,($^2S$)} products can form a triplet and a singlet state. The triplet state can lead to the destruction of H$_2$S$^+$ through \mbox{reaction} with H atoms.
The singlet state, however, produces very excited states of the reactants. Thus, it  
only leads to inelastic collisions but not not to
the destruction of H$_2$S$^+(^2A')$. In consequence, here we only 
consider the ground triplet electronic state of the system. In addition,
the \mbox{H$_3^+$ + S\,($^3P$)} channel is  about 2.4\,eV
above the \mbox{H$_2$ + SH$^+$} asymptote, and will not be included in the present study.

In order to study the regions where several electronic states intersect,
we performed a explicitly correlated internally contracted multireference
configuration interaction \mbox{(ic-MRCI-F12)} calculation
\citep{Shiozaki-Werner:13,Werner-Knowles:88,Werner-Knowles:88b} including
the Davidson correction  \citep[\mbox{icMRCI-F12+Q};][]{Davidson:75}.
The \mbox{ic-MRCI-F12} calculations were carried out using  state-averaged
complete active space self-consistent field \mbox{(SA-CASSCF)} 
orbitals with all the CAS configurations as the reference configuration
state functions.  We used a triple zeta
correlation consistent basis set for explicitly correlated 
wave functions \citep[\mbox{cc-pVTZ-F12};][]{Peterson-etal:08}. 
In order to avoid orbital flipping between core and valence orbitals.
\mbox{SA-CASSCF} calculations with three lowest triplet states were
carried out including the core and valence orbitals as 
active space (18 electrons in 11 orbitals).
For the \mbox{ic-MRCI-F12} calculation, the core orbitals was kept doubly occupied,
resulting in about $2.5 \times 10^6$ $(9 \times 10^7)$
contracted (uncontracted) configurations.
All {\sl ab initio} calculations were performed with MOLPRO
 \citep{MOLPRO-WIREs}.

Our \mbox{ic-MRCI-F12} calculations show that the crossings with electronic excited states
are  2\,eV above the energy of the reactants. The energy interval below 2\,eV is enough to study \mbox{reaction \ref{reac-2}}. In these low-energy regions, \mbox{RCCSD(T)-F12a} calculations  were also performed. They are in  good agreement with the  \mbox{ic-MRCI-F12} results
and the t$_1$ diagnostic is always below 0.03. This 
allows us to conclude that for energies below 2\,eV,  the \mbox{RCCSD(T)-F12a} method performs well, presents a simple convergence, and being size consistent, is well adapted to the present case. This method
is the same one employed in the inelastic collision calculations by \cite{Dagdigian:19}.

We performed extensive \mbox{RCCSD(T)-F12a} calculations in all accessible regions 
to properly describe the six-dimensional phase space. 150000 \textit{ab initio} points were
fitted to a multidimensional analytic function, that generates the six-dimensional PES
 represented as
\begin{eqnarray}\label{electronic-diabatic-matrix}
H= H^{diab} + H^{MB}
\end{eqnarray}
\citep{Aguado-etal:10,Sanz-Sanz-etal:13,Zanchet-etal:18,Roncero-etal:18}, where 
$H^{diab}$ is an electronic diabatic matrix in which
each diagonal matrix element describes a rearrangement channel -- six in this case, three equivalent for \mbox{SH$^+$ + H$_2$ channels}, and three equivalent for \mbox{H$_2$S$^+$ + H} fragments (we omitted the  \mbox{H$_3^+$ + S} channel) --
 as an extension of the reactive force field  approach \citep{Farah-etal:12}. In each diagonal term, the molecular fragments (SH$^+$, H$_2$ and H$_2$S$^+$) are described by 2 or 3 body fits \citep{Aguado-Paniagua:92},
and the interaction among them is described by a sum of atom-atom terms plus the long range interaction.
The non diagonal terms of $H^{diab}$ are described as previously \citep{Zanchet-etal:18,Roncero-etal:18}
and the parameters are fitted to approximately describe the saddle points along the minimum energy path in the right geometry.

In the reactants channel, the leading long range
interaction $\mbox{SH}^+(X^3\Sigma^-) + \mbox{H}_2(X^1\Sigma_g^+)$ corresponds
to  charge-quadrupole and charge-induced dipole
interactions \citep{Buckingham:67}:
\begin{eqnarray}
&&V_{\mbox{charge}}(\mathbf{r}_{HH},\mathbf{R}) = 
  \Theta_2 (r_{HH}) P_2(\cos \theta_2) R^{-3} \nonumber\\ 
&-&  \left[ \frac12 \alpha_0(r_{HH})
    + \frac{1}{3} \left( \alpha_\parallel(r_{HH})-\alpha_\perp(r_{HH}) \right)P_2(\cos \theta_2) 
  \right] R^{-4} 
\end{eqnarray}
and the dipole-quadrupole interactions \citep{Buckingham:67}:
\begin{eqnarray}
&&V_{\mbox{dipole}}(\mathbf{r}_{SH}, \mathbf{r}_{HH},\mathbf{R}) = 
  3 \mu_1 (r_{SH}) \Theta_2 (r_{HH}) \nonumber\\
&\times&  \left[ \cos \theta_1 P_2( \cos \theta_2) + \sin \theta_1 \sin \theta_2 \cos \theta_2 \cos \phi  \right]
  R^{-4},
\end{eqnarray}
where $\Theta_2 (r_{HH})$ is the cuadrupole moment of $\mbox{H}_2(X^1\Sigma_g^+)$, 
$\alpha_0(r_{HH})$, $\alpha_\parallel(r_{HH})$, and $\alpha_\perp(r_{HH})$
are the average, parallel, and perpendicular polarizabilities of $\mbox{H}_2(X^1\Sigma_g^+)$, respectively, and $\mu_1 (r_{SH})$ is the dipole moment of $\mbox{SH}^+(X^3\Sigma^-)$. $P_2( \cos \theta)$ represents the Legendre polynomial of degree 2.
The dependence of the molecular properties of H$_2$ with the interatomic distance $r_{HH}$ is obtained from  \citet{Velilla-etal:08}. The dipole moment of $\mbox{SH}^+$ depends on the origin of coordinates. Since $\mbox{SH}^+(X^3\Sigma^-)$ dissociates in $\mbox{S}^+(^4S) + \mbox{H}(^2S)$, we select the origin of coordinates in the $S$ atom, so that the dipole moment tends to zero when R goes to infinity. 

In the products channel, the long range interaction
$\mbox{H}_2\mbox{S}^+\,(X^2 A'')\,+\,\mbox{H}\,(^2S)$ corresponds
to the isotropic charge-induced dipole and charge-induced quadrupole 
dispersion terms
$$
V_{\mbox{disp}}(R) = -\frac{9}{4} R^{-4} -\frac{15}{4} R^{-6}.
$$

These long range terms diverge at $R$=0. To avoid this behavior, we replace $R$  by ${\cal R}$:
$$
{\cal R} = R + R_0 e^{-(R-R_e)} \quad\quad{\rm with}\quad\quad R_0= 10\, {\rm  bohr.}
$$

In Eq.~(\ref{electronic-diabatic-matrix}), $H^{MB}$ is the many-body term, which is described
by permutationaly invariant polynomials following the method of Aguado an collaborators \citep{Aguado-Paniagua:92,Tablero-etal:01,Aguado-etal:01b}.
This many-body term improves the accuracy of  the PES, especially in the region of the reaction barriers (as shown in Fig.~\ref{fig:PES}). Features of the stationary points are listed in Table~\ref{stationary-points}.

\begin{table}
\begin{center}
\caption{\label{stationary-points} RCCSD(T)-F12a and fit stationary points on the  PES.}
\begin{tabular}{llrr}
\hline
\hline
  Stationary point & Geometry    & Energy/cm$^{-1}$ & Energy/eV    \\
\hline
Reactants & SH$^+ + $ H$_2$        &    0.0    &  0.0    \\ 
Minimum 1 & SH$^+ - $ H$_2$  & $-$950.2 & $-$0.1178\\ 
TS12      & SH$^+ $ $\cdot \cdot $ H$_2$  & $-$579.5 & $-$0.0719\\
Minimum 2 & SH$^+ - $ H$_2$  & $-$937.9 & $-$0.1163\\ 
TS13      & SH$^+ $ $\cdot \cdot$ H $\cdot \cdot$ H    & 4843.9   &  0.6006  \\
Minimum 3 & H$_2$S$^+ - $ H    & 4766.5   &  0.5910  \\
Products  & H$_2$S$^+ + $ H          & 5422.3   &  0.6723  \\ 
\hline
\end{tabular}
\end{center}
\end{table}

\subsection{Determination of reactive collision rates\label{appendix-collisions}}

We studied the reaction dynamics using a quasi-classical trajectory (QCT)
method with the code miQCT \citep{Zanchet-etal:18,Roncero-etal:18}. In this method,
the initial vibrational energy of the reactants is included using the
adiabatic switching method (AS)
\citep{Grozdanov-Solovev:82,Johnson:87,Qu-Bowman:16,Nagy-Lendvay:17}.
Energies are listed in Table~\ref{vibrational-energies}.
The initial distance between the center-of-mass of the reactants (\mbox{H$_2$ + SH$^+$} or \mbox{H$_2$S$^+$ + H})
is set to 85 bohr, and the initial impact parameter is set randomly within a disk,
the radius of which is set
 according to a capture model \citep{Levine-Bernstein:87} using
the corresponding long-range interaction. The orientation among the two reactants
is set randomly.

\begin{table}
\begin{center}\label{vibrational-energies}
\caption{$E_v$ of reactants and products,  and
 adiabatic switching energies for the QCT initial conditions.}
\begin{tabular}{llrr}
\hline
  System(vibration) & Exact $E_v$ (eV)    & AS energy (eV) & \\
\hline
H$_2$\,($v$\,=\,0) & 0.270 & 0.269 \\
H$_2$\,($v$\,=\,1) & 0.786 & 0.785 \\
H$_2$\,($v$\,=\,2) & 1.272 & 1.272 \\
H$_2$\,($v$\,=\,3) & 1.735 & 1.730 \\
SH$^+$\,($v$\,=\,0) & 0.157 & 0.157\\
H$_2$S$^+$\,($v$\,=\,0) & 0.389 & 0.388 \\
\hline
\end{tabular}
\end{center}
\end{table}

A first exploration of the reaction dynamics is done at fixed collision energy,
for \mbox{H$_2$\,($v$\,=\,0, 1, 2 , 3) + SH$^+$\,($v$\,=\,0)} and
\mbox{H + H$_2$S$^+$($v$\,=\,0)}, and the reactive cross
section is calculated as in \citet{Karplus-etal:65}
\begin{eqnarray}\label{qctXsection}
\sigma_{vj}(E)= \pi b_{max}^2 P_r(E)
 \quad {\rm with }\quad
  P_r(E)= {N_r\over N_{tot}},
\end{eqnarray}
where
$N_t$ is the maximum number of trajectories with initial impact parameter lower
than $b_{max}$, the maximum impact parameter for which the reaction takes place,
and $N_r$ is the number of trajectories leading to products. 
Fig.~\ref{fig-sigma} shows results for $N_t$\,$>$\,20000 and all energies and initial reactant and vibrational states.

For the  \mbox{SH$^+$\,($v$\,=\,0, $j$\,=\,0)\,+\,H$_2$\,($v$,\,$j$\,=\,0)} reaction there is a strong dependence on the initial vibrational state. For H$_2$\,($v$\,=\,0), there is nearly no reactive
event, and only at 1\,eV there are some reactive trajectories.
For H$_2$\,($v$\,=\,2 and 3), however, the reaction  shows a relatively large  cross section, that decreases with increasing collision energy,
as expected for exoergic reactions. Energies below \mbox{10-100\,meV} are dominated
by long range interactions, leading to an increase in the
maximum impact parameter, $b_{max}$, consistent with the variation of the
cross section.

\begin{figure}[t]
\centering
\label{fig-sigma}
\includegraphics[width=8.5cm]{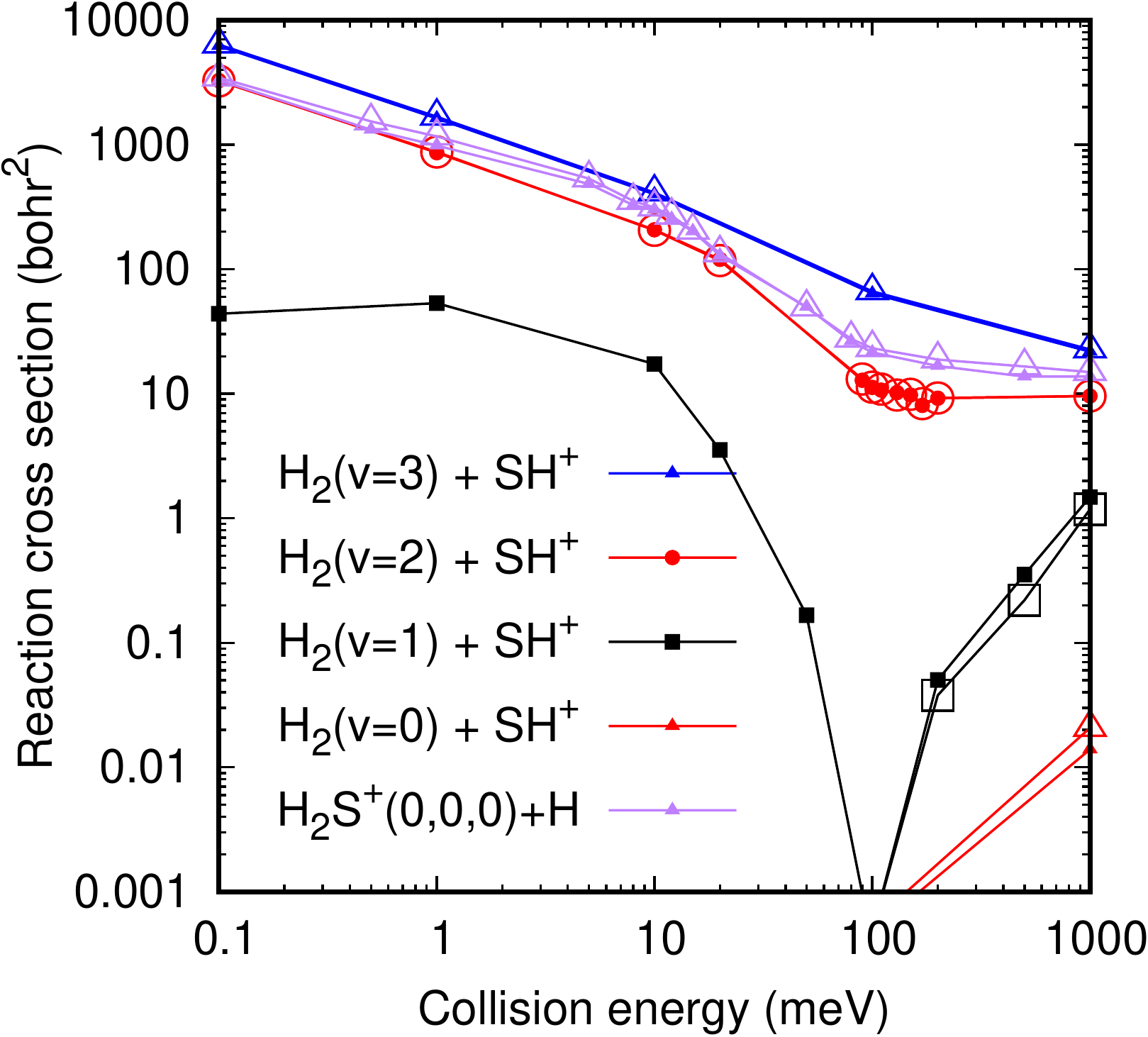}
\caption{Reaction cross section (in bohr$^2$) as a function of collision energy (in meV)
    for the  \mbox{SH$^+$\,($v$\,=\,0, $j$\,=\,0) + H$_2$\,($v$\,=\,1, 2, 3, $j$\,=\,0)} and  
    \mbox{H$_2$S$^+$\,($v$\,=\,0, $j$=0)\,+\,H} collisions.
    Filled symbols are obtained counting all trajectories leading to products,
    while open symbols correspond to the ZPE corrected ones.}
\end{figure}

Reaction \mbox{SH$^+$\,($v$\,=\,0, $j$\,=\,0) + H$_2$ ($v$\,=\,1, $j$\,=\,0)} shows an
unexpected behavior that deserves some discussion. At energies below 40\,meV, the cross
section is large and decreases with increasing energy. In the
40-200\,meV range, the reactive cross section drops to zero, showing a threshold
at 200\,meV that is consistent with the endothermicity of the reaction.

In order to  analyze the reaction mechanism for H$_2$\,($v$\,=\,1) below 40\,meV,
we carried out an extensive analysis of the trajectories. A typical
one is presented in Fig.~\ref{fig-lowEtraj} for 10\,meV. The H$_2$ and SH$^+$
reactants are attracted to each other by long range interactions, until
they get trapped in the $^3W_1$ wells, as it is shown by the evolution of
$R$, the distance between center-of-mass of the two molecules.
The trapping lasts for 8\,ps, thus allowing  several collisions
between H$_2$ and SH$^+$ and permitting the energy transfer between them.
The H$_2$ molecule ultimately breaks, and leaves  SH$^+$ with less vibrational
energy. This can be inferred from the decrease in the amplitudes of the SH$^+$ distance.
The energy of the H$_2$S$^+$ product is below the ZPE (see Table~\ref{vibrational-energies}). This is a clear indication of ZPE leakage in the QCT method, due to the energy transfer promoted by the long-lived collision complex.

\begin{figure}[t]
\centering
\includegraphics[width=8.cm]{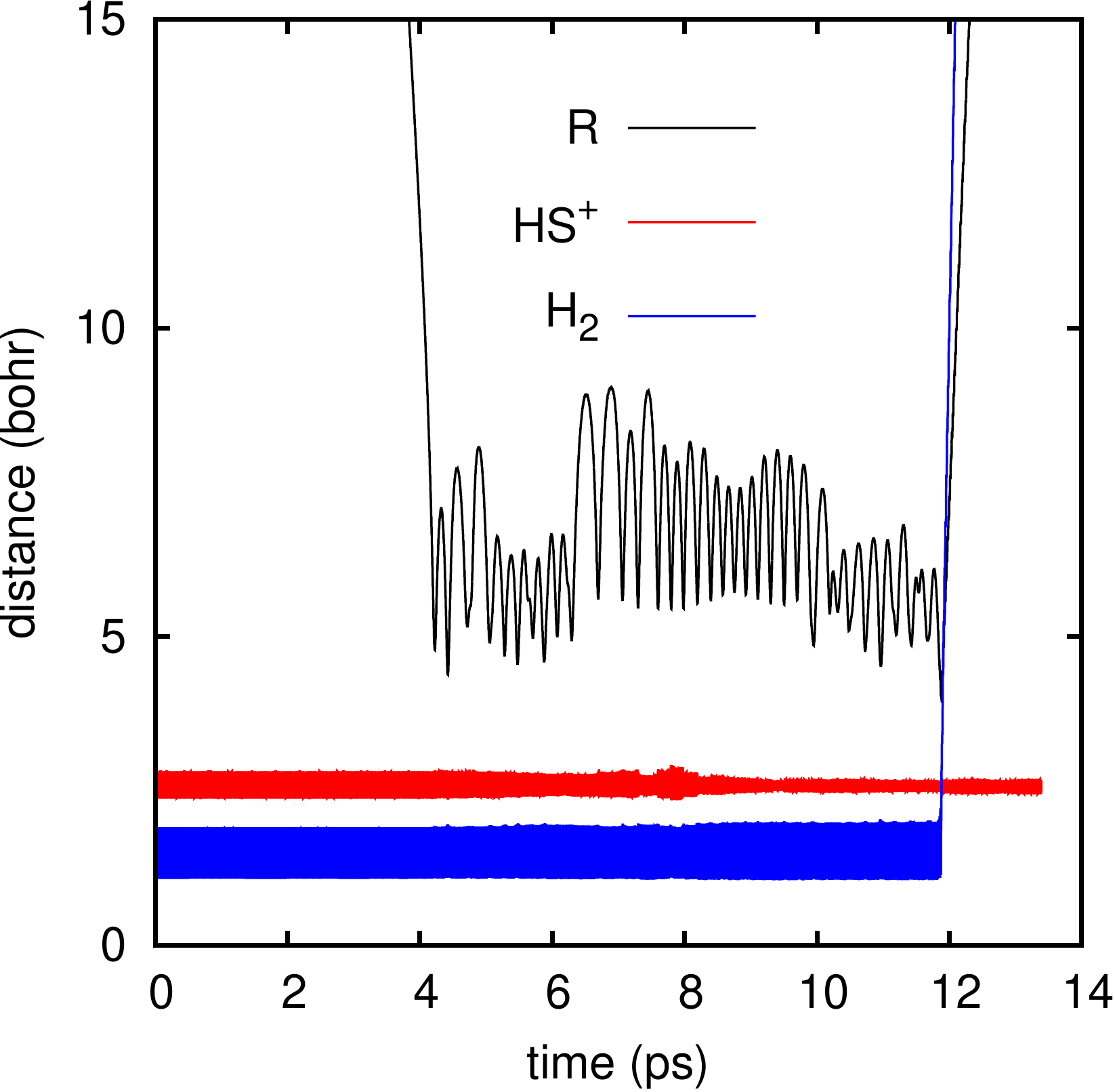}
\caption{H-H, SH$^+$ and $R$  distances (in bohr) versus time (in ps),
for a typical reactive trajectory for the 
\mbox{SH$^+$\,($v$\,=\,0,\,$j$\,=\,0) + H$_2$($v$\,=\,1,\,$j$\,=0)} collision at 10\,meV.}
\label{fig-lowEtraj}
\end{figure}

Several methods exist that correct the ZPE leakage. One is the gaussian binning
 \citep{gaussian-binning1,gaussian-binning2,gaussian-binning3,gaussian-binning4}.
Here we have applied a simplification of this method, which 
assigns a weight  ($w$) for each trajectory as
\begin{eqnarray}
  w=\left\lbrace
  \begin{array}{ccc}
    1 &{\rm for} & E_{vib} > ZPE\\
    e^{-\gamma (E_{vib}-ZPE)^2} & {\rm for} &  E_{vib} < ZPE
  \end{array}
  \right.
  ,
\end{eqnarray}
where $E_{vib}$ is the vibrational energy of reactants (adding those of H$_2$ and SH$^+$)
or H$_2$S$^+$ products at the end of each trajectory. These new weights
are used to calculate $N_r$ and N$_{tot}$ in Eq.~\ref{qctXsection}. 
\mbox{ZPE-corrected} results are shown in Fig.~A.3 with open symbols. This plot shows that all values are nearly the same as those calculated simply by counting trajectories as an integer 
(as done in the normal binning  method; see filled symbols in Fig.~A.3).
The only exception is the case of \mbox{SH$^+$ + H$_2$\,($v$\,=\,1)} below 400\,meV, which becomes zero 
 when considering the ZPE of fragments at the end of the trajectories.

The reaction thermal rate in specific initial vibrational state of reactants
are calculated running a minimum of 10$^5$ trajectories per temperature, with fixed vibrational states of reactants, assuming a Boltzmann distribution over translational and rotational degrees
of freedom, and following the ZPE-corrected method as:
\begin{eqnarray}\label{macrocanonical-rate}
  k_v(T)= \sqrt{{8 k_B T\over \pi\mu}}\quad \pi \, b^2_{max}(T)\quad P_r(T).
\end{eqnarray}
 The results of these calculations are shown in Fig.~\ref{fig-rates}.

\subsection{On the radiative associations of H$_n$S$^+$}\label{Apendix:destruction}

\cite{Herbst-etal:89} and \cite{Millar-Herbst:90} proposed that the radiative association
\mbox{H$_n$S$^+$ + H$_2$ $\rightarrow$ H$_{n+1}$S$^+$ + $h\nu$} is  viable process at low gas temperatures. Although this chemical route is widely used in astrochemical models, here we question the viability of this process. 
The lower multiplicity (L) PESs of \mbox{H$_2$S$^+$\,($^2A''$)} and \mbox{H$_3$S$^+$\,($^1A$)} are  L\,=\,1/2 and 0 respectively. These are shown in Fig.~\ref{fig:PES}, together with the minimum multiplicity electronic state of H$_4$S$^+$ (bottom panel). This state does not have a deep well or any  higher multiplicity state that could connect to higher states of reactants and products.

For of H$_3$S$^+$ formation through  radiative association, this process assumes that a H$_3$S$^+(^3A)$$^*$ complex forms in a triplet state, the high spin state H considered here. According to our 
calculations, such a complex is formed after low-energy \mbox{H$_2$\,($v$\,=\,0, 1) + SH$^+$} reactions (below 40\,meV). 
The complex is formed in the $^3W_1$ well, corresponding to geometries
very far from those of the low spin well, the $^1W$ well. Therefore, a radiative
spin flip and decay through phosphorescence is not possible. 
\citet{Herbst-etal:89} proposed a second step, in which the spin flips from the triplet
to the singlet state, followed by a radiative association, finally leading to
the H$_3$S$^+(^1A)$ product.

The origin of the spin flip must be the spin-orbit couplings, very relevant for S-bearing species, that favor the spin transition when singlet and triplet states are close in energy. \mbox{Using} the PESs calculated here, the lowest crossing region is at $\simeq$\,0.25\,eV, very close to that
of H$_2$($v$\,=\,0).  
At low temperatures, the H$_3$S$^+(^3A)$$^*$ \mbox{complex} formed by \mbox{H$_2$\,($v$\,=\,0) + SH$^+$ reactions} 
might allow a transition between the two electronic states with different spin.
However, the spin flip probability is proportional to the square of the overlap
$\vert\langle {\rm H_3S^+}\,(^3A)^* \,\vert \,{\rm H_3S^+}\,(^1A)^*\rangle \vert^2 $.
This probability is very small because the two wells, $^3W_1$ and  $^1W_1$, correspond to 
very different geometries. In consequence, we conclude that this radiative association mechanism must be negligible, especially at the high gas temperatures of PDR edges where the H$_3$S$^+$($^3A$)$^*$ complex is not formed. 

As an alternative, a spin flip in a direct collision (not forming a
  H$_3$S$^+(^3A)^*$ complex) may be more efficient and should be further investigated.
Indeed, experimental  measurements of the  \mbox{S$^+(^4S)$ + H$_2$\,($v$\,=\,0)} cross section  show a maximum at about 1\,eV of collisional energy  attributed to spin-orbit transitions leading to spin flip \citep{Stowe-etal:90}.

\section{Reaction $\rm S\,(^3P)\,+\,H_2\,({\it{v}}) \rightleftarrows SH \,+\, H$}\label{Apendix:S_H2}

This reaction involves open shell reactants,  \mbox{S\,($^3P$)}, and products, 
\mbox{SH\,($^2\Pi$)}. Neglecting spin flipping, there are three states that correlate to S($^3P$), two of them connect to the SH\,($^2\Pi$). These two electronic states are of $^3A'$ and $^3A''$ symmetry, and have been studied in detail by \cite{Maiti04}. Here we use the adiabatic PES calculated by \cite{Maiti04}.
 Reaction \mbox{S + H$_2$ $\rightarrow$ SH + H} is endothermic by $\simeq$\,1.02\,eV (without zero-point energy corrections), very similar to  the endothermicity of 
 reaction \mbox{S$^+$\,+\,H$_2$\,$\rightarrow$\,SH$^+$\,+\,H}
 \citep{Zanchet13a,Zanchet19}. The main difference
 is the presence of a barrier, of \mbox{$\simeq$\,78 meV} (\mbox{$\simeq$\,905\,K}) with respect to the \mbox{SH + H} asymptote.

\begin{figure}[ht]
\centering
\includegraphics[width=8.5cm]{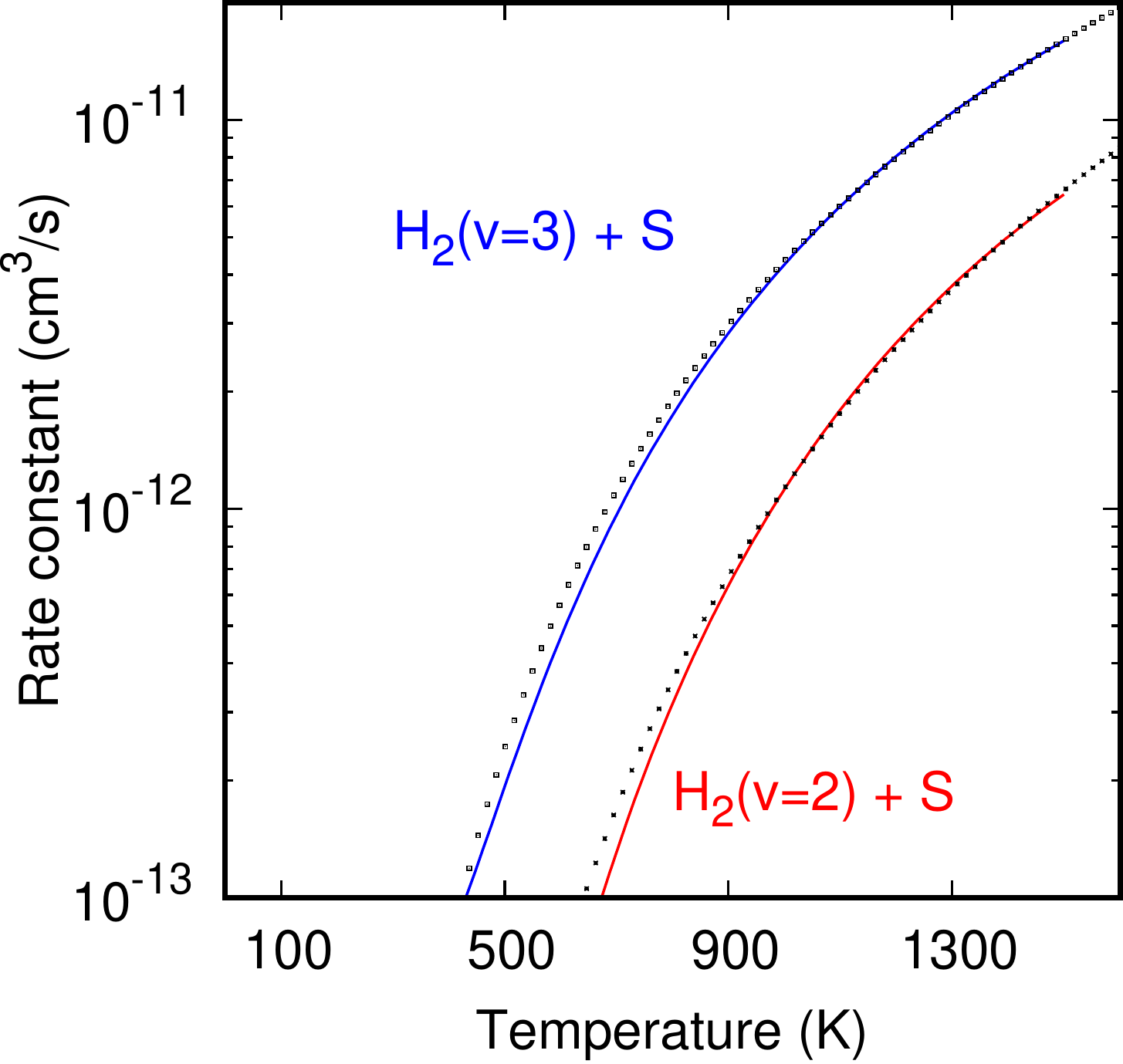}
\caption{Calculated rate constants as a function of temperature
for reaction \mbox{S($^3P$)\,+\,H$_2$($v$)\,$\rightarrow$\,SH + H}.
Dotted curves are fits of the form $k(T)$\,=\,$\alpha$\,($T$/300)$^{\beta}$\,exp$(-\gamma/T)$.
Rate coefficients  are listed in Table~\ref{rates}.}
\label{fig-S_H2}
\end{figure}

 We performed quantum wave packet calculations for the reactions 
 \mbox{S + H$_2$\,($v$\,=\,2, 3, $j$=0)} and \mbox{SH\,($v$\,=\,0, $j$=0) + H}. We used MADWAVE3 \citep{Gomez06,Zanchet09} to calculate the reaction probabilities for the initial vibrational state of the diatomic reactant (in the ground state rotational state, $j$\,=\,0).
We employed the usual partial wave expansion to calculate the reaction cross section. 
We calculated only few total angular momenta of the triatomic system, 
\mbox{$J$\,=\,0, 10 and 20}. The other $J$ needed in the partial wave expansion were obtained using  the $J$-shifting-interpolation method  \citep[see][]{Zanchet13a}. The initial-state-specific rate constants  are obtained by numerical integration of the  cross section using a Boltzmann  distribution  \citep{Zanchet13a}. The resulting reaction rate constants are shown in Figs.~\ref{fig-S_H2} and ~\ref{fig-SH_H}. The numerical values of the rate constants are fitted to the
 usual analytical Arrhenius-like expresion (shown as dotted curves).  We note that the shoulder in the rate  constants of reaction \mbox{SH\,($v$=0) + H} requires two functions in the temperature range of \mbox{200-800\,K}.  Rate coefficients are tabulated in Table~\ref{rates}.

\begin{figure}[h]
\centering
\includegraphics[width=8.5cm]{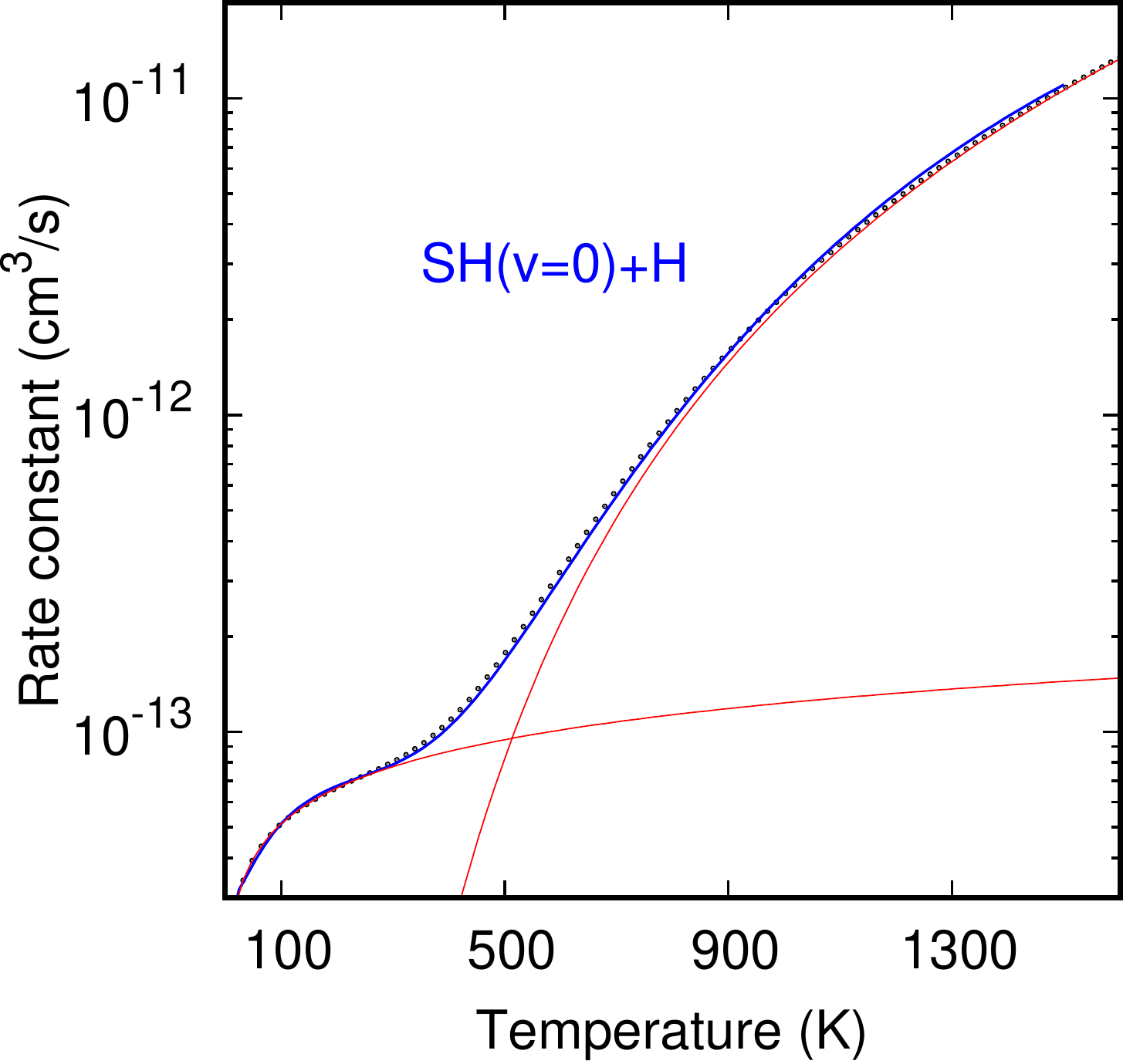}
\caption{Calculated rate constants as a function of temperature
for reaction \mbox{SH\,($v$=0)\,+\,H\,$\rightarrow$\,S + H$_2$}. The best fit to the calculated
rate requires two  Arrhenius-like  expressions (one for low temperatures and one for high temperatures). Rate coefficients of these fits are listed in Table~\ref{rates}.}
\label{fig-SH_H}
\end{figure}

\clearpage

\section{SH and H$_2$S photoionization and photodissociation cross sections}        
        
Figure~\ref{fig:photo-cross-sect} shows the experimental SH and H$_2$S photoionization and
photodissociation cross sections (cm$^{-2}$) used in our PDR models. 
We integrate these  cross sections over  the specific FUV radiation field at each $A_V$ depth of the PDR to obtain the specific photoionization and photodissociation rates (s$^{-1}$).

\begin{figure}[h]
\centering   
\includegraphics[scale=0.30, angle=0]{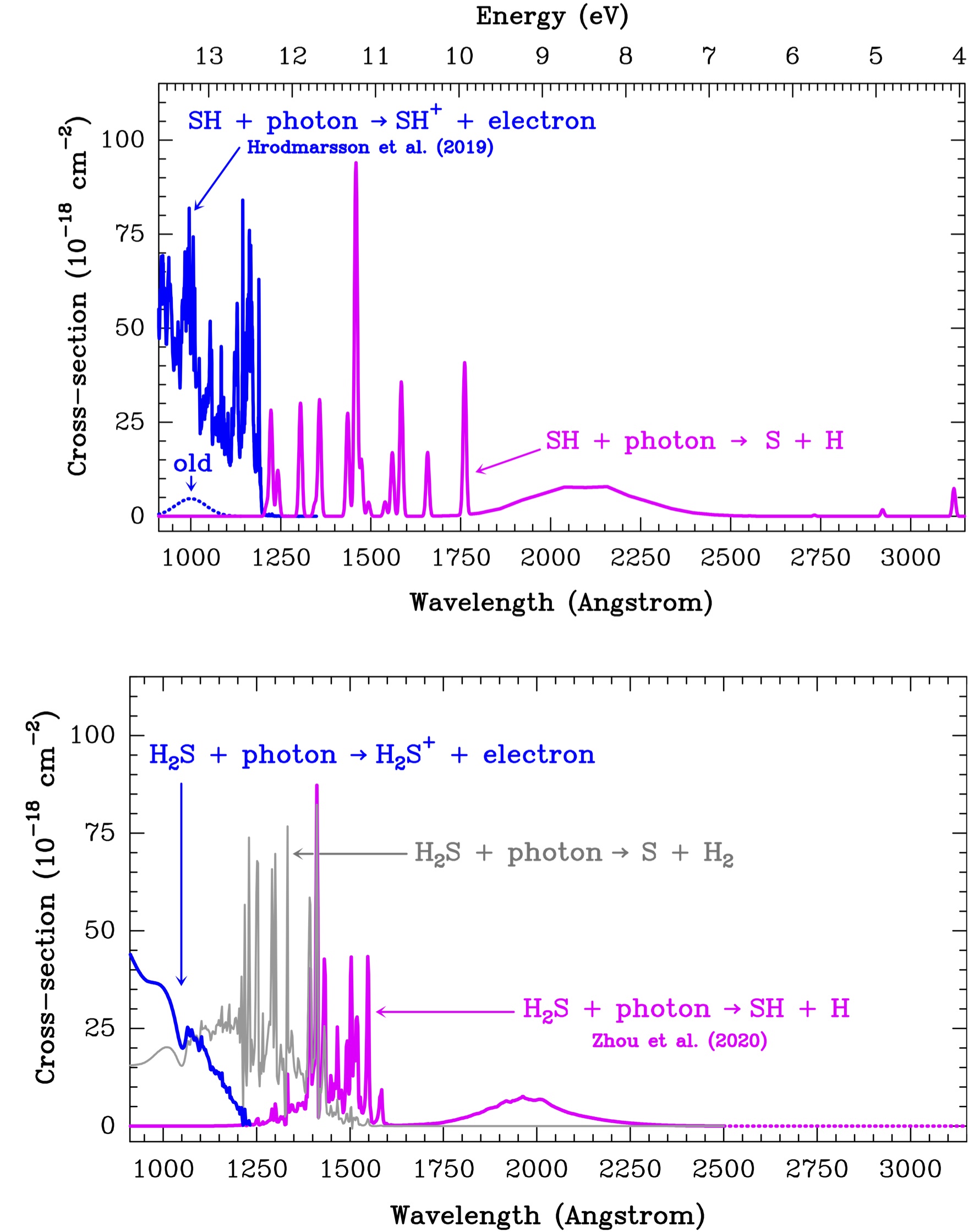}
\caption{Photoionization and photodissociation cross sections.
\mbox{\textit{Top panel}}: $\sigma_{\rm ion}$(SH)
\citep[blue curve from laboratory experiments by][]{Hrodmarsson19}. 
The pink curve is  $\sigma_{\rm diss}$(SH) \citep[][and references therein]{Heays17}.
\textit{Bottom panel}: $\sigma_{\rm ion}$(H$_2$S) (blue curve) and  $\sigma_{\rm diss}$(H$_2$S) \citep[gray and pink curves; from][]{Zhou20}.}
\label{fig:photo-cross-sect}
\end{figure}

\section{H$_2$S ortho-to-para ratio and $T_{\rm spin}$}

The OTP ratio is sometimes related to a nuclear-spin-temperature
 \citep[$T_{\rm spin}$, e.g.,][]{Mumma87} defined, for  H$_2$O or H$_2$S, as:

\begin{equation}\label{eq-OTP}
{{\rm{OTP}} =  \frac{3\sum{(2J+1)\,{\rm {exp}}(-E_o(J)/T_{\rm spin})}}{\sum{(2J+1)\,{\rm {exp}}(-E_p(J)/T_{\rm spin})}}}.
\end{equation} 
Here, $E_o(J)$ and $E_p(J)$ are the energies (in Kelvin) of $o$-H$_2$S and
 $p$-H$_2$S rotational levels (with the two ground rotational  states separated
 by $\Delta E$\,=\,19.8\,K). Figure~\ref{fig:OTP} shows the OTP ratio of the two H$_2$S nuclear spin isomers as a function of $T_{\rm spin}$. The OTP ratio we infer toward
 the DF position of the Bar, \mbox{2.9\,$\pm$\,0.3}, is consistent with the statistical ratio of 3/1, and implies  \mbox{$T_{\rm spin}$\,$\geq$\,30\,$\pm$\,10\,K}.

\begin{figure}[h]
\centering   
\includegraphics[scale=0.35, angle=0]{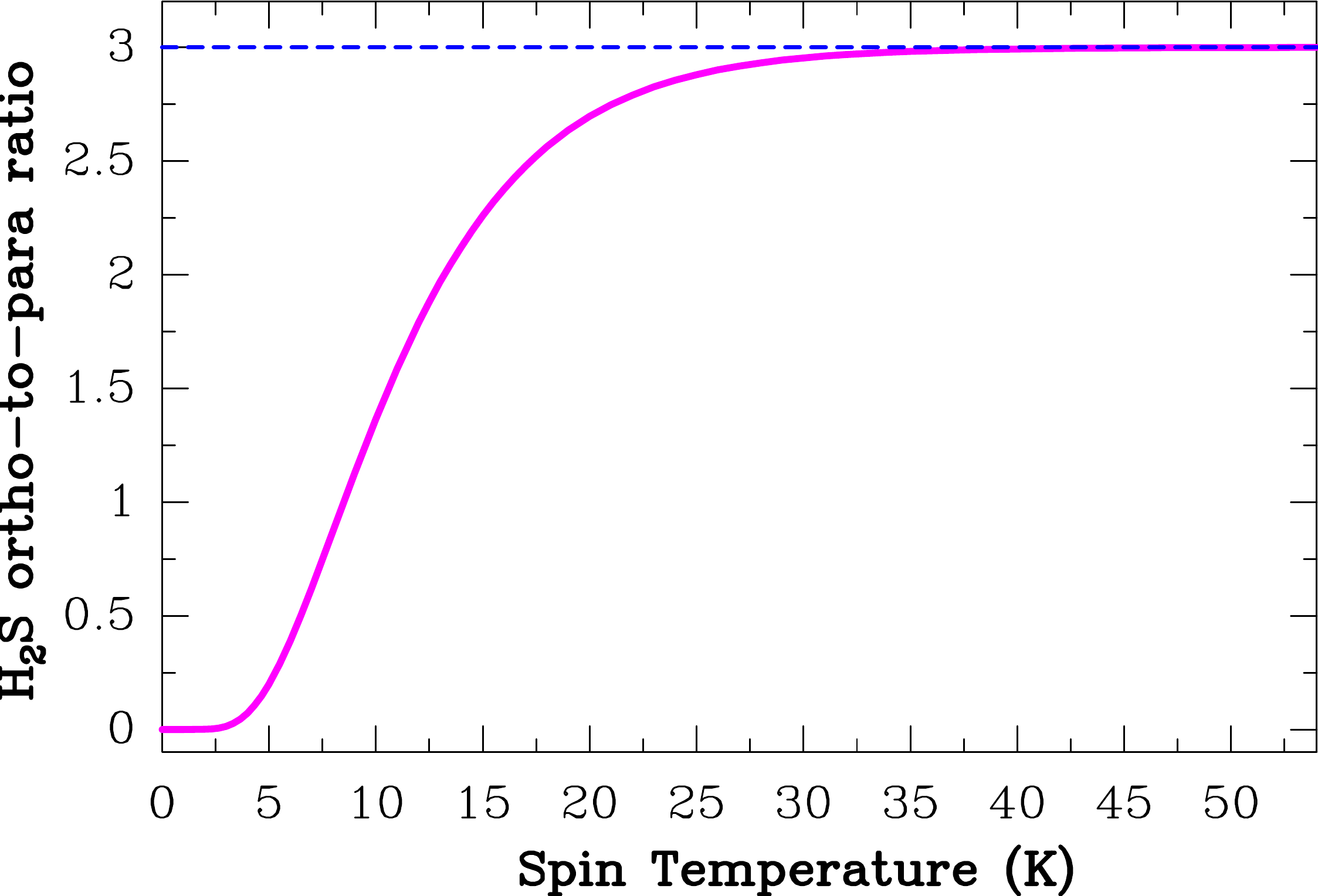}
\caption{OTP ratio of H$_2$S as a function of spin temperature (eq.~\ref{eq-OTP}).}
\label{fig:OTP}
\end{figure}

\section{Line parameters of IRAM\,30m, ALMA, and SOFIA observations}


\begin{table*}[!h] 
\begin{center}
\caption{Parameters of H$_{2}$S and H$_{2}^{34}$S lines detected with the IRAM\,30\,m telescope toward three positions of the Orion Bar.}  \label{Table_H2S}  
\resizebox{19cm}{!}{ 
\begin{tabular}{l r c c c c c c c c c c @{\vrule height 10pt depth 5pt width 0pt}}    
\hline\hline       

Position &  Species & Transition & Frequency & $E_{\rm u}$/k & $A_{\rm ul}$ &  $S_{\rm ul}$ & $g_{\rm u}$ &  $\displaystyle{\int} T_{\rm mb}$dv\,\,\,  &  v$_{\rm LSR}$ &  $\Delta$v &  $T_{\rm mb}$ \rule[-0.3cm]{0cm}{0.8cm}\ \\

  & & $J_{K_{\rm a},K_{\rm c}}$ & [GHz] & [K] & [s$^{-1}$] & & & [K km s$^{-1}$] & [km s$^{-1}$] & [km s$^{-1}$]  & [K] \\  

\hline
(+10, $-$10) & $o$-H$_{2}$S   & 1$_{1,0}$ -- 1$_{0,1}$     & 168.763 & \,\,\,8.1 & 2.68 $\times$ 10$^{-5}$ & 1.5 & 3 & 18.32\,(0.01) & 10.5\,(0.1)  & 2.5\,(0.1)     & 7.03   \\
             & $o$-H$_{2}^{34}$S  & 1$_{1,0}$ -- 1$_{0,1}$ & 167.911 & \,\,\,8.1 & 2.62 $\times$ 10$^{-5}$ & 1.5 & 3 & 1.22\,(0.01)  & 10.5\,(0.1)  &  2.0\,(0.1)    & 0.57   \\
             & $p$-H$_{2}$S   & 2$_{2,0}$ -- 2$_{1,1}$     & 216.710 & 84.0      & 4.87 $\times$ 10$^{-5}$ & 2.2 & 5 & 0.35\,(0.01)  & 10.4\,(0.1)  & 2.1\,(0.1)     & 0.16    \\
              \hline
(+30, $-$30) & $o$-H$_{2}$S       & 1$_{1,0}$ -- 1$_{0,1}$ & 168.763 & \,\,\,8.1 & 2.68 $\times$ 10$^{-5}$ & 1.5 & 3 & 17.16\,(0.02) &  10.3\,(0.1)    & 2.4\,(0.1)    & 6.85   \\
             & $o$-H$_{2}^{34}$S  & 1$_{1,0}$ -- 1$_{0,1}$ & 167.911 & \,\,\,8.1 & 2.62 $\times$ 10$^{-5}$ & 1.5 & 3 & 1.28\,(0.01)  &  10.4\,(0.1)    & 1.9\,(0.1)    & 0.63   \\
\hline
(+35, $-$55) & $o$-H$_{2}$S      & 1$_{1,0}$ -- 1$_{0,1}$ &  168.763 & \,\,\,8.1 & 2.68 $\times$ 10$^{-5}$ & 1.5 & 3 & 3.57\,(0.02)       &  \,\,\,9.6\,(0.1)  & 3.1\,(0.1)     & 1.08  \\
             & $o$-H$_{2}^{34}$S & 1$_{1,0}$ -- 1$_{0,1}$ &  167.911 & \,\,\,8.1 & 2.62 $\times$ 10$^{-5}$ & 1.5 & 3 & 0.18\,(0.02)       &  \,9.8\,(0.1)\,\,     & 2.7\,(0.3)\,\,\,  & 0.06  \\
\hline                                                                                                                                                        
                                                                           
\end{tabular} }                                                                                            
\end{center} 
\tablefoot{Parentheses indicate the uncertainty obtained by the Gaussian fitting programme.}                
\end{table*}      

\begin{table*}[!h] 
\vspace{2cm}
\begin{center}
\caption{Parameters of SH$^+$  targeted with ALMA toward the DF position.}  \label{Table_SHp}  
\resizebox{19cm}{!}{ 
\begin{tabular}{l l c c c c  c c c c @{\vrule height 10pt depth 5pt width 0pt}}    
\hline\hline       

Position &  Species & Transition & Frequency & $E_{\rm u}$/k & $A_{\rm ul}$ &  $\displaystyle{\int} T_{\rm mb}$dv\,\,\,  &  v$_{\rm LSR}$ &  $\Delta$v &  $T_{\rm mb}$  \rule[-0.3cm]{0cm}{0.8cm}\ \\

  & &  & [GHz] & [K] & [s$^{-1}$] &  [K km s$^{-1}$] & [km s$^{-1}$] & [km s$^{-1}$]  & [K]  \\  

\hline
(+10, $-$10) & SH$^+$ & $N_J$=1$_0$-0$_1$ $F$=1/2-1/2 & 345.858 & 16.6 & 1.14$\times$10$^{-4}$ & 0.36$^a$ (0.03) &  10.7 (0.2) & 2.7 (0.3)   & 0.12  \\
& SH$^+$ & $N_J$=1$_0$-0$_1$ $F$=1/2-3/2$^b$ & 345.944 & 16.6 & 2.28$\times$10$^{-4}$ & 0.70$^a$ (0.03) & 10.4 (0.1) & 2.5 (0.1)   & 0.26  \\
\hline
\end{tabular} }                                                                                            
\end{center} 
\tablefoot{$^a$Integrated over a  5$''$ aperture to increase the S/N of the  line profiles. $^b$Line integrated intensity map shown in Fig.~\ref{fig:SHp-ALMA}.}                  
\end{table*}      

\begin{table*}[!h] 
\vspace{2cm}
\begin{center}
\caption{Parameters of SH lines (neglecting HFS) targeted with SOFIA toward the DF position.}  \label{Table_SH}  
\resizebox{19cm}{!}{ 
\begin{tabular}{l l c c c c  c c c c @{\vrule height 10pt depth 5pt width 0pt}}    
\hline\hline       

Position &  Species & Transition & Frequency & $E_{\rm u}$/k & $A_{\rm ul}$ &  $\displaystyle{\int} T_{\rm mb}$dv\,\,\,  &  v$_{\rm LSR}$ &  $\Delta$v &  $T_{\rm mb}$  \rule[-0.3cm]{0cm}{0.8cm}\ \\

  & &  & [GHz] & [K] & [s$^{-1}$] &  [K km s$^{-1}$] & [km s$^{-1}$] & [km s$^{-1}$]  & [K]  \\  

\hline
(+10, $-$10) & SH     & $^2\Pi_{3/2}$ $J$=5/2$^+$--3/2$^-$ & 1382.911 & 66.4 & 4.72 $\times$ 10$^{-3}$   & $<$1.11$^a$ (0.20)  &  12.1$^a$ (0.8)  & 7.9$^a$ (1.3)  & 0.16   \\
			 & SH     & $^2\Pi_{3/2}$ $J$=5/2$^-$--3/2$^+$ & 1383.242 & 66.4 & 4.72 $\times$ 10$^{-3}$   & $<$0.34 (0.12) &  11.7 (0.5)      & 2.3 (0.8)      & 0.14   \\
\hline
\end{tabular} }                                                                                            
\end{center} 
\tablefoot{$^a$Uncertain fit.}                  
\end{table*}      

\end{appendix}

\end{document}